\documentclass[sigconf]{acmart}
\AtBeginDocument{%
  }

\usepackage{multirow}

\usepackage{subcaption}

\definecolor{col_product}{HTML}{6044e5}
\definecolor{col_brand}{HTML}{ed7014}
\definecolor{col_variety}{HTML}{078c4b}

\copyrightyear{2026}
\acmYear{2026}
\setcopyright{cc}
\setcctype{by}
\acmConference[CHI '26]{Proceedings of the 2026 CHI Conference on Human Factors in Computing Systems}{April 13--17, 2026}{Barcelona, Spain}
\acmBooktitle{Proceedings of the 2026 CHI Conference on Human Factors in Computing Systems (CHI '26), April 13--17, 2026, Barcelona, Spain}
\acmDOI{10.1145/3772318.3791309}
\acmISBN{979-8-4007-2278-3/2026/04}

\usepackage{acmart-taps}

\begin{document}

\title[Evaluating How Image Quality Affects Product Captioning with Vision-Language Models]{``It's trained by non-disabled people'': Evaluating How Image Quality Affects Product Captioning with Vision-Language Models}

%% Authors
\author{Kapil Garg}
\orcid{0000-0003-4593-4766}
\affiliation{%
	\institution{University of California, Irvine}
	\city{Irvine}
	\state{California}
	\country{USA}
	\postcode{92697}
}
\email{kapilg@uci.edu}

\author{Xinru Tang}
\orcid{0000-0001-6426-1363}
\affiliation{%
\department{Department of Informatics}
	\institution{University of California, Irvine}
	\city{Irvine}
	\state{California}
	\country{USA}
	\postcode{92697}
}
\email{xinrut1@uci.edu}

\author{Jimin Heo}
\orcid{0009-0004-4177-7083}
\affiliation{%
    \department{Computer Science}
	\institution{University of California, Irvine}
	\city{Irvine}
	\state{California}
	\country{USA}
	\postcode{92697}
}
\email{heoj4@uci.edu}

\author{Dwayne R. Morgan}
\orcid{0009-0006-0674-5662}
\affiliation{%
	\institution{University of California, Irvine}
	\city{Irvine}
	\state{California}
	\country{USA}
	\postcode{92697}
}
\email{dwaynem@uci.edu}

\author{Darren Gergle}
\orcid{0000-0003-4052-0214}
\affiliation{%
	\institution{Northwestern University}
	\streetaddress{633 Clark St}
	\city{Evanston}
	\state{Illinois}
	\country{USA}
	\postcode{60208}
}
\email{dgergle@northwestern.edu}

\author{Erik B. Sudderth}
\orcid{0000-0002-0595-9726}
\affiliation{%
    \department{Computer Science}
	\institution{University of California, Irvine}
	\city{Irvine}
	\state{California}
	\country{USA}
	\postcode{92697}
}
\email{sudderth@uci.edu}

\author{Anne Marie Piper}
\orcid{0000-0003-3085-3277}
\affiliation{%
	\institution{University of California, Irvine}
	\city{Irvine}
	\state{California}
	\country{USA}
	\postcode{92697}
}
\email{ampiper@uci.edu}

%%
%% Short authors' names
\renewcommand{\shortauthors}{Kapil Garg et al.}
%% No italics, no superscripts
%% If needed use a foot or author note to identify equal contribution

%TC:ignore
\begin{abstract}
Vision-Language Models (VLMs) are increasingly used by blind and low-vision (BLV) people to identify and understand products in their everyday lives, such as food, personal care items, and household goods. Despite their prevalence, we lack an empirical understanding of how common image quality issues---such as blur, misframing, and rotation---affect the accuracy of VLM-generated captions and whether the resulting captions meet BLV people's information needs. Based on a survey of 86 BLV participants, we develop an annotated dataset of 1,859 product images from BLV people to systematically evaluate how image quality issues affect VLM-generated captions. While the best VLM achieves 98\% accuracy on images with no quality issues, accuracy drops to 75\% overall when quality issues are present, worsening considerably as issues compound. We discuss the need for model evaluations that center on disabled people's experiences throughout the process and offer concrete recommendations for HCI and ML researchers to make VLMs more reliable for BLV people.
\end{abstract}
%TC:endignore




% teaser image
%TC:ignore
\begin{teaserfigure}
	\setkeys{Gin}{width=\linewidth}  
	\captionsetup[subfigure]{justification=centering}
	\centering
	\begin{subfigure}[t]{0.16\textwidth}
		\centering
		\caption*{\footnotesize Blur}
		\includegraphics[width=\linewidth]{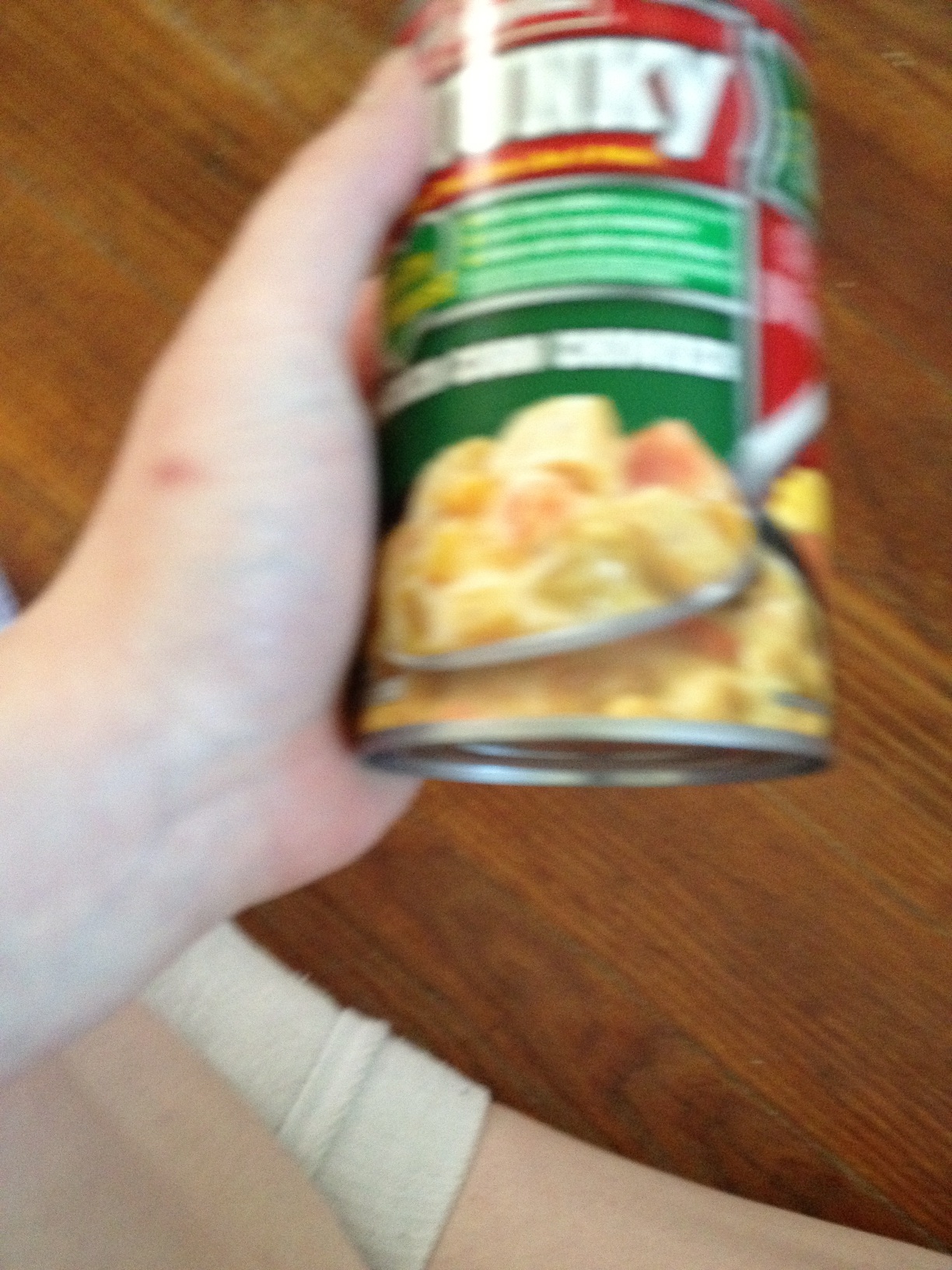}
		\caption*{\footnotesize Campbell’s Chunky Chicken Corn Chowder}
        \Description{Blurred picture of Campbell's Chunky Chicken Corn Chowder.}
	\end{subfigure}
	\hfill
	\begin{subfigure}[t]{0.16\textwidth}
		\centering
		\caption*{\footnotesize Framing}
		\includegraphics[width=\linewidth]{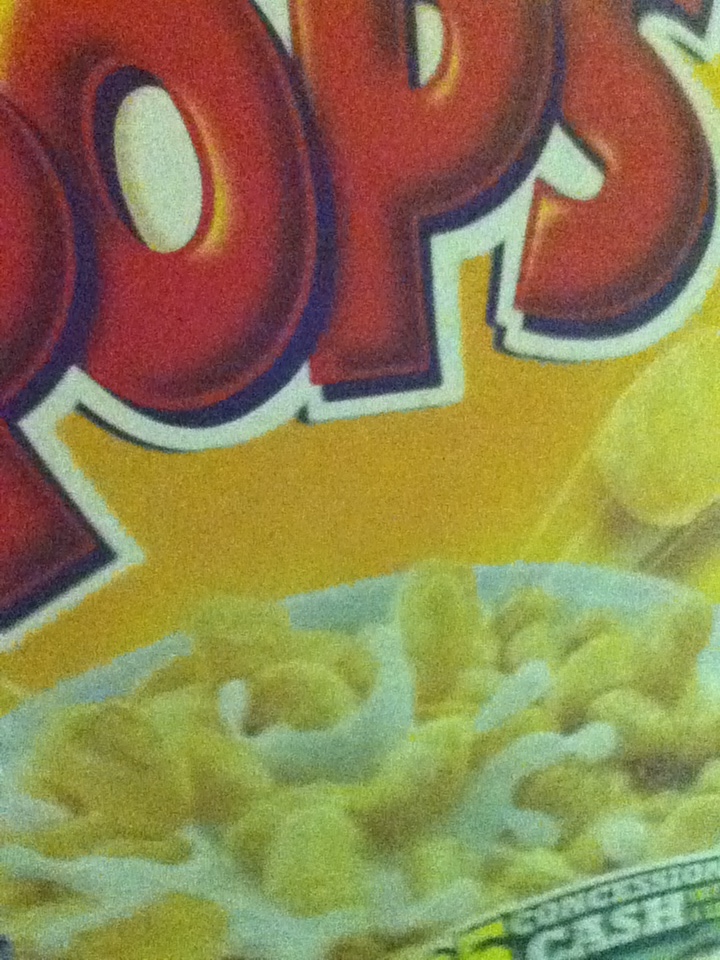}
		\caption*{\footnotesize Kellogg's Corn Pops Cereal}
        \Description{Misframed picture of a box of Kellogg's Corn Pops Cereal.}
	\end{subfigure}
	\hfill
	\begin{subfigure}[t]{0.16\textwidth}
		\centering
		\caption*{\footnotesize Blur, Rotation}
		\includegraphics[width=\linewidth]{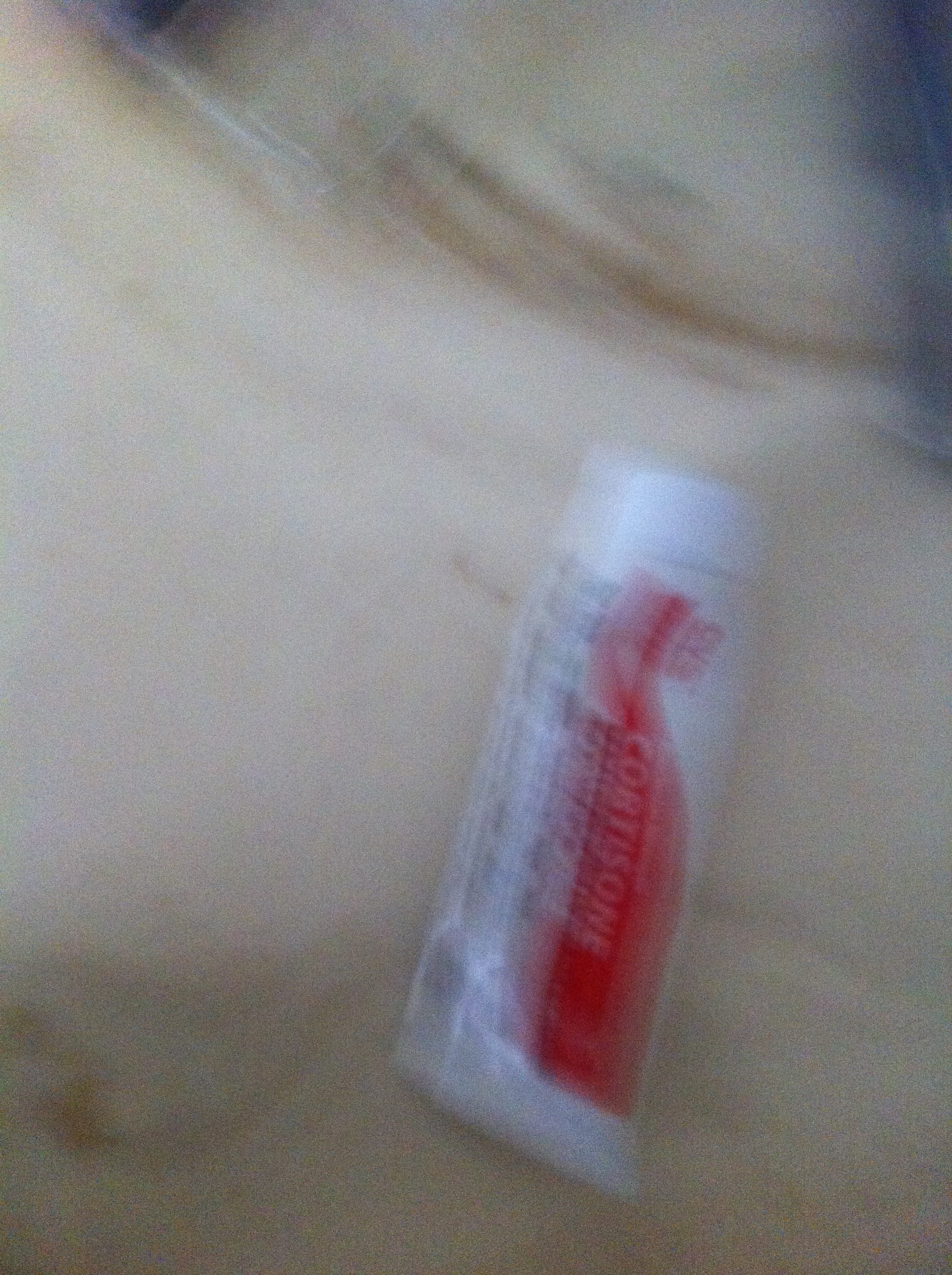}
		\caption*{\footnotesize CVS Cortizone Cream}
        \Description{Blurred and rotated picture of CVS Cortizone cream.}
	\end{subfigure}
	\hfill
	\begin{subfigure}[t]{0.16\textwidth}
		\centering
		\caption*{\footnotesize Framing, Rotation}
		\includegraphics[width=\linewidth]{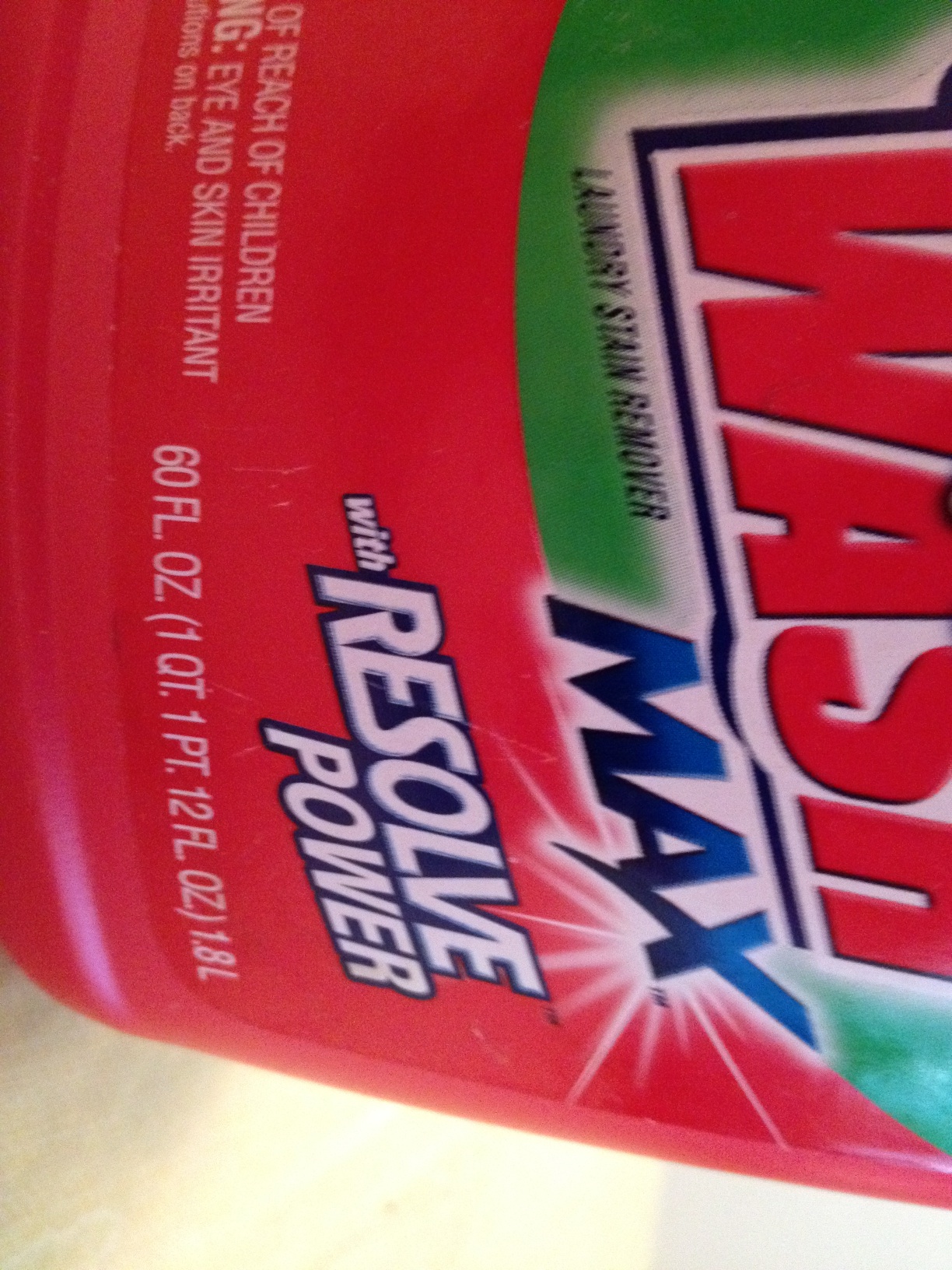}
		\caption*{\footnotesize Spray 'N Wash Max Laundry Stain Remover}
        \Description{Misframed and rotated picture of Spray ’N Wash Max Laundry Stain Remover.}
	\end{subfigure}
	\hfill
	\begin{subfigure}[t]{0.16\textwidth}
		\centering
		\caption*{\footnotesize Blur, Framing, Rotation}
		\includegraphics[width=\linewidth]{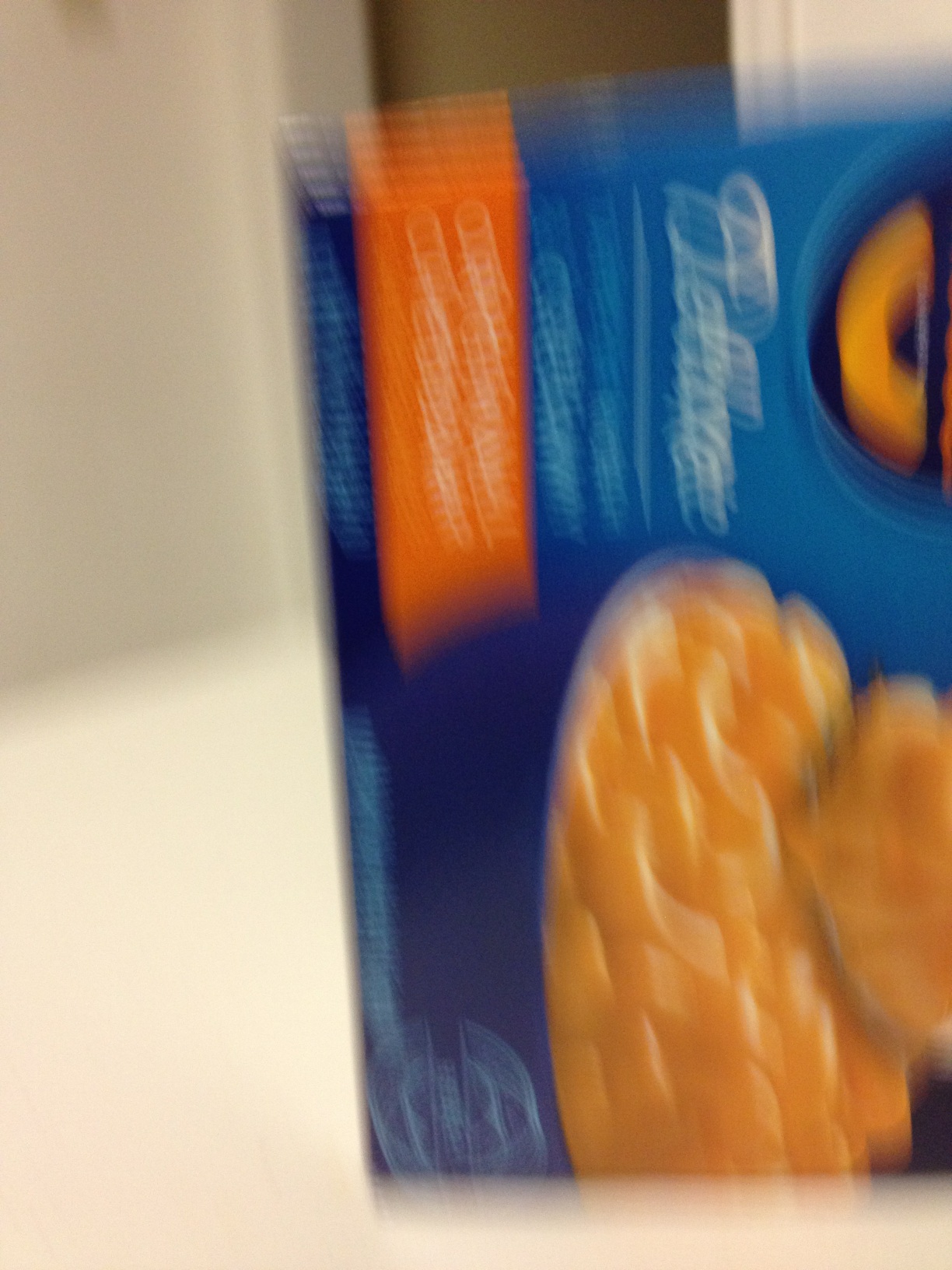}
		\caption*{\footnotesize Kraft Deluxe Mac and Cheese}
        \Description{Blurred, misframed, and rotated picture of Kraft Deluxe Mac and Cheese.}
	\end{subfigure}
	\hfill
	\begin{subfigure}[t]{0.16\textwidth}
		\centering
		\caption*{\footnotesize Blur, Framing, Rotation}
		\includegraphics[width=\linewidth]{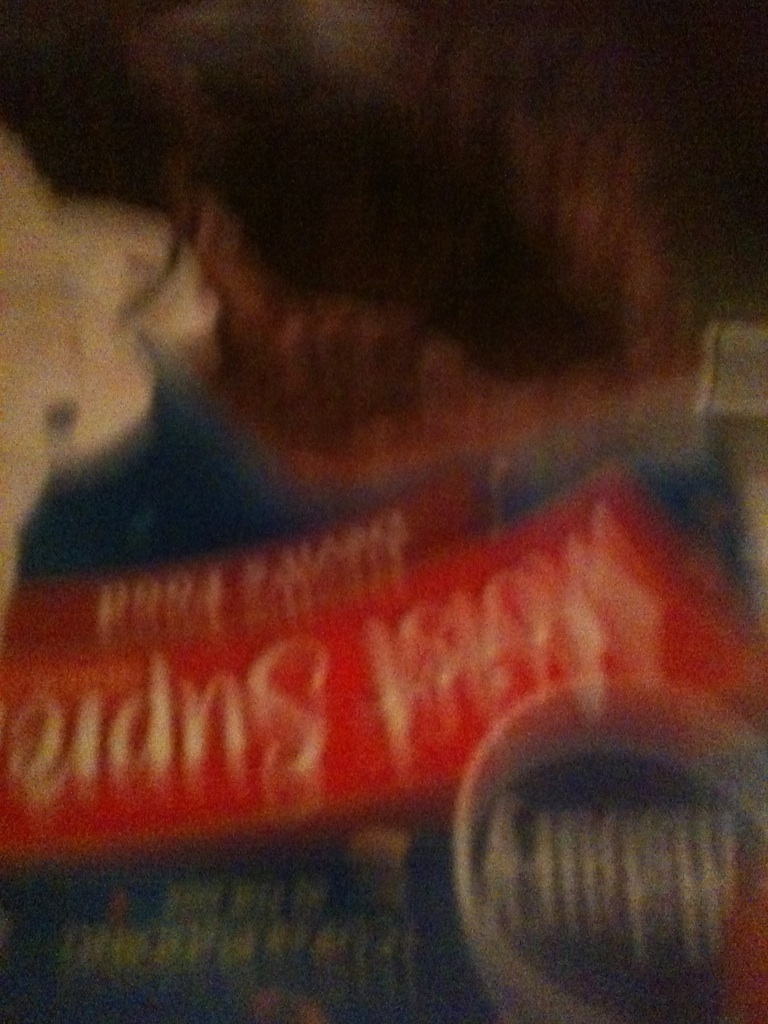}
		\caption*{\footnotesize Pillsbury Moist Supreme Devil's Food Cake Mix}
        \Description{Blurred, misframed, and rotated picture of Pillsbury Moist Supreme Devil’s Food Cake Mix.}
	\end{subfigure}
	\hfill
	\caption{Example images taken by blind and low-vision (BLV) people featuring common household products. While all of the products in these images are visually recognizable by sighted people, common image quality issues, such as blur, framing, and rotation, make them difficult for vision-language models (VLMs) to recognize. None of the VLMs tested in this study (GPT-4.1, Gemini 2.5 Flash, Llama 3.2 90B, and Molmo 72B) fully and accurately recognized the products in these images.}
	\Description[A row of six product images is displayed side by side. Each image has a different image-quality issue written above it, and a description of the product in the image below.]{Six images of products are arranged in a row, side-by-side, with text on top and bottom. The text at the top describes the photo's image quality issue, while the text at the bottom states what the product is.  In order from left to right: (1) Blurred picture of Campbell's Chunky Chicken Corn Chowder; (2) Misframed picture of a box of Kellogg's Corn Pops Cereal; (3) Blurred and rotated picture of CVS Cortizone cream; (4) Misframed and rotated picture of Spray ’N Wash Max Laundry Stain Remover; (5) Blurred, misframed, and rotated picture of Kraft Deluxe Mac and Cheese; and (6) Blurred, misframed, and rotated picture of Pillsbury Moist Supreme Devil’s Food Cake Mix.}
	\label{fig:teaser}
\end{teaserfigure}
%TC:endignore

\maketitle

% introduction
\section{Introduction}
Blind and low-vision (BLV) people regularly use automated (e.g., Microsoft Seeing AI, Be My AI, TapTapSee) and human-powered (e.g., Aira, Be My Eyes) tools to understand visual information \cite{tang2025every, alharbi2024misfitting, lee2020emerging, xie2025visual}. While Vision-Language Model (VLM)-based\footnote{We use ``VLMs'' to refer to tools like ChatGPT or Gemini that integrate vision-language models and are colloquially known as ``AI''. We use ``AI'' in the survey study (Section~\ref{sec:study-1}) since participants may be unfamiliar with ``VLMs'' versus ``AI''.} image captioning research has focused on many types of content (e.g., social media photos, scenes, objects) \cite{macleod2017understanding, mohanbabu2024contextaware, chang2024worldscribe, gonzalezpenuela2024investigating}, one widely-studied use case is to support BLV people in identifying products, such as packaged foods and household goods. As such, AI tools and their underlying VLMs are becoming more integral to how BLV people perform a variety of everyday tasks, including grocery shopping, cooking, cleaning, and personal care \cite{tang2025every, xie2025visual, alharbi2024misfitting}. Yet, we know little about the real-world experiences of BLV people using these tools for product identification or how well VLMs accurately identify products in naturalistic images, where objects of interest may be blurry, out of frame, or rotated.

Despite enthusiasm for VLM-based captioning tools in identifying and understanding products, three challenges complicate their real-world use and evaluation. First, extensive prior work has studied and introduced VLM-based captioning tools to help BLV people understand objects and products in their environment \cite{xie2025visual, tang2025every, alharbi2024misfitting}. However, we know less about the factors (e.g., privacy, accuracy, safety) that shape their decision to turn to automated systems rather than humans, and about their experience with captioning errors using existing tools (e.g., Be My AI, Seeing AI). Second, VLM-based image captioning tools perform best when BLV users take and upload high-quality photos, a known challenge for BLV people \cite{chiu2020assessing, gurari2018vizwiz, davis2020quality}. Prior work identifies various image quality issues (e.g., blur, rotation, framing, lighting) \cite{gurari2018vizwiz, davis2020quality}, automatically detects such distortions \cite{chiu2020assessing}, and introduces techniques to help BLV people take better photos \cite{theodorou2021disability, sharma2023disabilityfirst, hong2022blind, morrison2023understanding, lee2019revisiting, vazquez2014assisted, jayant2011supporting, ahmetovic2020recog}. However, limited prior work has examined how BLV people assess image quality issues with existing VLM tools and perceive their impact on captions \cite{hong2024understanding}. Third, interview studies with BLV people indicate that pervasive image quality issues affect whether images are captioned accurately \cite{zhao2018face, alharbi2024misfitting}; however, the relationship between image quality factors and the accuracy of resulting product captions has yet to be systematically analyzed. Prior datasets examine the prevalence of image quality issues, but evaluation of these issues remains coarse (e.g., determining whether an image is captionable or not) \cite{gurari2018vizwiz, chiu2020assessing, davis2020quality}. Moreover, existing evaluation approaches for image captions focus on how well a generated caption aligns with a reference text. This can result in false positives, where a caption appears reasonable even when it contains serious errors or omits critical information. Understanding how pervasive image quality issues affect the captions generated by state-of-the-art VLMs is critical, given that these models are used in a wide range of assistive technologies and research prototypes \cite{chang2024worldscribe, huh2023genassist, mohanbabu2024contextaware, chang2024editscribe, herskovitz2024programally, van2024making}.

%(e.g., mislabeling a key medicine ingredient). %Moreover, reference-based and reference-free metrics for evaluating caption quality may not work well with low-quality images, as they assess alignment between captions and may mask critical errors when identifying products. These measures could suggest a ``good'' result even when captions contain serious errors.

To help bridge these gaps in the literature, this paper examines challenges in using VLM tools to identify and understand products through two complementary efforts. First, we report results from a survey of 86 BLV participants that detail their experiences and perspectives on captioning product images with existing VLM-based tools. More than half of survey respondents emphasized using only AI tools (over human assistance) when personal privacy matters most, and roughly two-thirds said they would most often use AI when reading a food label, identifying personal care products or toiletries, and identifying an unknown item in their home. Taking a good photo remains the hardest part of the process for many participants (echoing \cite{lee2019revisiting, hong2024understanding}). Even with current tools that provide photo-taking guidance (e.g., SeeingAI, Be My AI), detecting and resolving image quality issues remains challenging. Moreover, the most frequently encountered error in product image captions is missing critical information, such as product brand names and ingredients, which can be obscured when images are of poor quality. 

Building on our survey findings, we then develop a structured, annotated dataset of 1,859 naturalistic product images (based on the VizWiz dataset \cite{gurari2020captioning, chiu2020assessing}) and use it to evaluate how robust four top-performing VLMs---GPT-4.1, Gemini 2.5 Flash, Llama 3.2 90B, and Molmo 72B---are to common image quality issues. All VLMs were proficient at product identification for high-quality images (i.e., without blur, framing, rotation, or other issues) taken by BLV people, with accuracy rates of 95\% or better for GPT and Gemini. Performance across all VLMs drops substantially for low-quality images, with the best model, GPT, achieving only 75\% accuracy. Accuracy is even lower when images have multiple image-quality issues, with GPT dropping to 69\% accuracy; see Figure~\ref{fig:teaser}. Our regression analysis confirms that all models are sensitive to image quality issues and specific content (e.g., cans with rounded labels, nutritional facts text panel) that reduce performance, and it also identifies which image quality issues specific models are more susceptible to and should be a focus for improving their performance.

This paper makes three primary contributions to the accessibility and HCI literature. First, we provide further empirical evidence of BLV people's preferences and experiences with VLM-based tools for product image captioning, underscoring the continued need for improvements in real-world product captioning applications. Second, we discuss the complexities of disability-centered approaches to model evaluation, including task and data selection, annotation procedures, and determining which models and metrics to use. Our work not only benchmarks the performance of four widely-used VLMs, which underlie many modern-day accessibility tools, but it also provides an example of how to approach the evaluation of VLMs that center on BLV people's information needs, answering prior calls to understand and address disability bias in AI models and systems (e.g.,  \cite{Gadiraju2023offensive, theodorou2021disability, silverman2025empowering, park2025autistic}). Third, we provide concrete recommendations for making VLMs more reliable for BLV people at all stages of the development pipeline, including data curation, improving model performance, and addressing captioning errors.

% background
\section{Background}
% briefly mention human approaches, and dismiss
Describing images for BLV people has been a long-standing research area in HCI. Historically, on-demand human assistance was the primary means by which BLV people accessed visual information about their environment. These include remote interpretation services, such as Aira \cite{aira}, that connect the caller with a trained visual interpreter; crowdsourcing-based systems, including VizWiz \cite{bigham2010vizwiz} and Be My Eyes \cite{be-my-eyes}, which ask a paid worker or volunteer to describe an image or video; and friends, colleagues, or family members. In the last decade, advances in computer vision have enabled machines to provide such descriptions (e.g., \cite{vinyals2015show}). For example, early versions of Seeing AI from Microsoft combined various deep learning techniques for computer vision and natural language processing to describe images \cite{linn2016decades}.\footnote{Architecture details for Seeing AI are sparse, but the original system's release date suggests it lacked attention-based mechanisms found in modern VLMs.} More recently, vision-language models (VLMs) have exploded in prevalence and capability, with many tools that support image description, like ChatGPT, Gemini, and Be My AI, all using variants of these models. Given their ubiquity, our work focuses on understanding these technologies in the context of BLV people's need for product identification, and their limitations when describing degraded images.

\subsection{How BLV People Use VLMs for Image Understanding}
Image captioning is a well-studied task in computer vision that aims to generate descriptive text for images and has led to extensive work within accessible computing \cite{gurari2020captioning,mohanbabu2024contextaware, stangl2021going, lee2022imageexplorer, macleod2017understanding}. It is often studied alongside other visual tasks such as visual question answering \cite{bigham2010vizwiz, cao2022whats}, object recognition \cite{kacorri2017people, morrison2023understanding, theodorou2021disability}, and image obfuscation \cite{alharbi2022understanding}. With the introduction of VLMs, researchers are exploring many new applications of image captioning for BLV people, such as context-aware captions for web images \cite{mohanbabu2024contextaware}, assisting with image editing \cite{chang2024editscribe}, and real-time scene interpretation of live environments \cite{gonzalezpenuela2024investigating, zhao2024vialm, chang2024worldscribe, chang2025probing}. Among these applications, object recognition is a core aspect of visual access tasks \cite{zeng2020vision} and represents a critical need among BLV people \cite{brady2013visual}. Significant efforts have been dedicated to helping BLV people identify objects \cite{gamage2023what}, including personal belongings \cite{theodorou2021disability, morrison2023understanding} and specific products \cite{hong2022blind}. 

% talk more broadly, too
More broadly, a substantial body of work has investigated how VLMs perform in object recognition. Modern VLMs are highly performant on zero-shot image identification benchmarks, such as ImageNet \cite{deng2009imagenet} and MS COCO \cite{lin2014microsoft, chen2015microsoft}, which cover a broad range of objects \cite{liu2024revisiting}. When VLMs fail, recent work suggests that failures are not due to inference-time (e.g., prompts; decoding strategies) or training-time issues (e.g., learning objective) but rather to limited data frequency for the objects the model is trying to identify \cite{zhang2024why}. Besides lacking knowledge of image content, VLMs can also fail when the input image is distorted. While significant work has studied how to {\em measure image quality issues} in photographs (e.g., \cite{yang2022maniqa, golestaneh2022noreference, agnolucci2024arniqa, fang2023sqad, ma2023modelagnostic}), relatively little has focused on the {\em impact of quality issues on captioning}. Initial studies have examined the negative impact of visual variations \cite{fan2025unveiling} and the effect of synthetic image degradation \cite{hendrycks2019benchmarking, qiu2024benchmarking} on captioning output, but the literature on systematically understanding the impact of real-world image distortions on captioning accuracy is limited.

\subsection{Understanding and Addressing Image Quality Issues}
A key issue in using VLMs for BLV people's visual needs lies in the photos they take. From analyzing VizWiz images, \citet{gurari2018vizwiz} found that blind users often struggle to take high-quality photographs, and many visual questions go unanswered because images fail to capture the relevant objects \cite{gurari2018vizwiz}. While these ``low-quality'' images are often treated as edge cases (labeled as ``other'' \cite{brady2013visual}, excluded in analysis \cite{gurari2020captioning}, or treated as a direction for future work  \cite{chang2024worldscribe}), they make up a significant portion of the photos taken by blind individuals \cite{davis2020quality}. Image quality has been identified as a major challenge in both model development \cite{gurari2020captioning} and user interactions \cite{zhao2018face}, leading to issues with annotation \cite{simons2020hope, gurari2017crowdverge, yang2018visual, bhattacharya2019why} and poor model performance \cite{zhao2018face}. For example, \citet{davis2020quality} analyzed 265 medication package images from the VizWiz dataset and found that only 46\% were legible. The prevalence of low-quality images has made image quality assessment a stand-alone task in developing image captioning tools for BLV individuals \cite{chiu2020assessing}.

% identified two image-quality–related issues from VizWiz data: (1) blind users often struggle to take high-quality photographs; and (2) many visual questions from blind users go unanswered because images fail to capture the relevant objects \cite{gurari2018vizwiz}.

 % and 40\% contained clear indicators for the information sought, 40\% had minimal background clutter, and just 5\% provided sufficient details to give a clear answer to users' questions. 

Recognizing the importance of image quality, tool designers have made considerable efforts to support BLV people in taking photos that both VLMs and humans can caption, with training and instruction playing a crucial role in data collection for model development \cite{sharma2023disabilityfirst, kacorri2017people, morrison2023understanding, theodorou2021disability}. Various techniques have been explored to improve data collection, such as using video feeds to capture objects \cite{morrison2023understanding, theodorou2021disability}, taking sequential photos of objects \cite{kacorri2017people}, and sending notifications when objects are out of frame \cite{morrison2023understanding}. While training may help, BLV users still find it hard to properly orient objects or avoid unintentionally capturing private content in the background \cite{sharma2023disabilityfirst, theodorou2021disability}. They may also be uncertain about how to fix photos, even when they know objects are poorly framed \cite{hong2022blind}. Across this literature, the emphasis is on having BLV people produce ``high-quality'' images for recognition, rather than systematically understanding how image-quality issues affect their experiences with VLMs accuracy when high-quality photos are not possible.
% The fact that BLV people may use ``garbled'' OCR output to determine photo quality \cite{hong2024understanding} suggests that current tools provide rather minimal cues about image quality issues.

\subsection{BLV People's Perspectives on AI Errors}
There is growing awareness among BLV people regarding AI tools and errors, leading to many creative and adaptive strategies to identify them \cite{alharbi2024misfitting, adnin2024look, gonzalezpenuela2024investigating, tang2025every, tang2025this}. Yet identifying errors can still be difficult for BLV people. For example, when using a prototype object recognizer to identify common food items (e.g., soda, bags of chips, canned foods), BLV participants were only able to identify half of the object recognition errors, even with successive attempts, potentially due to objects' similarity in shape and size \cite{hong2024understanding}. Moreover, most platforms provide little support for helping BLV users understand errors, such as confidence rates or multiple likely image descriptions \cite{adnin2024look, alharbi2024misfitting}. In addition, external factors, such as low-quality images or unreliable internet connectivity, often exacerbate perceived inaccuracies in image captioning \cite{zhao2018face}. When users encounter delays or fail to receive meaningful responses, they may view the system as inaccurate or untrustworthy, even if the underlying model is functioning properly \cite{zhao2018face}. As more products integrate VLMs into accessibility applications for BLV people (e.g., \cite{chang2024worldscribe, huh2023genassist}), it is critical to understand how robust they are to issues of accuracy in everyday tasks---such as identifying household products or goods---where details matter and inaccuracies can affect one's health and safety.

%Similar to the ``Not Knowing What You Don’t Know'' effect described by \citeauthor{bigham2017effects} for web accessibility \cite{bigham2017effects}, seeking visual information from VLMs can lead to a comparable experience due to the variety of errors involved. 

% studies
\section{Study 1: Understanding BLV People's Preferences, Experiences, and Challenges with AI-based Captioning of Product Images}
\label{sec:study-1}
To understand how image quality issues relate to errors during captioning, we first study BLV people's experiences using VLM-based tools to identify and understand products, such as household goods and foods. We extend prior work on how BLV people use AI tools for object recognition \cite{hong2022blind,hong2024understanding} by including the specific kinds of products, what information they are seeking, and errors that occur; the tradeoffs between using AI and human assistance based on privacy risks \cite{stangl2022privacy, stangl2023dump}, social norms \cite{tang2025every, lee2021independence}, speed, and other factors, as related to product identification; and the impact of image quality on their trust and confidence in the AI tool's output. %We focus on using VLMs to identify and understand products for two reasons: (1) this is a core use case of VLM-based accessibility tools, such as Be My AI and Seeing AI; and (2) VLMs are trained on images of products and their text, making them well-suited for providing captions of products.
% compared to general scenes and social images.

\subsection{Method}
We conducted an online survey with 86 BLV people who use AI tools for image captioning. To clarify the distinction between varying kinds of captioning support, we first asked about the general use of (1) {\em human-assistance} through remote sighted interpreting services that provide crowdsourced support (e.g., Be My Eyes) or a trained visual interpreter (e.g., Aira); (2) {\em accessibility-specific AI tools} (e.g., Microsoft Seeing AI, Be My AI, TapTapSee, Access AI); and (3) {\em general-purpose AI tools} (e.g., OpenAI's ChatGPT, Google's Gemini, Anthropic's Claude). Then, we focused on their use of AI tools to identify and understand products, which we defined as, ``packaged items and objects, such as foods, toiletries, cleaning supplies, and other household goods.'' Finally, the survey covered their preferences for using AI tools versus human alternatives, and their experiences using AI to understand products (e.g., taking photos, image quality issues, captioning errors).  %The survey then covered five sections of questions: (1) the use of AI versus human captioning across various scenarios (e.g., shopping in physical stores, identifying products in the home, reading food labels); (2) the use of AI versus human captioning when certain factors matter most (e.g., efficiency, data privacy, personal privacy, safety); (3) experiences with captioning products, comprised of open-ended questions asking about captioning mistakes and potentially harmful situations; (4) their experience with specific image quality issues (e.g., blur, rotation, framing) when using AI to caption product images and challenges with taking good photos; and (5) the perceived frequency of errors in product image captioning when using the AI tool they use most often. Finally, the survey asked about demographic information.

We revised the survey over three iterations. First, two researchers took the entire survey multiple times to check for language and length. This led to revisions to the survey structure and question wording. Then, we deployed the survey to 10 participants, including an open-ended question at the end that allowed participants to share any confusion or suggestions for improving the survey. This round resulted in two questions being removed and the rewording of others. Following these corrections, we distributed the survey to another 10 participants. No major issues were noted at this stage, and we proceeded with the final deployment. The final survey took approximately 10 minutes to complete. Complete survey questions are provided in the supplementary material.

The survey was hosted via Google Forms, which is accessible to screen reader users, and was open in March 2025. Participants were recruited through email lists maintained by the research team, as well as those of the National Federation of the Blind (NFB) \cite{nfb} and the American Foundation for the Blind (AFB) \cite{afb}. Interested participants signed up through a pre-survey screener. Eligible participants must identify as blind or low-vision, be age 18 or older, use a screen reader to access digital content, speak English, and reside in the United States. Given the focus of our study, participants must have regularly used at least one AI tool (e.g., Be My AI, Seeing AI, ChatGPT, Gemini, Claude) for image captioning. Upon confirming eligibility and excluding any bot-like responses, participants were invited to take the survey using their unique email address. Participants provided consent before beginning. Each participant received a \$20 Amazon gift card after completing the survey. The survey study was approved by our university IRB.

We received 97 survey responses, which the research team reviewed for duplicates and quality issues (e.g., spam-like responses or those lacking variation). To mitigate bot responses, we required participants to enter the email address to which the survey invitation was sent; responses with invalid email addresses were removed. In total, eleven responses were removed, resulting in a final sample of 86. More respondents in our sample identified as women ($n=58$, 67.4\%) than men ($n=24$, 27.9\%) or non-binary ($n=4$, 4.7\%). Most participants were aged 39--49 ($n=44$, 51.2\%) or 50--64 ($n=25$, 29.1\%), with smaller groups reporting age 18--29 ($n=9$, 10.5\%) and 65 or older ($n=8$, 9.3\%). Roughly 67.4\% ($n=58$) of participants identified as white, with some identifying as Asian ($n=16$, 18.6\%), Black or African American ($n=9$, 10.5\%), and/or Native American or Alaska Native or Native Hawaiian ($n=3$, 3.5\%). About 7\% ($n=6$) identified as Hispanic, Latino, or Spanish. More than 80\% ($n=72$) of our sample had earned a bachelor's degree or higher.

\subsection{Analysis}
We present descriptive statistics of our survey below. Where appropriate, we compare the experiences of BLV users and the effects of different tools or image quality issues on image captioning. We use a Mann-Whitney U Test for inferential statistics because the data being compared are ordinal (Likert scale) \cite{mann1947test}. Finally, we present excerpts of quotes from open-ended responses that provide additional context for our interpretations.

\subsection{Results}
%TC:ignore
\begin{table*}[ht]
	\centering
	\caption{Total number of survey respondents who used various accessibility-specific and general-purpose VLM-based tools for identifying products in photographs they took. Participants often used multiple tools for their visual information needs.}
	\label{tbl:tool-use}
	\begin{tabular}{@{}llr@{}}
		\toprule
		\textbf{Type}                          & \textbf{Tool}                  & \multicolumn{1}{l@{}}{\textbf{Percent of Respondents (Count)}} \\ \midrule
		\multirow{7}{*}{Accessibility-focused} & Be My AI                       & 76.7\% (66)                                                    \\
		                                       & Microsoft Seeing AI            & 69.8\% (60)                                                    \\
		                                       & AI captioning in screen reader & 51.2\% (44)                                                    \\
		                                       & Access AI                      & 30.2\% (26)                                                    \\
		                                       & TapTapSee                      & 26.7\% (23)                                                    \\
		                                       & Google Lookout                 & 11.6\% (10)                                                    \\
		                                       & WayAround                      & 7.0\% \ (6)                                                    \\ \midrule
		\multirow{6}{*}{General-purpose}       & ChatGPT                        & 38.4\% (33)                                                    \\
		                                       & Ray-Ban Meta Glasses           & 29.1\% (25)                                                    \\
		                                       & Google Gemini                  & 15.1\% (13)                                                    \\
		                                       & Microsoft Copilot              & 12.8\% (11)                                                    \\
		                                       & Claude AI                      & 4.7\% \ (4)                                                    \\
		                                       & Clarifai                       & 1.2\% \ (1)                                                    \\ \bottomrule
	\end{tabular}
    \Description{The table shows the percentage and number of blind survey participants who used different visual captioning tools. It has three columns and is organized with horizontal lines separating the header row from the two main tool categories. Tools are grouped into two categories: Accessibility-focused and General-purpose.}
\end{table*}
%TC:endignore

Most participants (76.7\%, $n=66$) reported using AI tools to identify and understand products at least weekly, and half (50.0\%, $n=43$) used remote, sighted visual interpreting applications for these purposes at least weekly. The top accessibility-focused tools used by our respondents to identify and understand products included Be My AI (76.7\%, $n=66$), Microsoft Seeing AI (69.8\%, $n=60$), and AI captioning built into screen readers (51.2\%, $n=44$); see Table~\ref{tbl:tool-use}. Among general-purpose AI tools, participants reported using ChatGPT (38.4\%, $n=33$), Ray-Ban Meta Glasses (29.1\%, $n=25$), and other tools. While we expected most users to regularly use AI tools (given our recruitment criteria), a majority of participants continue to rely on human assistance for product identification. We detail their preferences and challenges with AI tools below.

\subsubsection{AI Captioning Is Preferred for Identifying Food, Personal Products, and Items in the Home}
\begin{figure*}[t]
	\centering
	\includegraphics[width=\linewidth]{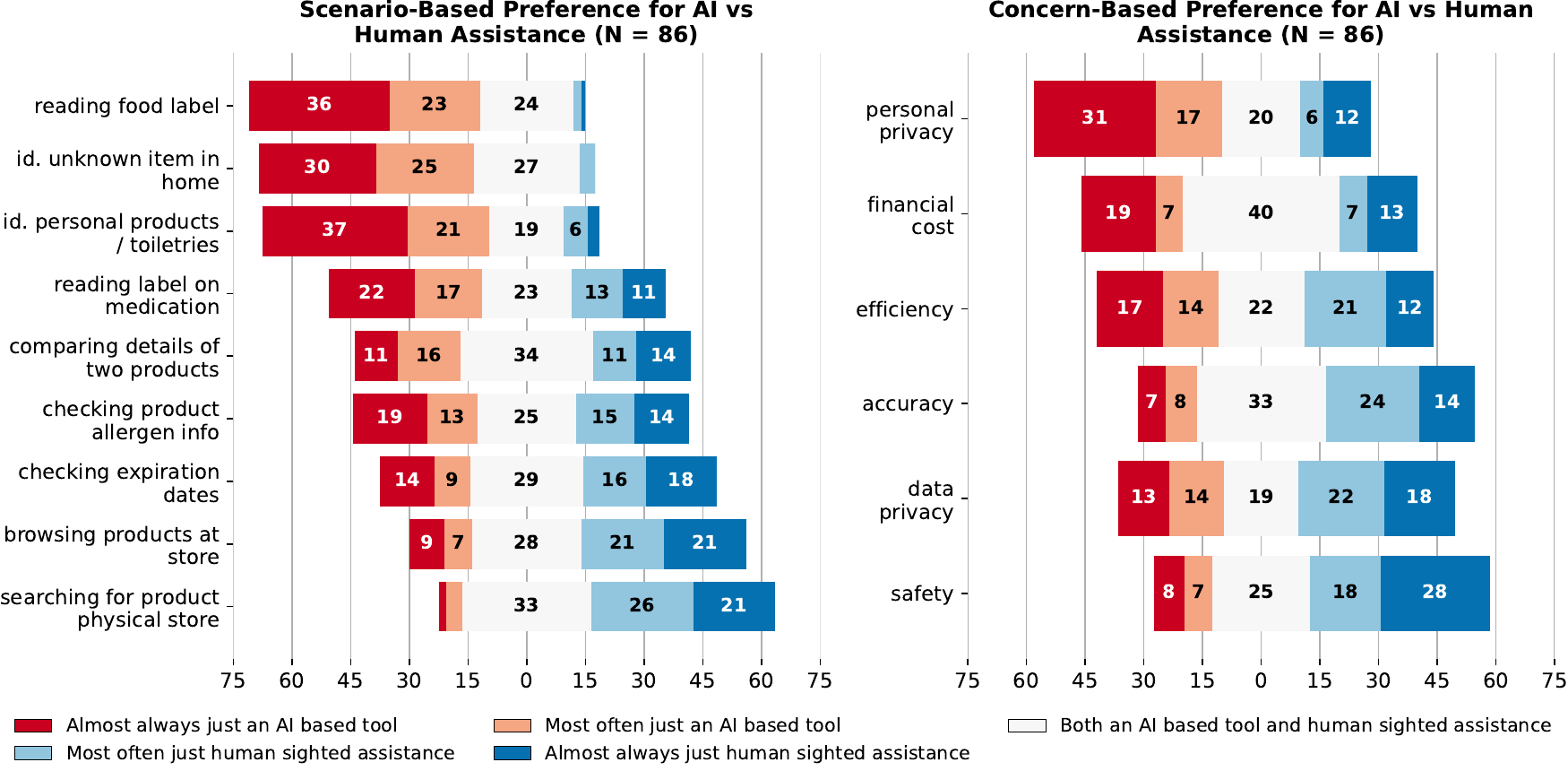}
	\caption{Divergent stacked bar charts show the distribution of reported responses for scenario-based preferences for AI vs human assistance (\emph{left}) and concern-based preferences for AI vs human assistance (\emph{right}). The x-axis shows the number of participants indicating each response. Bar labels 5 and under are hidden due to bar size.}
	\Description[Two horizontally stacked bar charts are positioned side by side. The left compares user preferences for AI versus human assistance in different scenarios, and the right compares user preferences based on concerns.]{The left chart is a divergent stacked bar chart that presents user preferences across nine scenarios, organized from top to bottom. For each scenario, a horizontal bar is divided into four colored sections, each representing the degree of preference. The top of the chart features situations where AI is more favored (e.g., reading food labels). Downwards, the chart shows a gradual shift towards scenarios where preferences become more balanced, and, towards the end, lean towards human assistance (e.g., searching for a product at a store). The horizontal axis represents the percentage of responses, ranging from -75 to 75, centered at 0. The right chart mirrors the structure of the left chart, displaying the distribution of user concerns with AI and human assistance. The concerns at the top are those most prominent with AI, with personal privacy being at the highest. Moving down the chart, the concerns are more balanced with AI and humans, and towards the end, where human assistance is deemed more important, such as safety.}
	\label{fig:prefs}
\end{figure*}

Building on previous research showing that BLV people move across human and AI assistance for access \cite{tang2025every, adnin2024look, alharbi2024misfitting}, our findings reveal their preferences and the trade-offs they consider when choosing between the two sources; see Figure~\ref{fig:prefs}, left. When considering common scenarios for product identification, roughly two-thirds of the participants said they would almost always or most often only use AI when reading a label on a food item (68.6\%, $n=59$), identifying personal care products or toiletries (67.4\%, $n=58$), and identifying an unknown item in their home (64\%, $n=55$). Surprisingly, more than 45\% of participants ($n=39$) said they would almost always or most often rely on AI to read a medication label, despite multiple AI tools issuing warnings about such use. Fewer participants said they would mainly rely on AI when checking allergen information on products (37.2\%, $n=32$), comparing the details of two products side by side (31.4\%, $n=27$), or checking product expiration dates (27.6\%, $n=23$). Although there has been prior work on object recognition when grocery shopping \cite{zhao2016cuesee, winlock2010toward, lanigan2006trinetra}, half of the participants said they leaned towards just relying on human-sighted assistance when searching for a specific product at a physical store (54.7\%, $n=47$) or browsing in a physical store (48.8\%, $n=42$). The cost of searching in a large space was a key reason for this preference, with participants explaining that in a grocery store, ``a human can often infer or already know where to go. Would take longer with just AI.'' %and humans are better when ``there are too many choices, like searching for a particular spice in my spice rack, as I have at least 40 different jars.''

Echoing prior work that highlights concerns about social norms \cite{stangl2022privacy, stangl2023dump}, more than half of the participants (55.8\%, $n=48$) said they would most often or almost always only use AI to caption products when personal privacy matters most; see Figure~\ref{fig:prefs}, right. The most cited concerns include feeling embarrassed discussing personal matters with real people and the potential misuse of their personal information. In contrast, more participants indicated they would most often rely on human assistance when data privacy was most important (46.5\%, $n=40$). Several people noted the dilemma between personal and data privacy, saying, ``it’s a catch-22: go with AI-generated [services] where they store a photo, or a person who could be copying down my information,'' which led to varying priorities regarding the associated risks. People were seen as a greater direct risk due to the potential for bad human actors (e.g., ``If I ask a human, someone will know''), while AI tools presented a broader indirect risk (e.g., ``A human has a limited number of people they could potentially share the information with, but AI means more companies can access your data''). Like data privacy, most participants leaned towards using human-sighted assistance when safety (53.5\%, $n=46$) and accuracy (44.2\%, $n=38$) mattered most, as humans were perceived as more reliable, especially when there was a clear and specific need, such as counting or reading text.
% Preferences regarding efficiency and cost were mixed.

% scenario-based vs concern-based preferences

\subsubsection{Taking Photos Remains Time-Consuming and Challenging}
Despite research on supporting BLV people to take photos \cite{lee2019revisiting, jayant2011supporting, vazquez2014assisted, ahmetovic2020recog}, it remains a key challenge for image captioning. With the tool they used most, nearly half of the participants (47.7\%, $n=41$) said it took 2--4 minutes to get the desired information, followed by 0--1 minutes (27.9\%, $n=24$) or 5--9 minutes (19.8\%, $n=17$). Two-thirds of participants (67.1\%, $n=49$) said taking a good photo was the hardest part of the captioning process, and roughly half (45.9\%, $n=34$) said it took the longest. Multiple photos were often needed, with most participants (62.8\%, $n=54$) saying 2--3 photos; fewer needed just one photo (23.3\%, $n=20$) or more than four photos (10.5\%, $n=9$). For some participants, taking photos was difficult due to physical disabilities that made it hard to hold the camera steady. Participants described learning to take photos over time, including learning from how Aira interpreters guide them to angle their camera and adjust the environment for visual captioning.

% In contrast, others said the longest part was waiting for the AI tool to respond (23\%, n=17), asking questions if details were missing (13.5\%, n=10), reading through the AI tool's response (9.5\%, n=7), or submitting the photo and question to an AI tool (8.1\%, n=6). Most participants (67.1\%, n=49) rated taking a good photo as the hardest part of the process, followed by asking questions if details are missing (12.3\%, n=9) or waiting for the AI tool to respond (11\%, n=8). When asked how many photos they usually need to take to get the information they want, most participants (62.8\%, n=54) said 2-3 photos. Others said one photo (23.3\%, n=20) or more than four photos (10.5\%, n=9). Some participants described individual differences that affected the photo-taking experience, such as being blind and having a physical disability that made it difficult to hold the camera steady. Many suggested features to get better results, such as being able to submit multiple photos at once and having continuous voice feedback guiding their photo taking. Participants described learning to take photos over time, including learning from how Aira interpreters guide them to angle their camera and adjust the environment for visual captioning. Synthesizing feedback from multiple participants suggests a multi-pronged approach that would let the user know before taking the photo if the lighting conditions are ideal, provide continuous feedback about how to position the camera while taking the photo, and then afterward explain what went wrong and how to improve it.

\subsubsection{Current Tools Make It Difficult to Assess and Resolve Image Quality Issues}
\begin{figure*}[t]
	\centering
	\includegraphics[width=\linewidth]{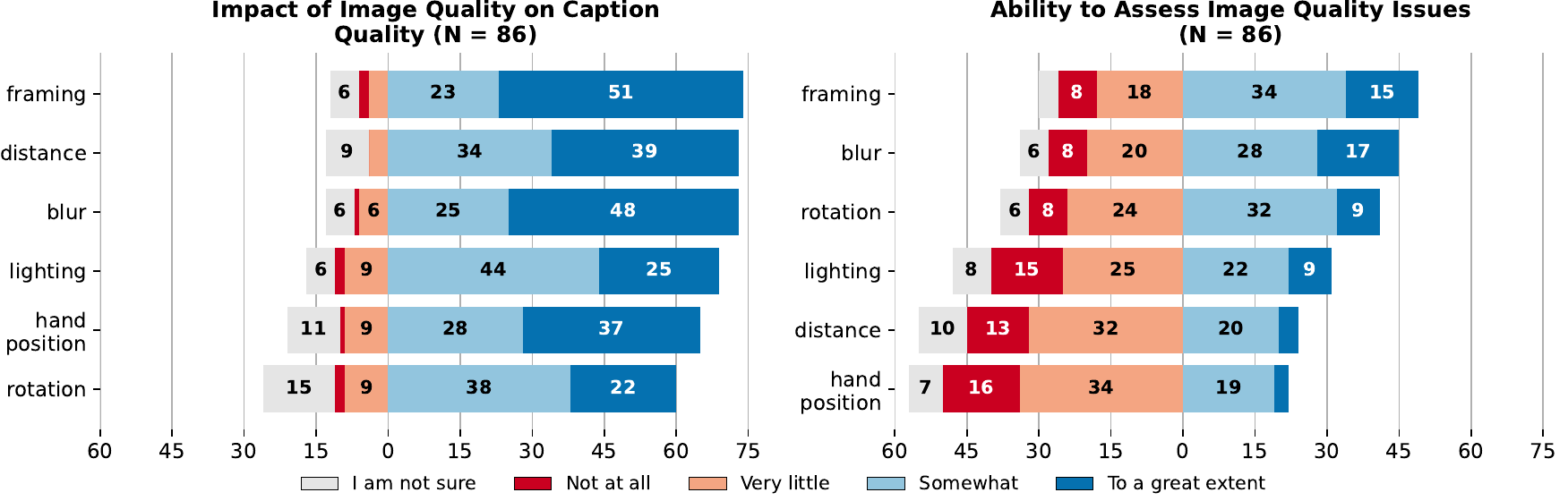}
	\caption{Divergent stacked bar charts show the distribution of reported responses for perceived impact of image quality issues on caption quality (\emph{left}) and the ability to assess a quality issue in their image (\emph{right}). The x-axis shows the number of participants indicating each response. Bar labels 5 and under are hidden due to bar size.}
	\label{fig:img-qual-conf}
	\Description[Two horizontally stacked bar charts are positioned side by side. The left chart compares the perceived impact of different image quality issues on the quality of generated captions. The right chart compares the ability to assess image quality issues when a photo is taken using the AI captioning tool.]{The left chart is a horizontal stacked bar chart displaying the perceived impact of six image quality issues on caption quality: framing, distance, blur, lighting, hand position, and rotation. The issues are listed from top to bottom on the vertical axis. For each image quality issue, a horizontal bar is segmented into five colored sections representing the degree of impact on caption quality (“I am not sure”, “Not at all”, “Very little”, “Somewhat”, “To a great extent”). Regarding all six issues, most respondents indicated that they had at least a 'somewhat' significant impact on caption quality. Hand position and rotation had the most cases where the participant was unsure whether the issue affected caption quality. The horizontal axis represents the percentage of responses, ranging from -60 to 75, centered at 0. The right chart is a horizontal stacked bar chart displaying the respondents’ ability to assess image quality issues using a selected captioning tool. The issues are listed from top to bottom on the vertical axis. For each image quality issue, a horizontal bar is segmented into five colored sections representing the degree of impact on caption quality. Regarding the top three issues–framing, blur, and rotation–respondents indicated the tools helped more, with more than half saying “Somewhat” or “to a great extent”, For the bottom three issues–lighting, distance, and hand position–the trend was reversed, with more than half saying “not at all” or “very little”. Distance and lighting were the factors most often causing uncertainty about whether the tool helped the participant assess that the factor was an issue in their picture.}
\end{figure*}

Difficulties during photo-taking can result in lower-quality photos, which then affect a VLM's caption quality. We asked participants about their perceived impact of image quality issues on the quality of AI-generated captions for products, on a 4-point scale from 1: ``not at all'' to 4: ``to a great extent'' with the option of ``I am not sure''; see Figure~\ref{fig:img-qual-conf}, left. Overall, BLV users perceived image quality issues of framing ($m=3.54$, $s=0.71$), blur ($m=3.5$, $s=0.69$), and distance to object ($m=3.45$, $s=0.6$) to affect caption quality the most, followed by hand placement and position ($m=3.35$, $s=0.74$), lighting ($m=3.15$, $s=0.71$), and rotation ($m=3.13$, $s=0.74$). A few respondents indicated ``I am not sure'', most often for rotation ($n=15$), hand position ($n=11$), and distance ($n=9$), suggesting that the impact of these might be more subtle than that of other quality issues. We also examined the differences between Seeing AI and Be My AI (but not other tools, due to the limited sample size). We found that framing was the only image quality issue whose perceived impact on caption quality was different across tools, being more impactful for Seeing AI than Be My AI ($m_{\text{Seeing AI}} = 3.79$ vs. $m_{\text{Be My AI}} =  3.39$; $p = 0.0076$, $U = 529.5$; $n_{\text{Seeing AI}} = n_{\text{Be My AI}} = 28$). This is not surprising given that Seeing AI has a feature specifically designed to support framing, which we discuss below.

Given the known challenges of taking good photos, multiple AI captioning tools include features to help BLV users understand image quality issues and adjust their camera position during captioning. We asked how well these tools helped participants assess image quality issues, using the same scale as the impact of image quality; see Figure~\ref{fig:img-qual-conf}, right. BLV people found framing ($m=2.75$, $s=0.90$), blur ($m=2.74$, $s=0.94$), and rotation ($m=2.58$, $s=0.85$) as the quality issues the tools helped them assess the best, followed by lighting ($m=2.35$, $s=0.96$), distance ($m=2.22$, $s=0.82$), and hand position ($m=2.12$, $s=0.80$). A few participants were unsure when asked whether the tools helped assess quality issues in their photographs, with the most common related to distance ($n=10$), lighting ($n=8$), and hand position ($n=7$), suggesting that the tools provide less support in addressing these issues when taking photos. We observed a significant difference between Seeing AI and Be My AI in how well they help assess whether an image is blurry (Be My AI more than Seeing AI; $m_{\text{Seeing AI}} = 2.27$ vs. $m_{\text{Be My AI}} =  2.92$; $p = 0.0147$, $U = 191.5$; $n_{\text{Seeing AI}} = 26, n_{\text{Be My AI}} = 24$), or if the product is obscured by hand positioning (Be My AI more than Seeing AI; $m_{\text{Seeing AI}} = 1.81$ vs. $m_{\text{Be My AI}} =  2.26$; $p = 0.0297$, $U = 198.5$; $n_{\text{Seeing AI}} = 26, n_{\text{Be My AI}} = 23$). While an in-depth analysis of why users perceive greater support for these two aspects is beyond the scope of the present paper, we present detailed user feedback below and note that Be My AI specifically instructs users to ask the system questions about whether an object is centered and focused \cite{be-my-ai-feedback}. Notably, the average scores for assessing each quality issue range from ``Very Little'' and ``Somewhat'', suggesting that both tools could do more to make quality issues apparent to BLV people. 

% Impact of Image Quality on Caption Quality + Confidence plot

BLV people's open-ended responses suggest that the built-in features for assessing image quality issues are only partially effective. Seeing AI, for example, emits beeps to help users move an object or product barcode into the camera's view. When asked about this feature, 14 people shared positive comments (e.g., ``Does a great job of letting me know when the object is in full view'' and ``This really helps me when adjusting the angle of the camera and increases my confidence''). However, 27 people shared negative or mixed experiences with this feature, mentioning that it is not always accurate regarding alignment, it is hard to rotate objects to find the barcode, making slight adjustments and holding one's hand steady is problematic, and the feedback can be misleading (e.g., forcing the camera to put a whole object in view when only a small portion is of interest). One person said, ``It's a game of hot and cold: it takes some trial and error every time to get it right, unless you have a good sense of where the barcode is.'' The remainder stated that they had not used this feature ($n=21$) or did not answer the question.

%``I found it more frustrating than helpful...The beeps don't really help other than I know it's not in the right spot or I'm getting closer, but it's still too challenging and time-consuming.'' Similarly, another said,  

%Multiple people said they would benefit from more guidance and detailed feedback (e.g., ``I wish it had more precise feedback on camera positioning''). 

While Be My AI offered more detailed feedback on photos, it also received mixed responses to its suggestions (e.g., asking users to take a new picture, contact a volunteer, or ask questions such as whether the photo is out of focus \cite{be-my-ai-feedback}). Of the 47 people who commented on the feedback feature, 26 had positive experiences. Others ($n=21$) provided mixed or negative comments, often noting that the feedback was limited, lacked clear guidance on resolving issues, and still required trial and error. One person said, ``It's good to know the photo is not clear enough, but tough to figure out sometimes if it's a lighting, placement, or angle issue.'' Another commented, ``It's really just overall not helpful...the devs still really don't get it. It's not enough to just say the photo's not of good quality; you have to tell someone how to fix it. Many of us have been blind since birth, and how to deal with photos completely escapes us.'' What's more, feedback was only given after the photo was taken, with multiple participants suggesting that the tool provide more detailed, real-time feedback on framing, lighting, and orientation.

%, including whether the object is in frame, if the lighting is suitable, and offering an orientation guide. 
%Additionally, some highlighted the limitations of providing feedback only after the photo is taken, noting the difficulty of ``not knowing what the final photo will turn out as before it is taken.''  Many suggested providing more detailed real-time feedback before taking the photo, including whether the object is in frame, if the lighting is suitable, and offering an orientation guide.
%Some users also mentioned auto-focus as a potentially useful feature. Although some of these features are available in specific applications (e.g., the document feature in Seeing AI alerts users if their documents are out of frame before auto-capturing), they have not yet become the standard.

Finally, we asked participants to rate their confidence in knowing {\em why} a photo is not good enough, even when the tool says it is not good enough to caption or returns a similar error. On average, participants rated themselves as between ``slightly confident'' and ``somewhat confident'' ($m=2.39$, $s=0.97$). Only six participants said they were ``very confident'' or ``extremely confident''  in knowing why an image was not good enough. There were no significant differences in confidence ratings between Be My AI and Seeing AI.

\subsubsection{Captions Frequently Lack Important Detail and Contain Inaccurate Information}
% frequency of errors
\begin{figure*}[ht]
	\includegraphics[width=.75\textwidth]{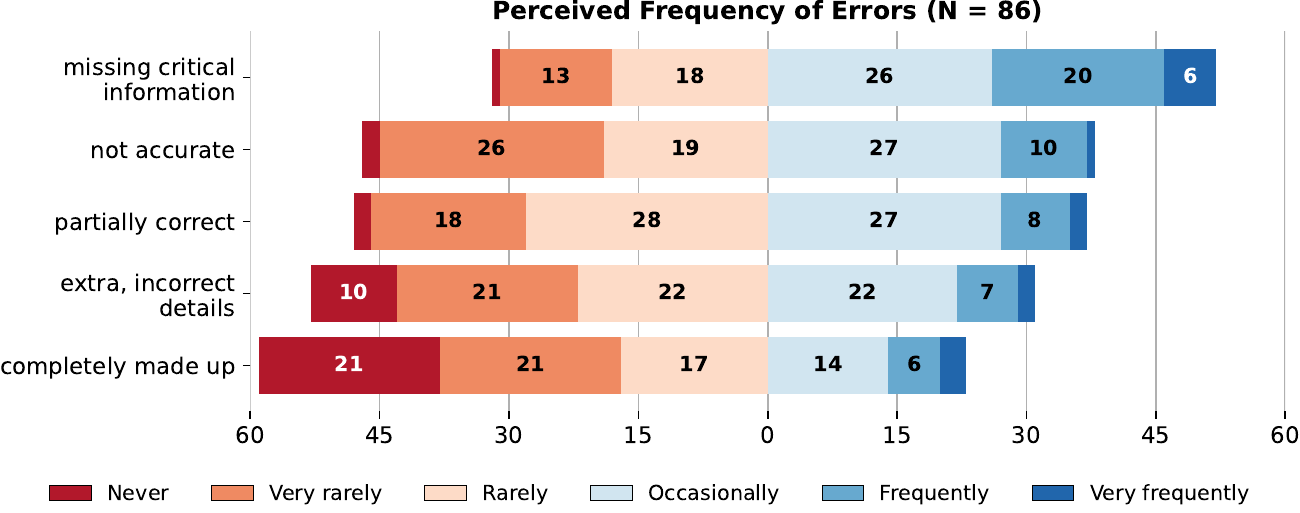}
	\caption{Distribution of perceived frequency of various types of errors in image captions when describing products. The x-axis shows the number of participants indicating each response. Bar labels 5 and under are hidden due to bar size.}
	\label{fig:verify-errors}
	\Description[One Horizontal stacked bar chart. The chart displays the perceived frequency of different types of errors.]{The horizontal stacked bar chart shows the perceived frequency of five error types, arranged vertically from those leaning towards 'very frequently' at the top, down to 'never' at the bottom. For each error type, a horizontal bar is segmented into six colored sections representing the perceived frequency of that error. From top to bottom, the kinds of errors are when the caption is missing critical information, not accurate, partially correct, has extra, incorrect details, or is completely made up. The horizontal axis represents the percentage of responses, ranging from -60 to 60, centered at 0. For each error type, the segments show the distribution of user responses, indicating how frequently they perceived that type of error to occur.}
\end{figure*}

Although prior work indicates BLV people expect error-prone output from AI tools \cite{adnin2024look, tang2025every, alharbi2024misfitting}, we know less about the specific kinds of errors they experience when captioning products and their relative frequency. Given this, we asked participants how frequently they experienced various types of errors with the AI tool they used most,
on a 6-point scale from 1: ``Never'' to 6: ``Very Frequently''; see Figure~\ref{fig:verify-errors}. Overall, participants reported the highest frequency of errors involving accurate captions that are \emph{missing critical information} ($m=3.82$, $ s=1.20$). Other frequently experienced errors were captions that are  \emph{not accurate} ($m=3.24$, $s=1.12$) and product captions that are only \emph{partially correct} ($m=3.32$, $s=1.06$). They somewhat less frequently experienced captions that \emph{include extra incorrect details} ($m=3.01$, $s=1.25$) and captions that are \emph{completely made up} ($m=2.66$, $s=1.41$). There were no significant differences in perceived frequency of errors between users of Seeing AI and Be My AI. When asked how frequently they verify captioning output regarding products with a human visual interpreter or another sighted person, more than half of the participants said ``rarely'' or less ($m=3.44$, $s=1.27$). This aligns with Hong and Kacorri's findings on the overall verification frequency \cite{hong2024understanding}.

Many participants reported captions \emph{missed critical information} and lacked details they were specifically seeking, especially regarding brand names, varieties, and ingredients. Respondents said, ``I'm trying to find out the color of a lipstick I want to wear, it may capture every bit of info other than the color name, which is very frustrating,'' and ``Many times, Seeing AI does not find the exact title of my yogurt.'' Others described receiving accurate, general information but lacking needed specificity, such as ``AI just says that the product is `beans' but doesn't specify what type of beans,'' and ``I was trying to find out if I was holding a pack of pork chops or neck bones... It would only tell me it was a package of meat.''

%``I was trying to differentiate between lotions, as I ordered three different kinds... It described the bottles but lacked...information I was looking for, like the kind of lotion.''

% and  %Users needed to prompt the AI again or take additional photos to capture this information, suggesting that not obtaining critical, detailed information may relate to image quality issues. 

%Participants also expressed concerns about the level of detail in AI descriptions. For example, one person shared an example where AI could describe the appearance of cheese but would not indicate if it was moldy, noting that ``I get there’s a yellow product in there with green spots, but it can’t really tell me what the product is and sometimes the color is off.'' 

In addition to captions frequently missing critical information, participants described captions that were \emph{not accurate} (e.g., recognizing a pregnancy test as a pen, a pair of boots as food item, protein bars as stuffing mix, green beans as spark plugs) as well as \emph{partially correct}, such as getting the product type correct but the specific details wrong (e.g., garlic powder as turmeric spice, frozen shrimp as frozen chicken, agave nectar as maple syrup). Partially correct captions can be more difficult for BLV people to assess and cause potentially life-threatening issues. One person recalled that Be My Eyes AI correctly identified a lotion bottle but got the specific variety wrong, saying it, ``left a horrible white cast on my skin, which I didn't notice until someone told me.'' Another explained that canned pears were identified as canned peaches and that, while both are canned fruits, such errors could be fatal: ``My husband is very allergic to peaches, and this probably would have meant a Benadryl shot for him if he'd gotten the wrong product.'' Similarly, one person said, ``It has told me completely different names of medicines than what is printed on the packet,'' highlighting yet another case where accurate details are essential. Respondents identified these errors based on their life experience (e.g., ``it seemed like it was taking me way too long to finish the prescriptions and then I called the pharmacy to verify'' and ``It told me a name of medicine I knew I never had.''), or by asking sighted people. Either way, people became cautious about using AI in life-threatening situations and turned to people or to more tested technologies (e.g., Script Talk for medications, which uses RFID technology \cite{scripttalk2025}).

Survey respondents also provided insight into how product design and packaging affect caption accuracy, confirming prior work regarding medical product packaging \cite{davis2020quality}. Many of these errors stem from package designs that are difficult to photograph effectively, particularly those with rounded or reflective surfaces (e.g., ``It won't read...all instructions on rounded bottles like eye drop bottles,'' and ``It usually takes more time and lots of rotating the can to piece together the information I'm looking for''). 

\subsection{Final Reflections}
Near the end of the survey, we asked BLV people what they would like to communicate to researchers and developers building these tools. Respondents across the board emphasized accuracy and precision, saying ``I need accuracy and precise captions,'' ``Be more specific!'' and ``Please, please be sure your tools are accurate. Especially if people are using it for life-reliant things like medicine.'' Others emphasized frustration (e.g., ``It's really frustrating that I have to go through so many hoops just to be able to find out what's in a box or can'') and that there is more work to be done, saying developers need to ``Take time to understand the specific use cases and needs that are unique to users who are blind or low vision.'' Another suggested that many of the issues BLV people are facing with such tools are because underlying models are typically ``trained by non-disabled people, [and] show implicit bias toward disabled people.'' Their final reflections underscore the importance of evaluating how the image quality issues BLV people contend with daily---which are often set aside in research---affect whether VLMs can accurately identify products with the level of detail that BLV people need for safe and effective use.
\section{Study 2: Evaluating VLM Caption Accuracy for Product Understanding}
Given the pervasive challenges with using VLMs to identify products, we systematically examine how image quality issues affect a VLM's ability to identify them correctly and in detail.

\subsection{Challenges in Evaluating VLMs’ Product Captioning Performance}
\label{ssec:data-metric-challenge}
We initially conducted experiments using \citeauthor{gurari2020captioning}'s VizWiz Image Captioning dataset \cite{gurari2020captioning}, but encountered two challenges. First, captions from crowdworkers varied in whether they correctly identified products and the level of detail provided, making it difficult to assess whether a VLM was performing poorly or if we lacked accurate product identification data to benchmark the model against. Second, existing metrics for measuring caption quality---like BLEU \cite{papineni2001bleu}, METEOR \cite{banerjee2005meteor}, ROUGE \cite{lin2004rouge}, CIDEr \cite{vedantam2015cider}, SPICE \cite{anderson2016spice}, and BERTScore \cite{zhang2020bertscore}---primarily measure {\em text alignment} and are unreliable for evaluating {\em correctness} of product information in captions. Adding images to such measures (e.g., Vilbertscore \cite{lee2020vilbertscore}, TIGEr \cite{jiang2019tiger}, SCAN \cite{lee2018stacked}) or using reference-free measures (e.g., CLIPScore \cite{hessel2021clipscore}) can help, but degraded images can be a confound in their scores. In short, reliance on these metrics could lead to false positives, where a caption appears reasonable even when it contains serious errors or omits critical information. These challenges motivated us to develop a dataset with verified annotations to determine whether products were correctly identified in captions.

\subsection{Method}
\subsubsection{Data Selection}
To create a dataset focused on products, we start with \citeauthor{gurari2020captioning}'s VizWiz Image Captioning dataset \cite{gurari2020captioning}, which includes five crowdworker-provided image captions on photos taken by blind people, and \citeauthor{chiu2020assessing}'s VizWiz Image Quality Assessment dataset \cite{chiu2020assessing}, which includes annotations on image quality issues by five crowdworkers. Using their training dataset (23,431 images), we first filter for images for which humans can confidently provide a caption, indicating that image quality issues are not severe enough to prevent image description. We select these data by including images for which two or fewer crowdworkers indicated the image was unrecognizable (conversely, three to five crowdworkers provided a caption). Upon inspecting the dataset, we noticed that most product images included text; therefore, we only included images in which crowdworkers identified text as a heuristic for product identification. This resulted in a filtered dataset of 14,398 images (61.5\% of the original dataset).

We then created two data subsets. First, we focused on {\em high-quality} product images without quality issues, serving as a benchmark for evaluating the performance of VLMs in product identification on natural images. We selected images for which 4 or 5 crowdworkers flagged no issues, and at most 1 person flagged each image quality issue, resulting in 2,599 images (11.1\% of the original). Second, we created a dataset of {\em low-quality images}, where 4 or 5 crowdworkers flagged the image having an {\em image quality} issue (blur, rotation, framing, obstruction, being too bright, or being too dark), resulting in 5,432 images (23.2\% of the original). This dataset corresponds to images for which human captioners felt confident providing a caption, despite identifying image quality issues that could potentially hinder their accuracy.

To assess how accurately VLMs identify products, we manually reviewed all images and identified those that appeared to contain products. Four researchers reviewed all images in each subset. They excluded images that did not include products, such as nondescript boxes or pictures of rooms in the home. We excluded images of computer screenshots, currency, printed papers, books, CDs, DVDs, clothing, and unpackaged electronic devices. These were, on the whole, difficult for annotators to verify objectively, such as identifying an article of clothing or the name of a book from a page of its text.\footnote{While these are real-world cases where objects are ambiguous and valuable to identify, we require images with clear, correct, and assessable annotations to understand how VLMs fail (our focus), where more ambiguous or hard to verify examples could create a confound in our analysis.} We also excluded any images where more than one product is pictured. For the high-quality images specifically, we also excluded any product images with even mild distortions (e.g., camera blur, lens flares) to ensure the subset was free of image-quality issues. This resulted in a high-quality subset of 729 images and a low-quality subset of 1,696 images.

\subsubsection{Data Annotation}
\label{study-2:data-annotation}
Our survey results showed that BLV people want specific product information when selecting foods, medicines, and personal products. To capture how well VLMs meet these needs, we developed a three-part annotation scheme consisting of: 
\begin{itemize}
	\item Product: the generic term for the product (e.g., cereal, soup, meal, medication).
	\item Brand: any detectable brand information (e.g., Betty Crocker, Kraft, Great Value, Kellogg's). 
	\item Variety: details about the type, flavor, or variety (e.g., peanut, low sodium)
\end{itemize}
A team of four researchers manually annotated each image using this structure. When annotating a product, researchers reviewed the image and crowdworkers. If unsure, researchers also searched online for product images or noted that they were unsure about the image, so another researcher could review it. The image was excluded from the dataset if researchers were uncertain about the pictured product. For example, we excluded images that showed only product barcodes or lacked the visible details required for product verification. To ensure the validity of product annotations, a second researcher then reviewed and confirmed agreement with each image and annotation. Any discrepancies were flagged for discussion, and if no agreement was reached, the image was removed from the dataset. To enable consistency in product naming, the research team aimed for the most specific name within the product (e.g., granola instead of cereal, Sprite instead of soda) and included both brand and sub-brand names when available. When possible, we included known flavor or ingredient details (e.g., vanilla soymilk; chicken with potatoes and green beans) and details related to dietary needs or potential allergies (e.g., zero-sugar Gatorade, peanut butter granola bars). This detailed, validated labeling is distinct from coarse object and product labels in prior work \cite{gurari2018vizwiz, kacorri2017people}. This process yielded a final dataset of 1,859 images annotated with product details, comprising 729 high-quality and 1,130 low-quality images. See Appendix~\ref{appendix:detailed-dataset-stats}, Table~\ref{tab:study2-dataset} for the number of images where 0--5 crowdworkers identified an image quality issue.

During data annotation, researchers also noted {\em product properties} that may affect caption quality, including when labels were rounded (e.g., on cans, bottles) or contained large panes of text (e.g., nutrition label, back of box recipes) by double-coding if a product had one or both properties. We identified 622 (33.5\% of our dataset) products with rounded labels, 126 (6.8\%) with large text panels, and 49 (2.6\%) with both characteristics (e.g., back of a can). 

Finally, two researchers observed that while the agreement of two crowdworkers on the presence of blur or framing issues effectively captured issue quality, it did not for rotation. Therefore, the researchers recoded rotation as an orientation beyond 45 degrees from the product's natural axis---depending on the product's top and bottom (such as a can) and text orientation---marking the image as rotated if both agreed. Tables~\ref{fig:all-examples-blur-framing} and ~\ref{fig:all-examples-rotation} include selected examples of product images showcasing various image quality issues; additional examples can be found in Figure~\ref{fig:teaser} and Appendix~\ref{appendix:additional-examples}.

\subsubsection{Generating Captions From VLMs}
\label{study-2:captioning}
% model selection
We used four different VLMs to generate captions for our dataset. We include GPT-4.1 since the three most commonly used AI tools in our survey---Seeing AI \cite{beatman20246}, Be My AI \cite{bemyeyes2023introducing}, and OpenAI's ChatGPT \cite{openai2025chatgpt}---all use a GPT-4 class model from OpenAI. We include Google's Gemini 2.5 Flash, another frequently used model. Finally, we include two recently released open-source models: Llama from Meta \cite{grattafiori2024llama}\footnote{The Ray-Ban Meta Glasses, the most used general-purpose AI tool after ChatGPT, are also powered by a version of Llama \cite{meta2025introducing}} and Molmo from the Allen Institute for AI \cite{deitke2025molmo-cvpr}, which both exhibit comparable performance to closed-source industry models on benchmarks. We include open-source models since BLV people in our survey and prior work \cite{stangl2022privacy, stangl2023dump} expressed concerns about data privacy when using LLMs, which open-source models can address when run locally. Moreover, open-source models provide access to the model architecture, training regime, and, in some cases, training data (e.g., Molmo), affording greater flexibility for improving performance than closed-source models. For GPT-4.1, we used OpenAI's API and selected the~\texttt{gpt-4.1-2025-04-14} \cite{openai2025gpt41} model checkpoint for reproducibility. For Gemini 2.5 Flash, we used Google's API for ~\texttt{gemini-2.5-flash} \cite{google2025gemini}; Google does not provide a more specific model checkpoint. For Llama and Molmo, we used \texttt{Llama-3.2\-90B-Vision-Instruct} \cite{meta2024metallama} and ~\texttt{Molmo-72B-0924} \cite{allenai2024allenai} from Hugging Face, with 4-bit quantization.\footnote{We tested the smaller~\texttt{Llama-3.2-11B-Vision-Instruct} and~\texttt{Molmo-7B-D-0924} with 16-bit precision, but found that the larger, quantized models performed better while fitting within our compute limitations. Prior work suggests that performance loss is marginal with 4-bit quantization \cite{dettmers2023case,frantar2023optq}.} Llama and Molmo were run locally on two NVIDIA RTX A6000 GPUs. For all models, we set ~\texttt{temperature = 1.0} and ~\texttt{top\_p = 0.95} to balance determinism and randomness of output generation\footnote{We tested various \texttt{temperature} (0--1.0) and \texttt{top\_p} (0--1.0) settings. \texttt{temperature} had little effect on product identification quality. In contrast, \texttt{top\_p} caused more noisy captions above 0.95 (the default for our models). These settings are similar to prior work that has used VLMs for image captioning \cite{chan2023ic3, nguyen2023improving}.}, and \texttt{max\_new\_tokens = 500} for generated tokens to allow for detailed captions. Before generating captions, images were converted to PNGs with the alpha channel removed---since some VLMs perform poorly with transparent images---but no additional processing was done (e.g., blur reduction; image super-resolution). For brevity, we refer to the VLMs as ``GPT'', ``Gemini'', ``Llama'', and ``Molmo'' in the following. 

% prompt development
We instructed each VLM to caption each image with the same prompt; see Appendix~\ref{appendix:captioning-prompt}. Our prompt was inspired by prior work using VLMs to describe images for BLV people \cite{mohanbabu2024contextaware, chang2024worldscribe, huh2024longform}, and developed following best practices \cite{openai2025prompt}. We focused on prompting the VLM to identify key features, such as the object, product type, brand names, and variety details essential to understanding the product in the image while abstaining from vague language.

\subsubsection{Dataset Coding}
\label{study-2:coding}
As the final step in our dataset creation process, we manually verified the correctness of each VLM-generated caption. We performed human coding due to the issues with existing captioning metrics (see Section~\ref{ssec:data-metric-challenge}) and because LLM-as-judges---while correlating well with human judgment for simple question-answer tasks \cite{zheng2023judging}---may be falsely lenient on more open-ended tasks, like slightly incorrect product descriptions (e.g., Coke Zero versus Diet Coke) \cite{thakur2025judging}. All VLM captions were anonymized to minimize potential bias during coding (i.e., Models A, B, C, D), and any image metadata (e.g., what quality issues were present) was concealed, except for the product annotations. The order of images was also randomized to reduce any ordering effects. 

Four researchers coded the accuracy of the four VLM captions for each of the 1,859 images in our dataset (7,436 captions in total). Before coding, the research team developed a coding scheme that allowed for minor spelling mistakes and term variation (e.g., soda vs. soft drink; chips vs. crisps) but was strict on key details (e.g., brand and sub-brand; ingredients when describing food variety). Captions were marked as incorrect if there were major hallucinations (e.g., 12 ounces reported as 12-pack, for a soda can) or contained errors that changed their meaning (e.g., grilled chicken instead of fried chicken). In this way, our evaluation measures both recall (the model gets all details) and precision (what the model says is largely correct). Each researcher coded a sample of 50 randomly selected images with all VLM captions (a total of 200 captions). IRR was computed using Krippendorff's alpha, with an agreement of 0.859. Following the training period, the four researchers independently coded the remaining images, marking ones they were unsure about as ``maybe''. The team reviewed and discussed these and other challenging cases. Our final dataset is available in the supplementary materials.

\subsubsection{Analytical and Statistical Approach}
% part 1 -- product identification accuracy
The first step in our analysis was to assess the overall accuracy of VLMs for identifying products across the range of image attributes and quality issues previously identified. We computed descriptive statistics to determine how often each VLM correctly identified products across different image quality types, image quality issues, and product properties.

The next stage of our analysis applied inferential statistics to determine how different types of image degradations and product properties influence each VLM's ability to accurately identify products. We modeled this relationship over a series of logistic regressions. We began with a single model that included all images and captions for each VLM. This allowed us to assess overall patterns in how the VLMs performed with degraded images, and to make direct statistical comparisons of performance across VLMs. The model predictors included binary variables capturing image quality dimensions (blur, framing, and rotation)\footnote{We excluded the variables for obstruction, too dark, and too bright from the analysis due to extreme class imbalance and an insufficient number of true cases, which violate the assumptions required for reliable model estimation. See Table~\ref{tab:study-2-perf-by-iq}, Low-Quality, Single Issue, Row ``Other Quality Issues''.}, product image properties (rounded label, text panel), and a categorical factor representing the VLMs (GPT, Gemini, Llama, Molmo). We binned each image quality variable as either true (if 2--5 crowdworkers reported the issue) or false (if 0--1 crowdworkers did). Because image quality issues co-occur \cite{chiu2020assessing}, we included all two- and three-way interactions among image quality issues, two-way interactions between image quality issues and product properties, and two-way interactions between image quality issues and the VLM factor. 

Our final analyses fit a set of independent logistic regression models for each VLM. This allows us to more clearly delineate and assess how a given VLM's performance degrades across different image quality issues. The predictors in these models include image quality dimensions (blur, framing, and rotation) and all two- and three-way interactions among them; product properties were excluded because they led to poorer model fit.

We used the Akaike Information Criterion (AIC) metric during model development to compare candidate models and determine final model parameterizations. The AIC metric assesses the balance between model fit and complexity, penalizing models with excessive numbers of parameters to avoid overfitting. We observed no outliers in the dataset, nor evidence of multicollinearity in the final models (all variance inflation factor (VIF) scores were less than five). Statistical modeling was performed using {R} (v 4.5.2) \cite{Rproject}.

Tables~\ref{tab:study-2-log-reg-all-models} and ~\ref{tab:study-2-log-reg-per-model} present the logistic regression coefficients as logits (i.e., log-odds). In the findings below, we report these as the percentage change in the odds of correctly identifying products (i.e., $100 * (\exp{(\beta)} - 1)$) for interpretability.

\subsection{Findings}
% accuracy and image quality
\aptLtoX[graphic=no,type=html]{\begin{table*}[ht]
	\centering
	\caption{VLM accuracy for identifying products, given different image quality issues present. All models perform well on high-quality images taken by BLV people. However, accuracy drops sharply as image quality issues compound.}
	\label{tab:study-2-perf-by-iq}
	\begin{tabular}{@{}llrrrrrrrrrr@{}}
		\toprule
		\textbf{Image Type} & \textbf{Image Quality Issue}
		& \multicolumn{2}{l}{\textbf{Num. Images}}
		& \multicolumn{2}{l}{\textbf{GPT}}
		& \multicolumn{2}{l}{\textbf{Gemini}}
		& \multicolumn{2}{l}{\textbf{Llama}}
		& \multicolumn{2}{l@{}}{\textbf{Molmo}} \\ \midrule
					
		% ------ High Quality ------ 
		High-Quality & None
		  & 729  & (100.0\%) 
		  & 718  & (98.5\%)  
		  & 698  & (95.7\%)  
		  & 628  & (86.1\%)  
		  & 633  & (86.8\%)  
		\\
		\midrule
					
		Low-Quality Overall & All Issues
		  & 1130 & (100.0\%) 
		  & 846  & (74.9\%)  
		  & 810  & (71.7\%)  
		  & 498  & (44.1\%)  
		  & 408  & (36.1\%)  
		\\
		\midrule
					
		% ------ Low Quality ------
		\multirow{5}{*}{\begin{tabular}[c]{@{}l@{}}Low-Quality,\\ Single Issue\end{tabular}}
		& Blur
		  & 143  & (12.7\%)  
		  & 112  & (78.3\%)  
		  & 113  & (79.0\%)  
		  & 71   & (49.7\%)  
		  & 74   & (51.7\%)  
		\\
					
		& Framing
		  & 250  & (22.1\%)  
		  & 209  & (83.6\%)  
		  & 191  & (76.4\%)  
		  & 139  & (55.6\%)  
		  & 124  & (49.6\%)  
		\\
					
		& Rotation
		  & 55   & (4.9\%)   
		  & 49   & (89.1\%)  
		  & 48   & (87.3\%)  
		  & 39   & (70.9\%)  
		  & 19   & (34.5\%)  
		\\
					
		& Other Quality Issues
		  & 12   & (1.1\%)   
		  & 11   & (91.7\%)  
		  & 10   & (83.3\%)  
		  & 7    & (58.3\%)  
		  & 5    & (41.7\%)  
		\\
					
		\cmidrule(l){2-12}
					
		& Single Issue Total
		  & 460  & (40.7\%)  
		  & 381  & (82.8\%)  
		  & 362  & (78.7\%)  
		  & 256  & (55.7\%)  
		  & 222  & (48.3\%)  
		\\
		\midrule
					
		\multirow{6}{*}{\begin{tabular}[c]{@{}l@{}}Low-Quality,\\ Multiple Issues\end{tabular}}
		& Blur and Framing
		  & 242  & (21.4\%)  
		  & 172  & (71.1\%)  
		  & 164  & (67.8\%)  
		  & 88   & (36.4\%)  
		  & 86   & (35.5\%)  
		\\
					
		& Blur and Rotation
		  & 75   & (6.6\%)   
		  & 43   & (57.3\%)  
		  & 50   & (66.7\%)  
		  & 27   & (36.0\%)  
		  & 14   & (18.7\%)  
		\\
					
		& Framing and Rotation
		  & 146  & (12.9\%)  
		  & 113  & (77.4\%)  
		  & 103  & (70.5\%)  
		  & 69   & (47.3\%)  
		  & 41   & (28.1\%)  
		\\
					
		& Blur, Framing, and Rotation
		  & 132  & (11.7\%)  
		  & 84   & (63.6\%)  
		  & 86   & (65.2\%)  
		  & 36   & (27.3\%)  
		  & 21   & (15.9\%)  
		\\
					
		& Other Co-Occurring Issues
		  & 75   & (6.6\%)   
		  & 53   & (70.7\%)  
		  & 45   & (60.0\%)  
		  & 22   & (29.3\%)  
		  & 24   & (32.0\%)  
		\\
					
		\cmidrule(l){2-12}
					
		& Multiple Issues Total
		  & 670  & (59.3\%)  
		  & 465  & (69.4\%)  
		  & 448  & (66.9\%)  
		  & 242  & (36.1\%)  
		  & 186  & (27.8\%)  
		\\
					
		\bottomrule
	\end{tabular}
	\end{table*}}{\begin{table*}[ht]
	\centering
	\caption{VLM accuracy for identifying products, given different image quality issues present. All models perform well on high-quality images taken by BLV people. However, accuracy drops sharply as image quality issues compound.}
	\label{tab:study-2-perf-by-iq}
	\begin{tabular}{@{}llr@{\ }rr@{\ }rr@{\ }rr@{\ }rr@{\ }r@{}}
		\toprule
		\textbf{Image Type} & \textbf{Image Quality Issue}
		& \multicolumn{2}{l}{\textbf{Num. Images}}
		& \multicolumn{2}{l}{\textbf{GPT}}
		& \multicolumn{2}{l}{\textbf{Gemini}}
		& \multicolumn{2}{l}{\textbf{Llama}}
		& \multicolumn{2}{l@{}}{\textbf{Molmo}} \\ \midrule
					
		% ------ High Quality ------ 
		High-Quality & None
		  & 729  & (100.0\%) 
		  & 718  & (98.5\%)  
		  & 698  & (95.7\%)  
		  & 628  & (86.1\%)  
		  & 633  & (86.8\%)  
		\\
		\midrule
					
		Low-Quality Overall & All Issues
		  & 1130 & (100.0\%) 
		  & 846  & (74.9\%)  
		  & 810  & (71.7\%)  
		  & 498  & (44.1\%)  
		  & 408  & (36.1\%)  
		\\
		\midrule
					
		% ------ Low Quality ------
		\multirow{5}{*}{\begin{tabular}[c]{@{}l@{}}Low-Quality,\\ Single Issue\end{tabular}}
		& Blur
		  & 143  & (12.7\%)  
		  & 112  & (78.3\%)  
		  & 113  & (79.0\%)  
		  & 71   & (49.7\%)  
		  & 74   & (51.7\%)  
		\\
					
		& Framing
		  & 250  & (22.1\%)  
		  & 209  & (83.6\%)  
		  & 191  & (76.4\%)  
		  & 139  & (55.6\%)  
		  & 124  & (49.6\%)  
		\\
					
		& Rotation
		  & 55   & (4.9\%)   
		  & 49   & (89.1\%)  
		  & 48   & (87.3\%)  
		  & 39   & (70.9\%)  
		  & 19   & (34.5\%)  
		\\
					
		& Other Quality Issues
		  & 12   & (1.1\%)   
		  & 11   & (91.7\%)  
		  & 10   & (83.3\%)  
		  & 7    & (58.3\%)  
		  & 5    & (41.7\%)  
		\\
					
		\cmidrule(l){2-12}
					
		& Single Issue Total
		  & 460  & (40.7\%)  
		  & 381  & (82.8\%)  
		  & 362  & (78.7\%)  
		  & 256  & (55.7\%)  
		  & 222  & (48.3\%)  
		\\
		\midrule
					
		\multirow{6}{*}{\begin{tabular}[c]{@{}l@{}}Low-Quality,\\ Multiple Issues\end{tabular}}
		& Blur and Framing
		  & 242  & (21.4\%)  
		  & 172  & (71.1\%)  
		  & 164  & (67.8\%)  
		  & 88   & (36.4\%)  
		  & 86   & (35.5\%)  
		\\
					
		& Blur and Rotation
		  & 75   & (6.6\%)   
		  & 43   & (57.3\%)  
		  & 50   & (66.7\%)  
		  & 27   & (36.0\%)  
		  & 14   & (18.7\%)  
		\\
					
		& Framing and Rotation
		  & 146  & (12.9\%)  
		  & 113  & (77.4\%)  
		  & 103  & (70.5\%)  
		  & 69   & (47.3\%)  
		  & 41   & (28.1\%)  
		\\
					
		& Blur, Framing, and Rotation
		  & 132  & (11.7\%)  
		  & 84   & (63.6\%)  
		  & 86   & (65.2\%)  
		  & 36   & (27.3\%)  
		  & 21   & (15.9\%)  
		\\
					
		& Other Co-Occurring Issues
		  & 75   & (6.6\%)   
		  & 53   & (70.7\%)  
		  & 45   & (60.0\%)  
		  & 22   & (29.3\%)  
		  & 24   & (32.0\%)  
		\\
					
		\cmidrule(l){2-12}
					
		& Multiple Issues Total
		  & 670  & (59.3\%)  
		  & 465  & (69.4\%)  
		  & 448  & (66.9\%)  
		  & 242  & (36.1\%)  
		  & 186  & (27.8\%)  
		\\
					
		\bottomrule
        \Description{The table has seven columns and is organized with horizontal lines separating the header row, the overall results, and the results for each specific image quality issue or issue combination. The table starts with a High-Quality row that provides the aggregate performance across all 729 high-quality images. Following this are the low-quality images divided into three sections: low-quality overall (the overall performance on all 1130 low-quality images), low-quality images with a single issue (460 of the 1130 images with just one quality issue), and low-quality images with multiple issues (670 of 1130 images with more than one quality issue). The second section has a row for each quality issue and the total for that section. The third section has a row for each combination of quality issues and the total for that section.}
	\end{tabular}
	\end{table*}}

% accuracy and product properties
%TC:ignore
\aptLtoX[graphic=no,type=html]{\begin{table*}[ht]
	\centering
	\caption{VLM product identification accuracy is not always affected by rounded labels, like canned foods, or text panels, like nutrition labels. Compared to images with no rounded label or text panel, GPT, Gemini, and Llama show little to no performance loss in the rounded-label-only and text-panel-only conditions across high- and low-quality images; across all images, Molmo's performance drops when only text panels are present. Gemini, Llama, and Molmo all experience performance drops when both a rounded label and a text label are present (e.g., a nutrition label on a can) across all images.}
	\label{tab:study-2-prod-props}
	\begin{tabular}{@{}llrrrrrrrrrr@{}}
		\toprule
		\textbf{Image Type} & \textbf{Product Image Property}
		& \multicolumn{2}{l}{\textbf{Num. Images}}
		& \multicolumn{2}{l}{\textbf{GPT}}
		& \multicolumn{2}{l}{\textbf{Gemini}}
		& \multicolumn{2}{l}{\textbf{Llama}}
		& \multicolumn{2}{l@{}}{\textbf{Molmo}} \\ \midrule
								
		% ------ High Quality ------ 
		\multirow{5}{*}{High-Quality}
		& Overall (All Images)
		  & 729  & (100.0\%) 
		  & 718  & (98.5\%)  
		  & 698  & (95.7\%)  
		  & 628  & (86.1\%)  
		  & 633  & (86.8\%)  
		\\
								
		& Without Rounded Label or Text Panel
		  & 356  & (48.8\%)  
		  & 350  & (98.3\%)  
		  & 338  & (94.9\%)  
		  & 302  & (84.8\%)  
		  & 311  & (87.4\%)  
		\\
								
		& Rounded Label Only
		  & 334  & (45.8\%)  
		  & 330  & (98.8\%)  
		  & 324  & (97.0\%)  
		  & 293  & (87.7\%)  
		  & 293  & (87.7\%)  
		\\
								
		& Text Panel Only
		  & 30   & (4.1\%)   
		  & 29   & (96.7\%)  
		  & 29   & (96.7\%)  
		  & 26   & (86.7\%)  
		  & 22   & (73.3\%)  
		\\
								
		& Rounded Label and Text Panel
		  & 9    & (1.2\%)   
		  & 9    & (100.0\%) 
		  & 7    & (77.8\%)  
		  & 7    & (77.8\%)  
		  & 7    & (77.8\%)  
		\\
		\midrule
								
		% ------ Low Quality ------ 
		\multirow{5}{*}{Low-Quality}
		& Overall (All Images)
		  & 1130 & (100.0\%) 
		  & 846  & (74.9\%)  
		  & 810  & (71.7\%)  
		  & 498  & (44.1\%)  
		  & 408  & (36.1\%)  
		\\
								
		& Without Rounded Label or Text Panel
		  & 706  & (62.5\%)  
		  & 522  & (73.9\%)  
		  & 500  & (70.8\%)  
		  & 303  & (42.9\%)  
		  & 244  & (34.6\%)  
		\\
								
		& Rounded Label Only
		  & 288  & (25.5\%)  
		  & 222  & (77.1\%)  
		  & 211  & (73.3\%)  
		  & 137  & (47.6\%)  
		  & 123  & (42.7\%)  
		\\
								
		& Text Panel Only
		  & 96   & (8.5\%)   
		  & 71   & (74.0\%)  
		  & 72   & (75.0\%)  
		  & 41   & (42.7\%)  
		  & 29   & (30.2\%)  
		\\
								
		& Rounded Label and Text Panel
		  & 40   & (3.5\%)   
		  & 31   & (77.5\%)  
		  & 27   & (67.5\%)  
		  & 17   & (42.5\%)  
		  & 12   & (30.0\%)  
		\\
		\bottomrule
	\end{tabular}
    \Description{The table has seven columns and is organized with horizontal lines separating the header row and the four models’ performance. The table is divided into two rows–high-quality and low-quality images–each with 5 sub-rows of product image properties, each representing the performance for the type of images: overall performance, without rounded labels or text panels, rounded labels only, text panels only, and rounded labels and text panels. The third column, following the property column, includes the total count of images in each subgroup. The four columns following the image count column state the count correct and percentage correct for each model.}
	\end{table*}}{\begin{table*}[ht]
	\centering
	\caption{VLM product identification accuracy is not always affected by rounded labels, like canned foods, or text panels, like nutrition labels. Compared to images with no rounded label or text panel, GPT, Gemini, and Llama show little to no performance loss in the rounded-label-only and text-panel-only conditions across high- and low-quality images; across all images, Molmo's performance drops when only text panels are present. Gemini, Llama, and Molmo all experience performance drops when both a rounded label and a text label are present (e.g., a nutrition label on a can) across all images.}
	\label{tab:study-2-prod-props}
	\begin{tabular}{@{}llr@{\ }rr@{\ }rr@{\ }rr@{\ }rr@{\ }r@{}}
		\toprule
		\textbf{Image Type} & \textbf{Product Image Property}
		& \multicolumn{2}{l}{\textbf{Num. Images}}
		& \multicolumn{2}{l}{\textbf{GPT}}
		& \multicolumn{2}{l}{\textbf{Gemini}}
		& \multicolumn{2}{l}{\textbf{Llama}}
		& \multicolumn{2}{l@{}}{\textbf{Molmo}} \\ \midrule
							
		% ------ High Quality ------ 
		\multirow{5}{*}{High-Quality}
		& Overall (All Images)
		  & 729  & (100.0\%) 
		  & 718  & (98.5\%)  
		  & 698  & (95.7\%)  
		  & 628  & (86.1\%)  
		  & 633  & (86.8\%)  
		\\
							
		& Without Rounded Label or Text Panel
		  & 356  & (48.8\%)  
		  & 350  & (98.3\%)  
		  & 338  & (94.9\%)  
		  & 302  & (84.8\%)  
		  & 311  & (87.4\%)  
		\\
							
		& Rounded Label Only
		  & 334  & (45.8\%)  
		  & 330  & (98.8\%)  
		  & 324  & (97.0\%)  
		  & 293  & (87.7\%)  
		  & 293  & (87.7\%)  
		\\
							
		& Text Panel Only
		  & 30   & (4.1\%)   
		  & 29   & (96.7\%)  
		  & 29   & (96.7\%)  
		  & 26   & (86.7\%)  
		  & 22   & (73.3\%)  
		\\
							
		& Rounded Label and Text Panel
		  & 9    & (1.2\%)   
		  & 9    & (100.0\%) 
		  & 7    & (77.8\%)  
		  & 7    & (77.8\%)  
		  & 7    & (77.8\%)  
		\\
		\midrule
							
		% ------ Low Quality ------ 
		\multirow{5}{*}{Low-Quality}
		& Overall (All Images)
		  & 1130 & (100.0\%) 
		  & 846  & (74.9\%)  
		  & 810  & (71.7\%)  
		  & 498  & (44.1\%)  
		  & 408  & (36.1\%)  
		\\
							
		& Without Rounded Label or Text Panel
		  & 706  & (62.5\%)  
		  & 522  & (73.9\%)  
		  & 500  & (70.8\%)  
		  & 303  & (42.9\%)  
		  & 244  & (34.6\%)  
		\\
							
		& Rounded Label Only
		  & 288  & (25.5\%)  
		  & 222  & (77.1\%)  
		  & 211  & (73.3\%)  
		  & 137  & (47.6\%)  
		  & 123  & (42.7\%)  
		\\
							
		& Text Panel Only
		  & 96   & (8.5\%)   
		  & 71   & (74.0\%)  
		  & 72   & (75.0\%)  
		  & 41   & (42.7\%)  
		  & 29   & (30.2\%)  
		\\
							
		& Rounded Label and Text Panel
		  & 40   & (3.5\%)   
		  & 31   & (77.5\%)  
		  & 27   & (67.5\%)  
		  & 17   & (42.5\%)  
		  & 12   & (30.0\%)  
		\\
		\bottomrule
	\end{tabular}
    \Description{The table has seven columns and is organized with horizontal lines separating the header row and the four models’ performance. The table is divided into two rows–high-quality and low-quality images–each with 5 sub-rows of product image properties, each representing the performance for the type of images: overall performance, without rounded labels or text panels, rounded labels only, text panels only, and rounded labels and text panels. The third column, following the property column, includes the total count of images in each subgroup. The four columns following the image count column state the count correct and percentage correct for each model.}
	\end{table*}}
%TC:endignore

\subsubsection{VLMs Struggle to Identify Products on Low-Quality Images}
% general accuracy + why we think these issues happen
All VLMs struggled to provide accurate captions for degraded images; see Table~\ref{tab:study-2-perf-by-iq}. For high-quality images, GPT and Gemini performed well, recognizing 98.5\% and 95.7\% of products, respectively. Accuracy for open-source models was slightly less, with Llama correctly recognizing 86.1\% and Molmo 86.8\%. Performance across all VLMs dropped substantially for low-quality images, with the best model, GPT, achieving only 74.9\% accuracy. Gemini performed slightly worse than GPT (71.7\% accuracy), but Llama and Molmo fared much worse, at 44.1\% and 36.1\% accuracy, respectively.

What's more, accuracy is even worse when images have multiple distortions, with GPT dropping to 69.4\%, Gemini to 66.9\%, Llama to 36.1\%, and Molmo to 27.8\%. While identifying over two-thirds of products in images with image quality issues may not seem problematic, the stakes for misidentifying products are higher for BLV people, especially for products with health or safety-related issues. For instance, \citeauthor{davis2020quality} showed how medical packaging presents a challenging task for VLMs and is a case where knowing the correct medicine and dosage is critical \cite{davis2020quality}. Moreover, images with multiple degradations are common in our dataset, comprising nearly 60\% of all low-quality images and 36\% of the entire dataset.

% specific content that affects accuracy
Recognizing products with rounded labels is generally challenging \cite{davis2020quality}, as is identifying products from a large panel of text. However, we found that these product properties do not always affect the studied VLMs; see Table~\ref{tab:study-2-prod-props}. For high-quality images with only rounded labels or only text panels, GPT, Gemini, and Llama had little to no performance loss compared to high-quality images with neither (maximum drop of 1.6\%, for GPT on text panels only). For low-quality images, performance loss for these models was similar (maximum drop of 0.2\%, for Llama on text panels only). Molmo showed a larger drop in performance for text panels in high-quality images (87.4\% to 73.3\%) and low-quality images (34.6\% to 30.2\%). However, product images with {\em both} rounded labels and text panels had a greater impact on performance. While GPT remained unaffected, Gemini, Llama, and Molmo all dropped to 77.8\% for high-quality images and similarly for low-quality images (67.5\%, 42.5\%, and 30.0\%, respectively). We suspect that performance drops due to text panels occur because VLMs overfocus on visible text details, leading them to become conflicted between text and visual details, which in turn leads to incorrect inferences \cite{deng2025words}. Molmo did this frequently, including one instance in which it labeled a carton of ``O Organics almond milk'' as ``Horizon Organic'' because it read ``organics'', despite the carton's completely different design; see Appendix \ref{appendix:additional-examples}, Table \ref{fig:text-panels-rounded-labels}.

\subsubsection{Effects of Image Quality on Product Identification Accuracy Across VLMs}
% overall regression
\aptLtoX[graphic=no,type=html]{\begin{table}[t]
	\centering
	\caption{Logistic regression model across all images and VLMs, which shows us general challenges VLMs face when describing degraded images. The model coefficients represent logits (i.e., log-odds). p-value significant at: * 0.05; ** 0.01; *** 0.001.}
	\label{tab:study-2-log-reg-all-models}
	\begin{tabular}{@{}lrl@{}}
		\toprule
		\textbf{Independent Variable} & \multicolumn{2}{l@{}}{\textbf{Estimate}} \\
		\midrule
				
		(Intercept)                            & 3.6402  & *** \\
		Blur = True                            & -2.1414 & *** \\
		Framing = True                         & -1.8610 & *** \\
		Rotation = True                        & -1.5839 & *** \\ \midrule
				
		Rounded Label = True                   & 0.0938  &     \\
		Text Panel = True                      & -0.5707 & **  \\ \midrule
				
		Model = Gemini                         & -0.5987 & **  \\
		Model = Llama                          & -1.7674 & *** \\
		Model = Molmo                          & -1.8242 & *** \\ \midrule
				
		Blur and Framing = True                & 1.1892  & *** \\
		Blur and Rotation = True               & 0.5838  & *   \\
		Framing and Rotation = True            & 1.0371  & *** \\
		Blur, Framing, and Rotation = True     & -0.5610 & *   \\ \midrule
				
		Blur and Rounded Label = True          & 0.0202  &     \\
		Framing and Rounded Label = True       & -0.1693 &     \\
		Rotation and Rounded Label = True      & 0.5980  & **  \\ \midrule
				
		Blur and Text Panel = True             & -0.1057 &     \\
		Framing and Text Panel = True          & 0.8561  & *** \\
		Rotation and Text Panel = True         & -0.0923 &     \\ \midrule
				
		Blur = True and Model = Gemini         & 0.4816  & *   \\
		Blur = True and Model = Llama          & 0.2167  &     \\
		Blur = True and Model = Molmo          & 0.3438  &     \\ \midrule
				
		Framing = True and Model = Gemini      & -0.0436 &     \\
		Framing = True and Model = Llama       & 0.0839  &     \\
		Framing = True and Model = Molmo       & 0.0260  &     \\ \midrule
				
		Rotation = True and Model = Gemini     & 0.3053  &     \\
		Rotation = True and Model = Llama      & 0.2684  &     \\
		Rotation = True and Model = Molmo      & -0.5686 & **  \\
				
		\midrule
		Null deviance ($\text{df} = 7435$)     & 9026.8  &     \\
		Residual deviance ($\text{df} = 7408$) & 6966.6  &     \\
		AIC                                    & 7022.6  &     \\
		\bottomrule
	\end{tabular}
    \Description{The table has two columns, independent variable and estimate, and is organized with horizontal lines separating the header row, the results for each independent variable and parameter estimate, model's fit statistics. A total of 28 rows are present for different independent variables and their interaction effects. The bottom three rows detail fit statistics for the logistic regression model.}
	\end{table}}{\begin{table}[t]
	\centering
	\caption{Logistic regression model across all images and VLMs, which shows us general challenges VLMs face when describing degraded images. The model coefficients represent logits (i.e., log-odds). p-value significant at: * 0.05; ** 0.01; *** 0.001.}
	\label{tab:study-2-log-reg-all-models}
	\begin{tabular}{@{}lr@{\hspace{0.1em}}l@{}}
		\toprule
		\textbf{Independent Variable} & \multicolumn{2}{l@{}}{\textbf{Estimate}} \\
		\midrule
				
		(Intercept)                            & 3.6402  & *** \\
		Blur = True                            & -2.1414 & *** \\
		Framing = True                         & -1.8610 & *** \\
		Rotation = True                        & -1.5839 & *** \\ \midrule
				
		Rounded Label = True                   & 0.0938  &     \\
		Text Panel = True                      & -0.5707 & **  \\ \midrule
				
		Model = Gemini                         & -0.5987 & **  \\
		Model = Llama                          & -1.7674 & *** \\
		Model = Molmo                          & -1.8242 & *** \\ \midrule
				
		Blur and Framing = True                & 1.1892  & *** \\
		Blur and Rotation = True               & 0.5838  & *   \\
		Framing and Rotation = True            & 1.0371  & *** \\
		Blur, Framing, and Rotation = True     & -0.5610 & *   \\ \midrule
				
		Blur and Rounded Label = True          & 0.0202  &     \\
		Framing and Rounded Label = True       & -0.1693 &     \\
		Rotation and Rounded Label = True      & 0.5980  & **  \\ \midrule
				
		Blur and Text Panel = True             & -0.1057 &     \\
		Framing and Text Panel = True          & 0.8561  & *** \\
		Rotation and Text Panel = True         & -0.0923 &     \\ \midrule
				
		Blur = True and Model = Gemini         & 0.4816  & *   \\
		Blur = True and Model = Llama          & 0.2167  &     \\
		Blur = True and Model = Molmo          & 0.3438  &     \\ \midrule
				
		Framing = True and Model = Gemini      & -0.0436 &     \\
		Framing = True and Model = Llama       & 0.0839  &     \\
		Framing = True and Model = Molmo       & 0.0260  &     \\ \midrule
				
		Rotation = True and Model = Gemini     & 0.3053  &     \\
		Rotation = True and Model = Llama      & 0.2684  &     \\
		Rotation = True and Model = Molmo      & -0.5686 & **  \\
				
		\midrule
		Null deviance ($\text{df} = 7435$)     & 9026.8  &     \\
		Residual deviance ($\text{df} = 7408$) & 6966.6  &     \\
		AIC                                    & 7022.6  &     \\
		\bottomrule
	\end{tabular}
    \Description{The table has two columns, independent variable and estimate, and is organized with horizontal lines separating the header row, the results for each independent variable and parameter estimate, model's fit statistics. A total of 28 rows are present for different independent variables and their interaction effects. The bottom three rows detail fit statistics for the logistic regression model.}
	\end{table}}

% all examples 
%TC:ignore
\begin{table*}[ht]
	\Large
	\centering
	\caption{Examples of blurred (rows 1--2) and misframed (3--4) product images where VLMs struggle to correctly identify products. Captions had to include accurate product, brand, and variety information to be coded as correct. Captions were shortened for presentation purposes only, indicated by [...].}
	\label{fig:all-examples-blur-framing}
	\resizebox{\linewidth}{!}{%
		\begin{tabular}{@{}p{0.20\linewidth}p{0.11\linewidth}*{4}{p{0.2925\linewidth}}@{}}
			\toprule
			\textbf{Image} & \textbf{Annotation} & \textbf{GPT} & \textbf{Gemini} & \textbf{Llama} & \textbf{Molmo} \tabularnewline \midrule 
			% blur: lemonheads
			\parbox[t]{\linewidth}{\vspace{-0.75em}\includegraphics[width=\linewidth]{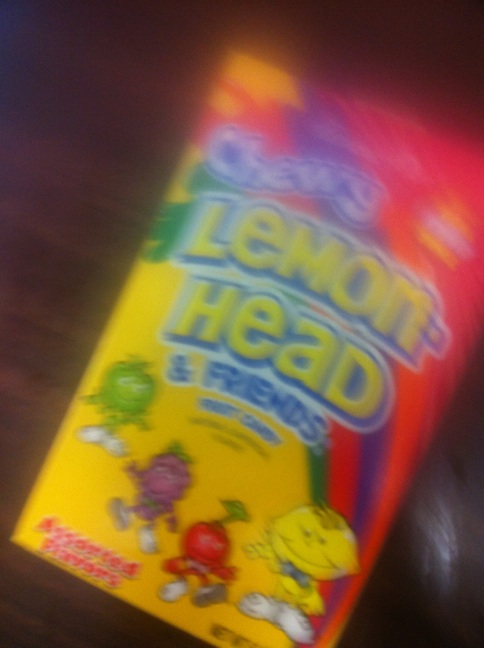}\Description{A slightly rotated and blurry chewy Lemonhead box. The whole box is pictured. The text ‘Chewy LemonHead & Friends' is readable.}} 
			& \raggedright \textbf{\textcolor{col_product}{Product:\newline candy\newline\newline}}\textbf{\textcolor{col_brand}{Brand:\newline Lemonhead \& Friends\newline\newline}}\textbf{\textcolor{col_variety}{Variety:\newline chewy}}
			& \raggedright \textbf{[Correct]} Box of \textbf{\textcolor{col_variety}{Chewy}} \textbf{\textcolor{col_brand}{Lemonhead \& Friends}} \textbf{\textcolor{col_product}{candy}}, assorted fruit flavors, with a yellow background and cartoon fruit characters on the front. 
			& \raggedright \textbf{[Correct]} A rectangular yellow box of \textbf{\textcolor{col_variety}{Chewy}} \textbf{\textcolor{col_brand}{Lemonhead \& Friends}} \textbf{\textcolor{col_product}{candy}}. The box has a yellow background with a vertical rainbow stripe on the right side. The words ``\textbf{\textcolor{col_variety}{Chewy}} \textbf{\textcolor{col_brand}{Lemonhead \& Friends}}'' are printed in large, colorful letters on the front of the box. [...]
			& \raggedright \textbf{[Incorrect]} Product Information:\newline
			- Object Type: Box\newline
			- Product Type: \textbf{\textcolor{col_product}{Candy}}\newline
			- Brand: Lemon Head\newline
			- Variety: Lemon [...]
			\newline\newline
			\textbf{Missing:}\newline
			- \textbf{\textcolor{col_brand}{brand: Lemonhead \& Friends}}\newline
			- \textbf{\textcolor{col_variety}{variety: chewy}}
			& \raggedright \textbf{[Incorrect]} Bag of Lemonhead \textbf{\textcolor{col_product}{candy}} featuring 4 fruity flavors, displayed against a wooden background.
			\newline\newline
			\textbf{Missing:}\newline
			- \textbf{\textcolor{col_brand}{brand: Lemonhead \& Friends}}\newline
			- \textbf{\textcolor{col_variety}{variety: chewy}} 
			\tabularnewline \midrule
									
			% blur: beans
			\parbox[t]{\linewidth}{\vspace{-0.75em}\includegraphics[width=\linewidth]{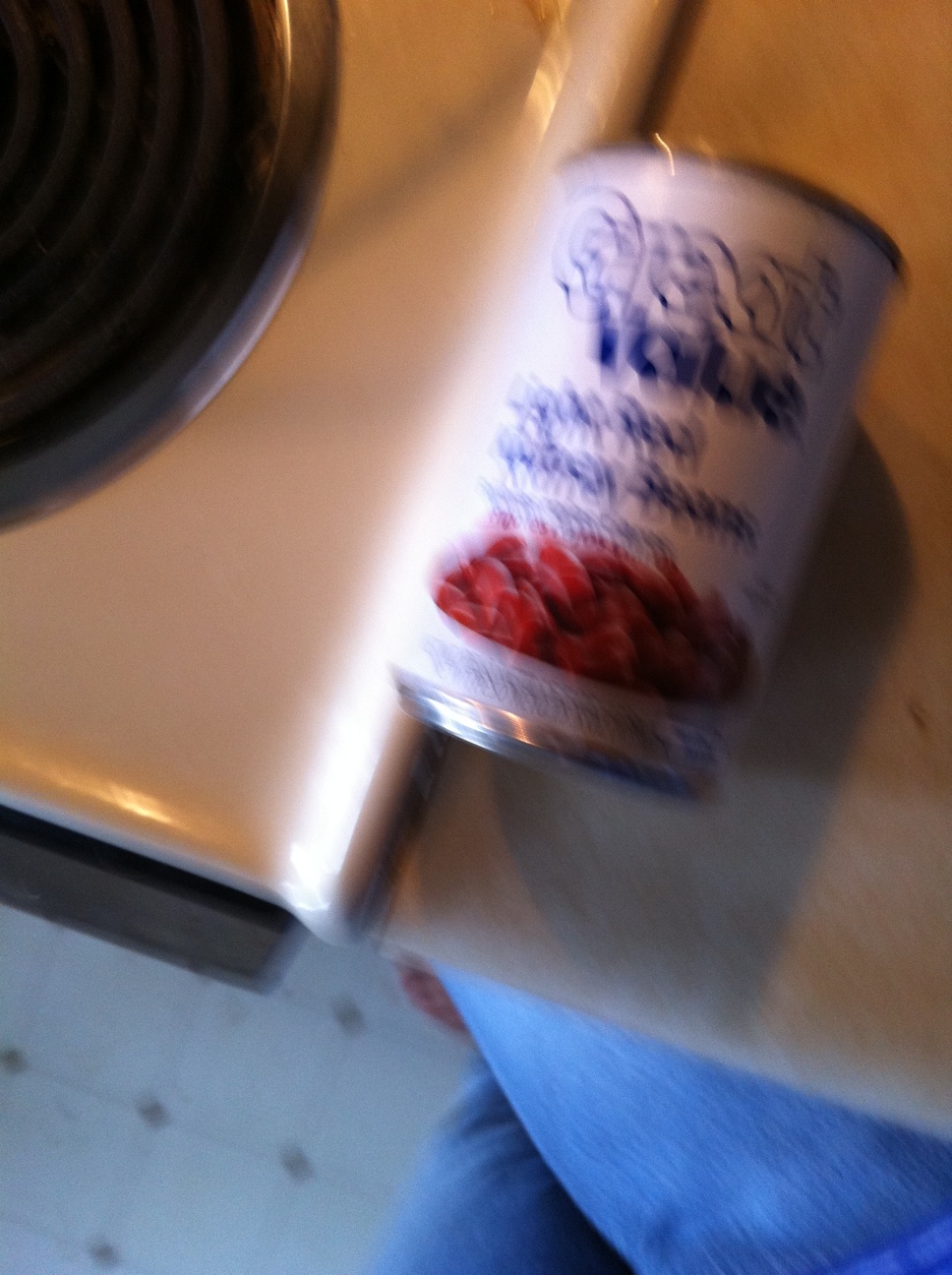}\Description{A blurry can of Great Value light red kidney beans.}} 
			& \raggedright \textbf{\textcolor{col_product}{Product:\newline kidney beans \newline\newline}}\textbf{\textcolor{col_brand}{Brand:\newline Great Value\newline\newline}}\textbf{\textcolor{col_variety}{Variety:\newline light red}}
			& \raggedright \textbf{[Correct]} Can of \textbf{\textcolor{col_brand}{Great Value}} \textbf{\textcolor{col_variety}{light red}} \textbf{\textcolor{col_product}{kidney beans}}, with a white label and image of kidney beans on the front. 
			& \raggedright \textbf{[Incorrect]} A metal can of Dole brand sliced strawberries in syrup. The can is white with red and blue lettering. The top of the can features the Dole logo in blue, with the words ``sliced red sweet strawberries'' beneath it in red. Below the text is an image of several red strawberries in syrup.
			\newline\newline
			\textbf{Missing:}\newline
			- \textbf{\textcolor{col_product}{product: beans}}\newline
			- \textbf{\textcolor{col_brand}{brand: Great Value}}\newline
			- \textbf{\textcolor{col_variety}{variety: light red kidney}}
			& \raggedright \textbf{[Incorrect]} The object is a can. The can is white with blue text and a red image on it. The can is a soup can, specifically a tomato soup can. The can is from the \textbf{\textcolor{col_brand}{Great Value}} brand. The can is 14.5 ounces. The can is white, rectangular, and has a red image of tomatoes on it.
			\newline\newline
			\textbf{Missing:}\newline
			- \textbf{\textcolor{col_product}{product: beans}}\newline
			- \textbf{\textcolor{col_variety}{variety: light red kidney}}
			& \raggedright \textbf{[Incorrect]} Cylindrical white can containing Honeysuckle Hill Turkey Chili, featuring blue text and an image of cooked ground meat
			\newline\newline
			\textbf{Missing:}\newline
			- \textbf{\textcolor{col_product}{product: beans}}\newline
			- \textbf{\textcolor{col_brand}{brand: Great Value}}\newline
			- \textbf{\textcolor{col_variety}{variety: light red kidney}}
			\tabularnewline \midrule
									
			% framing: tide pods
			\parbox[t]{\linewidth}{\vspace{-0.75em}\includegraphics[width=\linewidth]{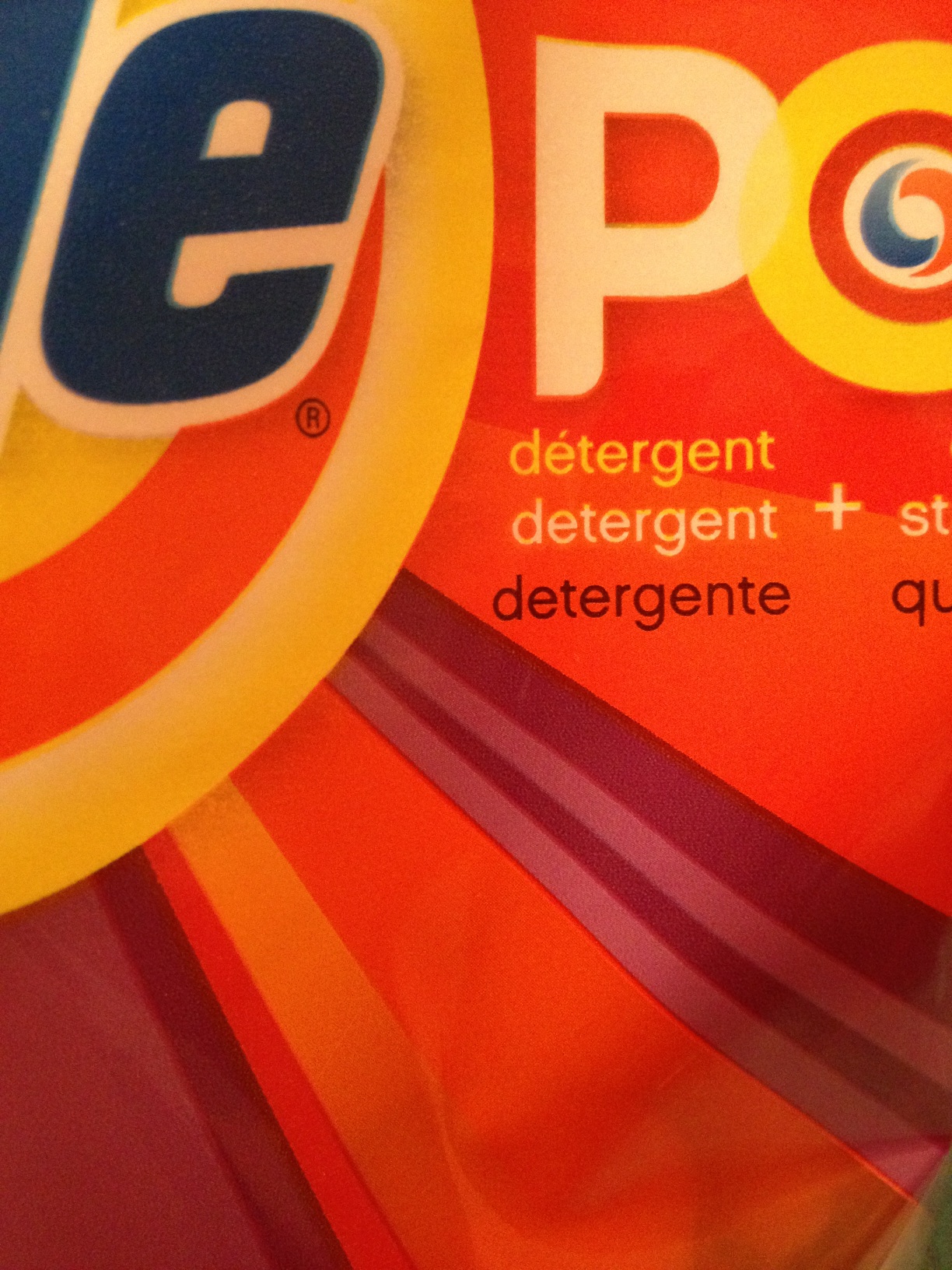}\Description{A close-up of a Tide Pods detergent package with visible texts including “e”, “PO”. The word “detergent” appears in three different languages.}} 
			& \raggedright \textbf{\textcolor{col_product}{Product:\newline laundry;\newline detergent\newline\newline}}\textbf{\textcolor{col_brand}{Brand:\newline Tide\newline\newline}}\textbf{\textcolor{col_variety}{Variety:\newline PODS}}
			& \raggedright \textbf{[Correct]} Plastic bag of \textbf{\textcolor{col_brand}{Tide}} \textbf{\textcolor{col_variety}{Pods}} \textbf{\textcolor{col_product}{laundry detergent}}, featuring the \textbf{\textcolor{col_brand}{Tide}} brand logo with orange, yellow, and purple color patterns and the word ``detergent'' in multiple languages.
			& \raggedright \textbf{[Correct]} A \textbf{\textcolor{col_brand}{Tide}} \textbf{\textcolor{col_variety}{PODS}} \textbf{\textcolor{col_product}{laundry detergent}} package. It is a large, rectangular, plastic pouch with a red background and orange, yellow, and blue stripes. The word ``d\'et\'ergent'' is printed in white letters on the front of the package.
			& \raggedright \textbf{[Incorrect]} The product is a red plastic container with a yellow and orange label, featuring the brand name \textbf{\textcolor{col_brand}{Tide}} in large white letters. [...] %The container has a distinctive shape with a curved top and straight sides, and it appears to be made of a smooth, glossy material.
			The label features the brand name prominently at the top, with the words \textbf{\textcolor{col_product}{``detergent''}} and ``detergente'' written in smaller text below it. [...]\newline\newline
			\textbf{Missing:}\newline
			- \textbf{\textcolor{col_product}{product: laundry}}\newline
			- \textbf{\textcolor{col_variety}{variety: PODS}}
			& \raggedright \textbf{[Incorrect]} A round \textbf{\textcolor{col_product}{laundry detergent}} container with red coloring, featuring the letter `e' on the left side and a bullseye-like circle with a blue wave in the center. The word `detergent' is written in three languages.\newline\newline
			\textbf{Missing:}\newline
			- \textbf{\textcolor{col_brand}{brand: Tide}}\newline
			- \textbf{\textcolor{col_variety}{variety: PODS}} 
			\tabularnewline \midrule
									
			% framing: sprite
			\parbox[t]{\linewidth}{\vspace{-0.75em}\includegraphics[width=\linewidth]{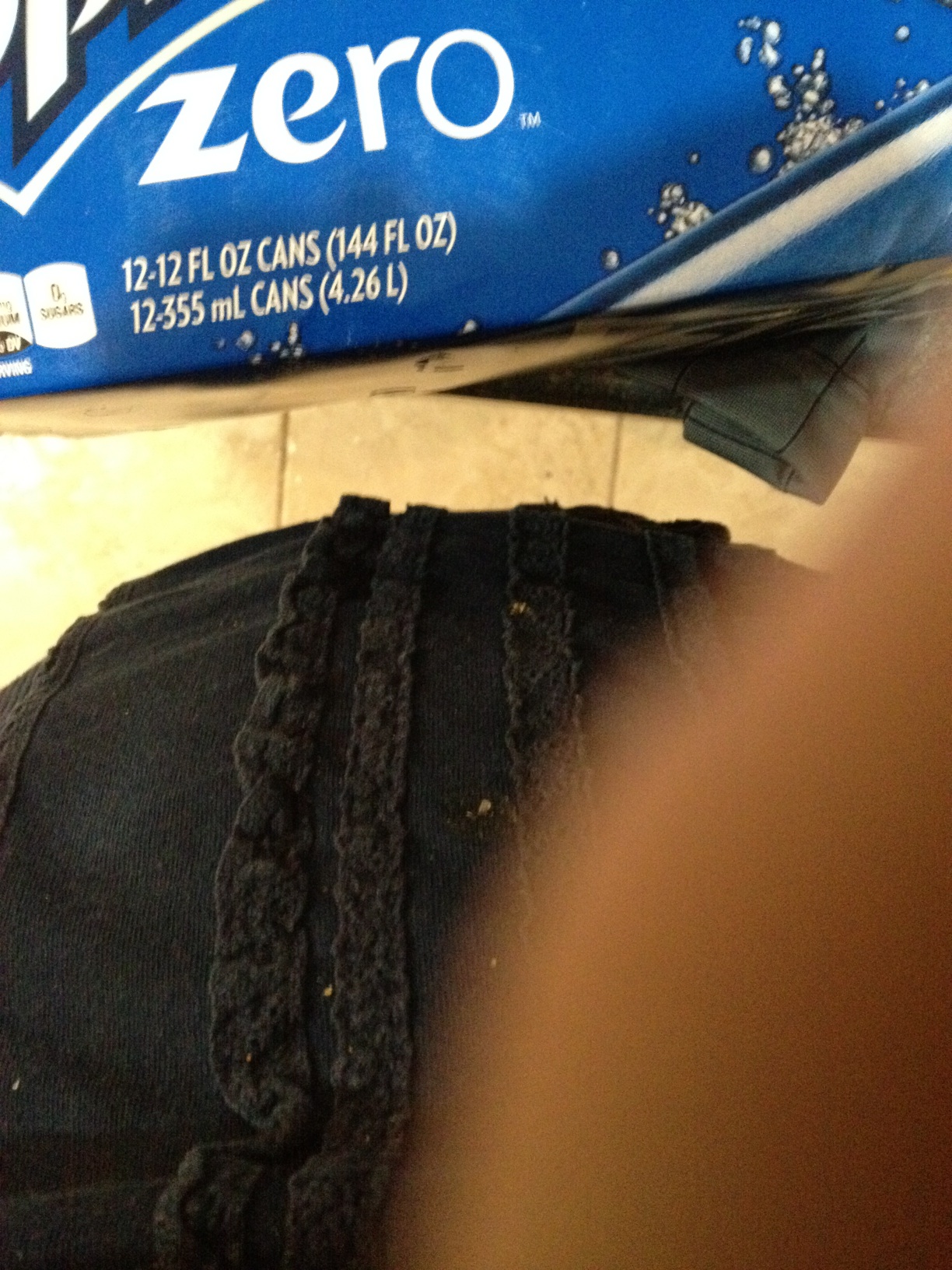}\Description{The bottom of a 12-pack box of Sprite Zero. Only the “p” is partially visible from “sprite”, while all of “zero” is visible underneath.}} 
			& \raggedright \textbf{\textcolor{col_product}{Product:\newline soda/soft drink\newline\newline}}\textbf{\textcolor{col_brand}{Brand:\newline Sprite\newline\newline}}\textbf{\textcolor{col_variety}{Variety:\newline zero; 12-pack}}
			& \raggedright \textbf{[Incorrect]} \textbf{\textcolor{col_variety}{Twelve-pack}} of Dr Pepper \textbf{\textcolor{col_variety}{Zero}} \textbf{\textcolor{col_product}{soda}} in 12 fluid ounce cans, blue packaging with white and light blue text and bubble graphics.\newline\newline
			\textbf{Missing:}\newline
			- \textbf{\textcolor{col_brand}{brand: Sprite}}
			& \raggedright \textbf{[Incorrect]} A \textbf{\textcolor{col_variety}{12-pack}} of 12-ounce cans of Pepsi \textbf{\textcolor{col_variety}{Zero}} Sugar. The packaging is a blue cardboard box with the word ``zero'' printed in white.\newline\newline
			\textbf{Missing:}\newline
			- \textbf{\textcolor{col_product}{Product: soda/soft drink}} \newline
			- \textbf{\textcolor{col_brand}{brand: Sprite}}
			& \raggedright \textbf{[Incorrect]} Product Description\newline
			- Object Type: Blue cardboard \textbf{\textcolor{col_variety}{12-pack}}\newline
			- Product Type: \textbf{\textcolor{col_product}{Soft drink}}\newline
			- Brand: Pepsi\newline
			- Variety: \textbf{\textcolor{col_variety}{Zero}} Sugar\newline
			- Visual Features:\newline
			+ Color: Blue\newline
			+ Size: 12 cans, 12 fluid ounces each\newline\newline
			\textbf{Missing:}\newline
			- \textbf{\textcolor{col_brand}{brand: Sprite}}
			& \raggedright \textbf{[Incorrect]} Pepsi \textbf{\textcolor{col_variety}{Zero}} \textbf{\textcolor{col_product}{soda}} \textbf{\textcolor{col_variety}{12-pack}}. Royal blue packaging with white text displaying brand name and product details. Case contains 12-12 fl oz cans (144 fl oz total). Zero sugar variety. Packaged in a cardboard box.\newline\newline
			\textbf{Missing:}\newline
			- \textbf{\textcolor{col_brand}{brand: Sprite}}
			\tabularnewline \bottomrule
		\end{tabular}%
	}
	\Description{Organized in six columns, separated by horizontal lines, the table presents four image examples, their annotation supplied by the researchers, and the outputs from four VLM models. The first column has a preview of the image. The second has annotations of products, including product, brand, and variety. The third through sixth include caption outputs from each VLM, with an indicator of whether it is correct and color coding for which annotations matched or were missed. The four images in the table are: (1) A slightly rotated and blurry chewy Lemonhead box. The whole box is pictured. The text ‘Chewy LemonHead & Friends' is readable; (2) A blurry can of Great Value light red kidney beans; (3) A close-up of a Tide Pods detergent package with visible texts including “e”, “PO”. The word “detergent” appears in three different languages; and (4) The bottom of a 12-pack box of Sprite Zero. Only the “p” is partially visible from “sprite”, while all of “zero” is visible underneath.}
\end{table*}
%TC:endignore

%TC:ignore
\begin{table*}[ht]
	\Large
	\centering
	\caption{Examples of rotated product images where VLMs struggle to correctly identify products. Captions had to include accurate product, brand, and variety information to be coded as correct. Captions were shortened for presentation purposes only, indicated by [...].}
	\label{fig:all-examples-rotation}
	\resizebox{\linewidth}{!}{%
		\begin{tabular}{@{}p{0.20\linewidth}p{0.11\linewidth}*{4}{p{0.2925\linewidth}}@{}}
			\toprule
			\textbf{Image} & \textbf{Annotation} & \textbf{GPT} & \textbf{Gemini} & \textbf{Llama} & \textbf{Molmo} \tabularnewline \midrule 
			% rotation: bigelow tea
			\parbox[t]{\linewidth}{\vspace{-0.75em}\includegraphics[width=\linewidth]{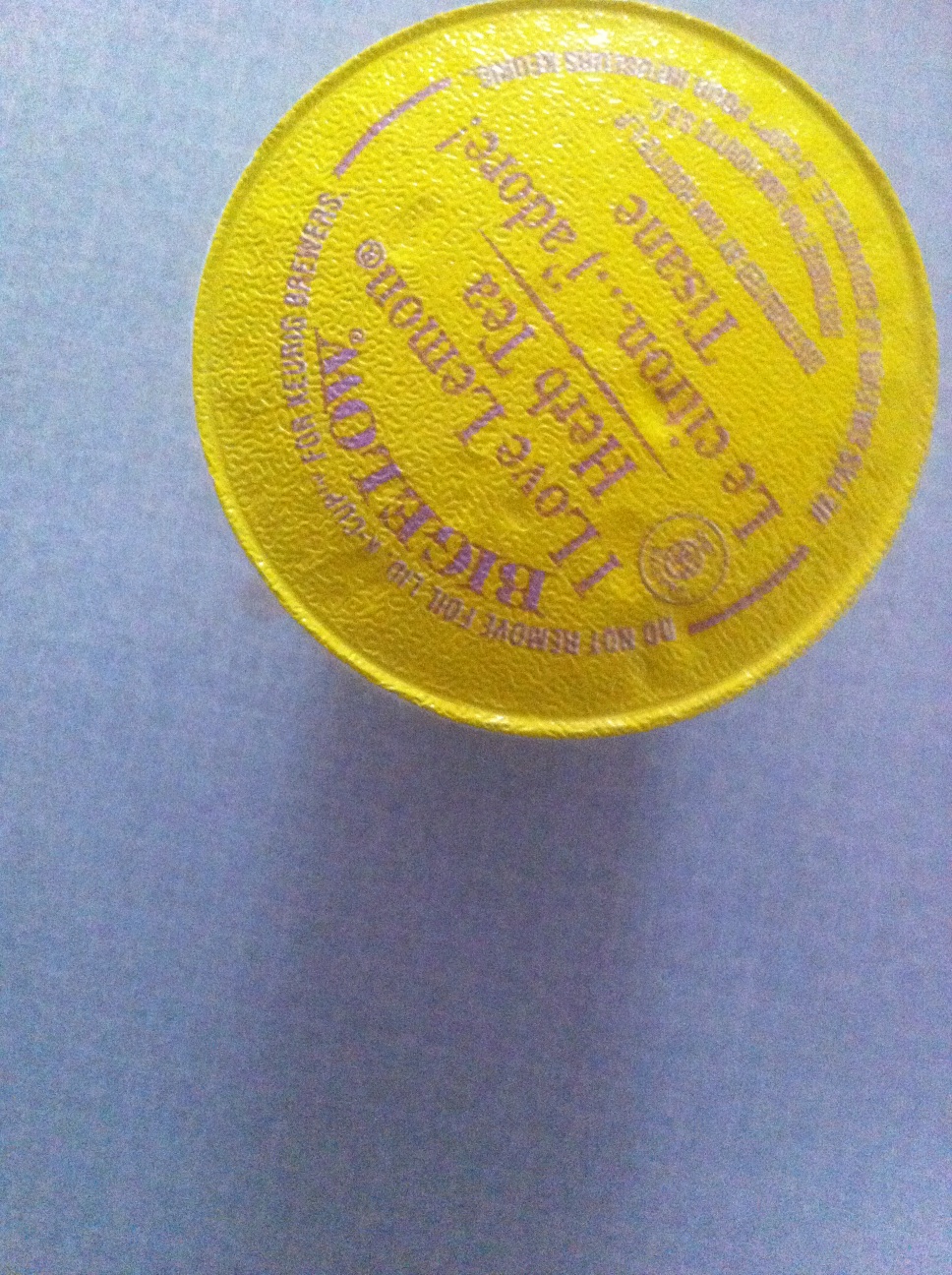}\Description{A yellow K-cup of Bigelow I Love Lemon herb tea. The K-cup is rotated counterclockwise about 135 degrees.}} 
			& \raggedright \textbf{\textcolor{col_product}{Product:\newline herb tea\newline\newline}}\textbf{\textcolor{col_brand}{Brand:\newline Bigelow\newline\newline}}\textbf{\textcolor{col_variety}{Variety:\newline I love lemon}}
			& \raggedright \textbf{[Incorrect]} Yellow plastic bottle cap from a Bickford's \textbf{\textcolor{col_variety}{I Love Lemon}} Iced Tea bottle. The text on the cap mentions the flavor (Lemon), product type (Iced Tea), and brand (Bickford's). The cap is round with textured yellow plastic.\newline\newline
			\textbf{Missing:}\newline
			- \textbf{\textcolor{col_product}{product: herb tea (incorrectly identified as iced tea)}}\newline
			- \textbf{\textcolor{col_brand}{brand: Bigelow}}
			& \raggedright \textbf{[Correct]} A yellow K-cup for \textbf{\textcolor{col_brand}{Bigelow}} \textbf{\textcolor{col_variety}{I Love Lemon}} \textbf{\textcolor{col_product}{Herb Tea}}. The lid is covered with a textured pattern. Purple text on the lid reads, ``Bigelow I Love Lemon Herb Tea,'' with ``Le citron... j'adore!'' printed below it.
			& \raggedright \textbf{[Incorrect]} Yellow plastic container with circular shape and slightly raised edges. The brand name ``BIGZIONY'' is written in purple letters at the bottom of the container. The container has a textured surface and features purple text, including ``Love, Lemon'' and ``High Tea'', which are likely product labels or descriptions. [...] \newline\newline
			\textbf{Missing:}\newline
			- \textbf{\textcolor{col_product}{product: herb tea}}\newline
			- \textbf{\textcolor{col_brand}{brand: Bigelow}}\newline
			- \textbf{\textcolor{col_variety}{variety: I love lemon}}
			& \raggedright \textbf{[Incorrect]} Blue rectangular box of Land O'Lakes butter. Yellow circular lid on top with purple print. Text includes `Land O'Lakes', `Since 1921', `Buttery', `Smooth', `Taste', `Quality', `From our family of farm owners to yours', `5 pounds', `One pound rolls', `454 grams', `Made in USA', `Land O'Lakes, Inc.', `Arden Hills, MN 55112'.\newline\newline
			\textbf{Missing:}\newline
			- \textbf{\textcolor{col_product}{product: herb tea}}\newline
			- \textbf{\textcolor{col_brand}{brand: Bigelow}}\newline
			- \textbf{\textcolor{col_variety}{variety: I love lemon}}
			\tabularnewline \midrule
						
			% rotation: chewy
			\parbox[t]{\linewidth}{\vspace{-0.75em}\includegraphics[width=\linewidth]{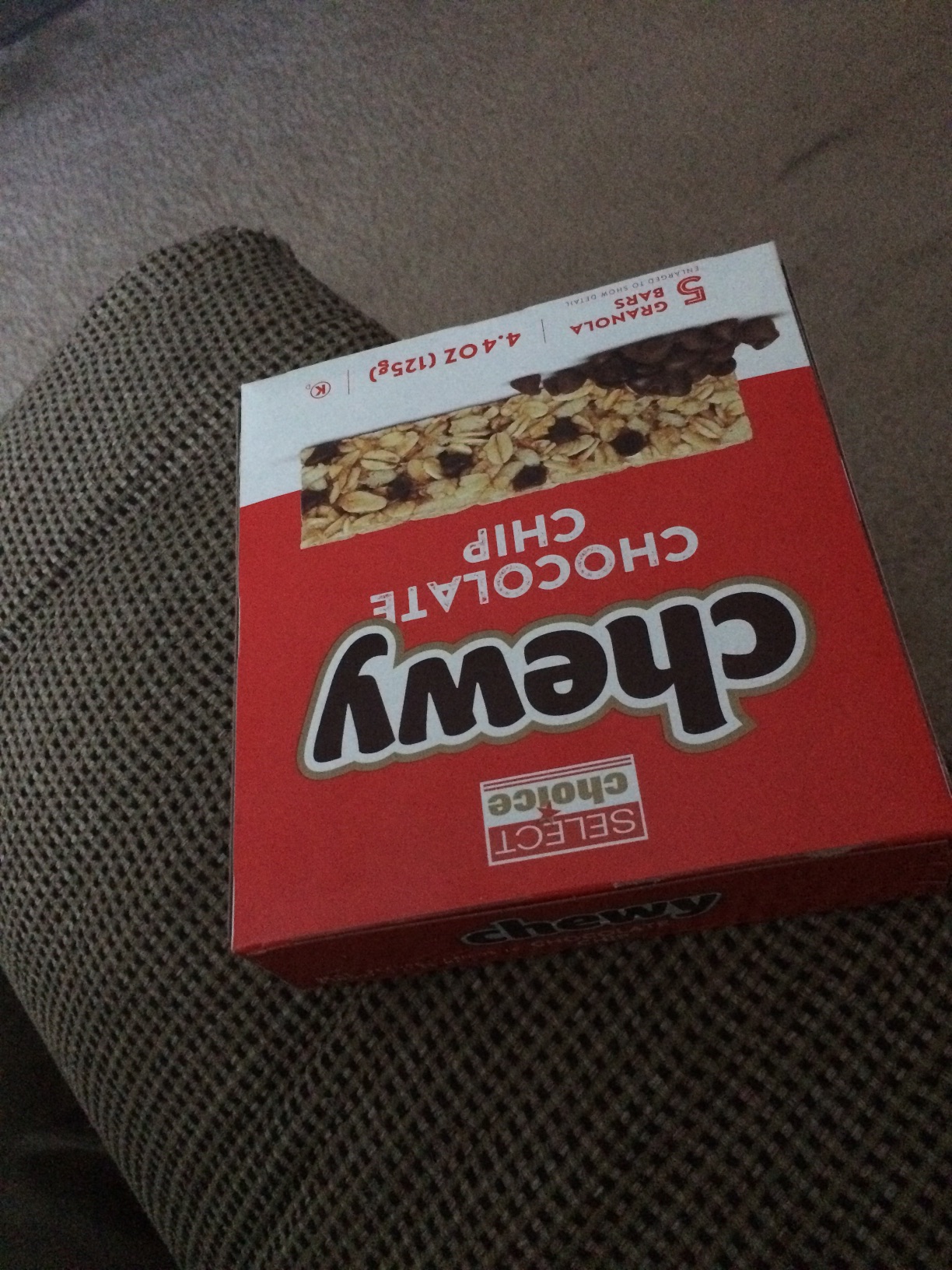}\Description{An upside-down box of Select Choice Chewy Chocolate Chip Granola Bars. The front of the box and all details are fully visible.}} 
			& \raggedright \textbf{\textcolor{col_product}{Product:\newline granola bars\newline\newline}}\textbf{\textcolor{col_brand}{Brand:\newline Select Choice\newline\newline}}\textbf{\textcolor{col_variety}{Variety:\newline chewy; chocolate chip}}
			& \raggedright \textbf{[Correct]} Box of \textbf{\textcolor{col_brand}{Select Choice}} \textbf{\textcolor{col_variety}{Chewy Chocolate Chip}} \textbf{\textcolor{col_product}{Granola Bars}}, 5 bars, each weighing 1.25 ounces. The box is mostly red with a picture of a granola bar containing oats and chocolate chips.
			& \raggedright \textbf{[Incorrect]} A rectangular red cardboard box of Quaker \textbf{\textcolor{col_variety}{Chewy Chocolate Chip}} \textbf{\textcolor{col_product}{granola bars}}, 5 bars, 4.4 ounces (125 grams). The box is 6 inches long and 5 inches wide. The front of the box shows the ``chewy'' logo in white letters. Below the logo, a white box with red text reads ``SELECT CHOICE.'' [...]\newline\newline
			\textbf{Missing:}\newline
			- \textbf{\textcolor{col_brand}{brand: Select Choice}}
			& \raggedright \textbf{[Incorrect]}  The product is a red and white box of \textbf{\textcolor{col_product}{granola bars}}. The box is rectangular, with a red base and a white top. The brand name ``\textbf{\textcolor{col_variety}{Chewy}}'' is written in brown letters on the front of the box. The variety is ``\textbf{\textcolor{col_variety}{Chocolate Chip}},'' which is written in white letters above the brand name. [...]\newline\newline
			\textbf{Missing:}\newline
			- \textbf{\textcolor{col_brand}{brand: Select Choice}}
			& \raggedright \textbf{[Incorrect]}  A cardboard box containing Quaker \textbf{\textcolor{col_variety}{Chewy Chocolate Chip}} \textbf{\textcolor{col_product}{granola bars}}. It has a red-and-white design and is oriented upside down. The box contains 12 bars.
			\newline\newline
			\textbf{Missing:}\newline
			- \textbf{\textcolor{col_brand}{brand: Select Choice}}
			\tabularnewline \bottomrule
		\end{tabular}%
	}
	\Description{Organized in six columns, separated by horizontal lines, the table presents two image examples, their annotation supplied by the researchers, and the outputs from four VLM models. The first column has a preview of the image. The second has annotations of products, including product, brand, and variety. The third through sixth include caption outputs from each VLM, with an indicator of whether it is correct and color coding for which annotations matched or were missed. The two images in the table are: (1) A yellow K-cup of Bigelow I Love Lemon herb tea. The K-cup is rotated counterclockwise about 135 degrees; and (2) An upside-down box of Select Choice Chewy Chocolate Chip Granola Bars. The front of the box and all details are fully visible.}
\end{table*}
%TC:endignore

% impact of blur
Our regression results reveal that image quality issues impact all VLMs when identifying products; see Table~\ref{tab:study-2-log-reg-all-models}. All image quality variables (blur, framing, and rotation) were statistically significant and negative, indicating that their presence increases the likelihood that the studied VLM would incorrectly identify a product. Blurred images were the most likely to be incorrect, reducing the odds of correct product identification by 88.3\%. We hypothesize that all four VLMs are trained on high-quality (i.e., non-blurry) images and never learn to handle blurred images during inference. In examples of blurred images, we observe discrepancies in identifying the product generally versus providing necessary details for BLV (see Table~\ref{fig:all-examples-blur-framing}, row 1--2)\footnote{Our examples focus on brands from English-speaking countries, primarily the U.S., which the studied models should perform the best on. While our dataset includes brands from other English-speaking countries (e.g., crisps in the U.K.), these examples are sparse and less likely to be in training data for models built by U.S. companies.} For example, GPT and Gemini correctly identify a box of Chewy Lemonhead \& Friends candy, while Llama only identifies ``Lemon Head'' (missing ``\& Friends'' sub-brand) and that it is candy (missing ``chewy'' variety). Molmo similarly misses sub-brand and variety details. This suggests that VLMs can capture large, easily readable text, such as brand labels, that is more resistant to distortion than fine-text details (e.g., food flavor). In another example, only GPT can correctly identify a can of Great Value light red Kidney Beans; Llama can identify the brand, but not ``kidney beans''; Gemini and Molmo identify nothing correctly.

% impact of framing
Framing was the second-most problematic image-quality issue across models, reducing the odds of correct product identification by 84.5\%. Specific examples from show that framing issues even affect the identification of common U.S. brands (e.g., Tide detergent, Sprite Zero), which almost certainly occur frequently in the internet-scale training data for these models; see Table~\ref{fig:all-examples-blur-framing}, row 3--4. What makes framing interesting is how well VLMs fill in or infer the rest of the content. Each VLM was varied in this regard. For example, GPT and Gemini could fill in ``Tid'' and ``DS'' for Tide Pods, while Llama could fill in ``Tide'' and Molmo filled in neither (despite recognizing it was laundry detergent). However, no models could fill in ``Sprite''.

% impact of rotation
% which had a greater impact on model accuracy than blur or rotation.
Finally, rotation was the least problematic image quality issue, reducing the odds of correct product identification by 79.5\%. Qualitatively, we found that rotation makes it harder for VLMs to understand fine text details---which often includes key details about the product---compared to larger attributes, like brand text and logos, or well-known varieties (e.g., Diet for Coke); see Table~\ref{fig:all-examples-rotation}. For product and variety details, we observed that GPT, Llama, and Molmo failed to identify the product (herb tea) and brand (Bigelow) of a K-Cup pod, whereas Gemini was correct. In the second example of Select Choice Chewy granola bars, all models identified the product (granola bars) and variety (chewy, with chocolate chips), but only GPT correctly recognizes the brand.

As shown earlier, co-occurring quality issues can negatively impact performance and are common in BLV people's photos \cite{chiu2020assessing}, complicating the challenge of using VLMs to identify products. The regression results reveal significant two-way interaction effects between blur and framing ($p < 0.001$), blur and rotation ($p < 0.05$), and framing and rotation ($p < 0.001$). The interaction plots reveal that when two image quality issues co-occur (e.g., blur and misframing), the drop in performance is less steep than when only one issue is present. We also observe a significant three-way interaction among blur, framing, and rotation ($p < 0.05$); inspection of this interaction plot reveals a similar pattern to the two-way interactions, where additional image quality issues reduce performance, but not to the same extent as a single issue. This suggests that once product images are sufficiently degraded, models struggle to identify them, regardless of further image degradation. Our qualitative observations echo these findings; see Appendix~\ref{appendix:additional-examples}, Table~\ref{fig:cooccurrance-failure-examples-mucinex}. For example, all four VLMs failed to identify a box of Mucinex Expectorant medication when the image is blurry, rotated 90 degrees, and half of the ``M'' in Mucinex is out of frame (despite the rest of the label being visible). Yet in a second image, moved ever so slightly so that the ``M'' in Mucinex is fully in view but still blurred and rotated 90 degrees, three of the four VLMs correctly identify it. Further disentangling how co-occurring image quality issues affect product identification is an important area for future work.

% text panels
As we saw earlier, rounded labels and text panels had varying effects on model performance; our regression results provide a clearer illustration. Only text panels caused a significant drop in performance, reducing the odds of correct product identification by 43.5\%. An interaction effect for framing by text panel was also significant ($p < 0.001$), with the interaction plot showing that framing generally reduces performance, but no text panel results in {\em poorer} performance when misframed. This suggests that text panels can provide the VLM with clues about the product (e.g., from a longer description of a frozen meal), even if other identifying features are not in clear view (e.g., the brand logo or meal title). While having a rounded label had an insignificant effect on product identification odds, the interaction effect for rotation by rounded labels was significant ($p < 0.01$), with the drop in performance being less steep than when only one variable is true (similar to image quality interactions). Appendix~\ref{appendix:additional-examples}, Table~\ref{fig:text-panels-rounded-labels} shows examples of these effects. For instance, no model correctly identified the ground beef as 90\% lean, 10\% fat, despite it being clearly visible in the upper left, and only Llama noticed the text. The rounded Manwich sloppy joe can partially shows the ``M'' from the logo and an image of prepared sloppy joe, but all models focused on the more visible tomatoes instead, inferring it was just tomato sauce.

% model-specific differences
Finally, our regression analysis shows model-wise differences in product identification performance. Compared to GPT, the best-performing model, all VLMs had significantly reduced performance (Gemini: 45.1\% reduced odds; Llama: 82.9\%; Molmo: 83.9\%). We found a significant interaction effect for blur by Gemini ($p < 0.05$). The interaction plots showed that Gemini's performance relative to GPT declines more slowly for blurred images, suggesting greater resistance to it. We also found a significant negative interaction between rotation and Molmo ($p < 0.01$). The interaction plot showed that the drop in performance is steeper when images are rotated, suggesting that Molmo is worse at handling rotations than GPT is. 

\subsubsection{Differences in What Each VLM Struggles With}
% individual VLM regressions
\aptLtoX[graphic=no,type=html]{\begin{table*}[ht]
	\centering
	\caption{Logistic regression results on a per-VLM basis that let us understand how image quality issues and product image properties affect the likelihood of correct identification. The model coefficients represent logits (i.e., log-odds). p-value significant at: * 0.05; ** 0.01; *** 0.001.}
	\label{tab:study-2-log-reg-per-model}
	\begin{tabular}{@{}lrlrlrlrl@{}}
		\toprule
		\textbf{Independent Variable} & 
		\multicolumn{2}{l}{\textbf{GPT}} & 
		\multicolumn{2}{l}{\textbf{Gemini}} & 
		\multicolumn{2}{l}{\textbf{Llama}} & 
		\multicolumn{2}{l@{}}{\textbf{Molmo}} \\ \midrule
		(Intercept)                            & 4.1068  & *** & 3.0659  & *** & 1.7902  & *** & 1.8236  & *** \\
		Blur = True                            & -2.7830 & *** & -1.7801 & *** & -1.8539 & *** & -1.7344 & *** \\
		Framing = True                         & -2.4973 & *** & -1.8848 & *** & -1.6231 & *** & -1.8539 & *** \\
		Rotation = True                        & -2.0067 & *** & -1.1406 & **  & -0.8992 & **  & -2.4627 & *** \\ \midrule
		Blur and Framing = True                & 2.0405  & *** & 1.2420  & *** & 1.0932  & *** & 1.0883  & *** \\
		Blur and Rotation = True               & 0.9562  &     & 0.4929  &     & 0.3790  &     & 0.8919  & *   \\ 
		Framing and Rotation = True            & 1.6500  & **  & 0.8037  &     & 0.6144  &     & 1.6175  & *** \\ \midrule
		Blur, Framing, and Rotation = True     & -0.9292 &     & -0.1690 &     & -0.5340 &     & -1.0584 &     \\ \midrule
		Null deviance ($\text{df} = 1858$)     & 1626.6  &     & 1801.4  &     & 2493.4  &     & 2550.3  &     \\
		Residual deviance ($\text{df} = 1851$) & 1342.9  &     & 1581.3  &     & 2083.4  &     & 1978.0  &     \\
		AIC                                    & 1358.9  &     & 1597.3  &     & 2099.4  &     & 1994.0  &     \\ \bottomrule
	\end{tabular}
	\Description{The table has five columns and is organized with horizontal lines separating the header row and the results for each independent variable and the four models' fit statistics. A total of 8 rows are present for the different independent variables and their interaction effects. The bottom three rows detail fit statistics for the logistic regression model.}
	\end{table*}}{\begin{table*}[ht]
	\centering
	\caption{Logistic regression results on a per-VLM basis that let us understand how image quality issues and product image properties affect the likelihood of correct identification. The model coefficients represent logits (i.e., log-odds). p-value significant at: * 0.05; ** 0.01; *** 0.001.}
	\label{tab:study-2-log-reg-per-model}
	\begin{tabular}{@{}l*{4}{r@{\hspace{0.1em}}l}@{}}
		\toprule
		\textbf{Independent Variable} & 
		\multicolumn{2}{l}{\textbf{GPT}} & 
		\multicolumn{2}{l}{\textbf{Gemini}} & 
		\multicolumn{2}{l}{\textbf{Llama}} & 
		\multicolumn{2}{l@{}}{\textbf{Molmo}} \\ \midrule
		(Intercept)                            & 4.1068  & *** & 3.0659  & *** & 1.7902  & *** & 1.8236  & *** \\
		Blur = True                            & -2.7830 & *** & -1.7801 & *** & -1.8539 & *** & -1.7344 & *** \\
		Framing = True                         & -2.4973 & *** & -1.8848 & *** & -1.6231 & *** & -1.8539 & *** \\
		Rotation = True                        & -2.0067 & *** & -1.1406 & **  & -0.8992 & **  & -2.4627 & *** \\ \midrule
		Blur and Framing = True                & 2.0405  & *** & 1.2420  & *** & 1.0932  & *** & 1.0883  & *** \\
		Blur and Rotation = True               & 0.9562  &     & 0.4929  &     & 0.3790  &     & 0.8919  & *   \\ 
		Framing and Rotation = True            & 1.6500  & **  & 0.8037  &     & 0.6144  &     & 1.6175  & *** \\ \midrule
		Blur, Framing, and Rotation = True     & -0.9292 &     & -0.1690 &     & -0.5340 &     & -1.0584 &     \\ \midrule
		Null deviance ($\text{df} = 1858$)     & 1626.6  &     & 1801.4  &     & 2493.4  &     & 2550.3  &     \\
		Residual deviance ($\text{df} = 1851$) & 1342.9  &     & 1581.3  &     & 2083.4  &     & 1978.0  &     \\
		AIC                                    & 1358.9  &     & 1597.3  &     & 2099.4  &     & 1994.0  &     \\ \bottomrule
	\end{tabular}
	\Description{The table has five columns and is organized with horizontal lines separating the header row and the results for each independent variable and the four models' fit statistics. A total of 8 rows are present for the different independent variables and their interaction effects. The bottom three rows detail fit statistics for the logistic regression model.}
	\end{table*}}

We now analyze each VLM separately to understand its susceptibility to image quality issues; see Table~\ref{tab:study-2-log-reg-per-model}. Our VLM-level regression shows that GPT and Llama are less affected by rotated images than by misframed or blurred images (GPT: 86.8\% versus 91.8\% and 93.8\% lower odds; Llama: 59.3\% versus 80.3\% and 84.3\% lower odds). This suggests that efforts to improve GPT and Llama's performance should prioritize blurred images, which are also the most prevalent in our dataset. On the other hand, Molmo is more susceptible to rotated images (91.5\% lower odds) than to blurred (82.4\%) or misframed (84.3\%) images, suggesting that additional training on rotated images is likely to yield the greatest benefit. Gemini was the only model that had relatively worse performance for misframing (84.8\% lower odds) than for blur (83.1\%) or rotation (68.0\%). 

All models showed a significant interaction between blur and framing, with positive coefficients (all $p < 0.001$). GPT also had a significant, positive interaction effect for framing by rotation ($p < 0.01$), while Molmo had significant, positive interaction effects for blur by rotation ($p < 0.05$) and framing by rotation ($p < 0.001$). Inspecting the interaction plots for these revealed that when both independent variables are true (e.g., blur and misframing), the drop in performance is less steep than when only one is true, similar to the interactions between image quality issues in our prior regression.

% discussion
\section{Discussion}
Despite their impressive capabilities for object recognition, our analysis reveals that VLMs struggle to provide detailed, accurate product captions that BLV people need when images have common quality issues (e.g., blur, framing, rotation). To our knowledge, this study is the first to systematically examine how image quality affects VLMs' ability to recognize products. While numerous studies have examined how VLMs can support BLV people's visual access needs, they largely sidestep image quality issues by asking for better photos (e.g., Seeing AI, Be My AI, \cite{mandal2023helping}) or leaving users to triangulate facts across multiple models \cite{chen2025surfacing}. While such adaptive practices are creative and skillful, the normalization of errors signals a dire need to improve how VLMs (and large AI models, broadly) are adapted to applications for BLV people. Based on our findings, we first discuss how our approach moves towards disability-centered VLM evaluation and development, arguing that while VLMs are designed for ``everyone'', particular attention needs to be paid to BLV people's specific use cases and how tools fail for them. Second, we argue that improving VLMs requires changes across the model and end-user tool development pipeline, and we propose research directions to improve VLM reliability through data curation, post-training procedures, and inference techniques to reduce errors.
 
\subsection{Towards Disability-Centered Model Evaluation of AI Systems}
Developing methods to evaluate model performance is an active area of research across HCI, AI, and ML communities. As such, accessibility researchers within these areas have begun to develop various approaches to disability-centered model evaluation that involve prompting \cite{Gadiraju2023offensive,park2025autistic}, metric assessment \cite{kapur2024reference}, interviews \cite{alharbi2024misfitting,tang2025every}, and more. A disability-centered approach not only depends on the creation of disability-first datasets (e.g., \cite{sharma2023disabilityfirst, theodorou2021disability}) but also on evaluation that centers on disability throughout. This includes questions of which data are focal to the study, how data are annotated to establish ``ground truth'', which tasks and models are selected for evaluation, and which criteria or metrics are used to assess model performance. Below, we describe these issues and the challenges of disability-centered model evaluation.

We began by understanding the information needs of BLV people within a common yet often challenging everyday task: using VLM-based AI tools to identify household products and goods. Our approach of using a survey complemented related interview studies \cite{tang2025every,alharbi2024misfitting, adnin2024look, xie2025visual} and allowed a relatively large sample of BLV people to share their experiences and issues with a diversity of AI tools for captioning images of products, surfacing unmet needs around details in images, and the difficulty of understanding and resolving common image quality issues. Our research team is all sighted, making it even more critical to understand and prioritize BLV people's perspectives from the start. %Developing our structured dataset then gives a scalable way to begin measuring how models fail on these specific objectives---the details BLV people want to know about products.

While related disability-centered approaches aim to support people with disabilities in generating ``good'' data for training systems \cite{hong2022blind,goodman2021toward}, our study examined the opposite side of this issue. We intentionally curated a disability dataset such that it targets important but understudied cases (i.e., product images with quality issues), thus aiming to interrogate cases that are central to BLV people's lived experiences but often set aside in research (i.e., labeled as others \cite{brady2013visual}, excluded in analysis \cite{gurari2020captioning}, or treated as a direction for future work \cite{chang2024worldscribe}). Rather than placing the burden on BLV users to consistently capture ``high-quality'' photos required for successful object recognition or training, future datasets should treat image quality variability as a central design consideration, in contrast to existing datasets that overwhelmingly focus on high-quality images (e.g., ImageNet \cite{deng2009imagenet} and MS COCO \cite{lin2014microsoft, chen2015microsoft}) that VLMs are optimized on. Including representative quality variations that reflect the real-world conditions under which BLV people capture images can help us develop VLMs that are more resistant to such variations from the start, rather than needing to fix them in post-training.

Although academic scholars and industry corporations have emphasized the pressing need for more disability-centered datasets \cite{sharma2023disabilityfirst, morrison2023understanding, theodorou2021disability, li2024want, bragg2021fate, gurari2018vizwiz, desai2023asl}, annotating these datasets with meaningful ``ground truth'' labels so that they can be used in benchmark studies and model evaluations such as the present paper remains a challenge, particularly when the phenomena of interest are inaccessible to the people who matter most \cite{hong2022blind,goodman2021toward}. Relying on crowdworkers is a common approach to annotation, but they may lack insight into disabled people's information needs and may apply varying standards of visual interpretation in BLV-focused datasets \cite{simons2020hope}. They are also often constrained by the time allotted to each annotation and tend to move on quickly when encountering difficult cases. Using other VLMs to synthetically generate annotations is a popular approach \cite{tan2024large, liu2023visual}, but it is likely to perpetuate inaccuracies or biases that the model already has (see distribution shift \cite{schroeder2025just}), rather than capturing important nuances. In other words, the most challenging use cases for machines require extensive human labor. In our case, four researchers spent more than three months reviewing, discussing, validating, and annotating low-quality images. While we developed a structured annotation framework based on BLV users' information needs, we were still limited by the information available in images, and could not reliably code expiration dates or product ingredient lists (other details that BLV people wanted captured and should be examined in future work).

Another challenge is selecting models to evaluate that align with disabled people's experiences and needs, and are amenable to further research. Our study selects a complementary set of VLMs: two closed-source models that power the AI image captioning tools BLV people use daily (e.g., Seeing AI, Be My AI), enabling industry relevance and application of our findings; and two open-source models because data privacy was an important concern for BLV people, and these models can be run locally, allowing greater control over privacy-sensitive data, as we discuss below. Open-source models also enable the understanding of training procedures, which can aid in interpreting evaluation results. %As our study showed, modern VLMs can accurately describe products in low-quality images (echoing prior work \cite{kacorri2017people}), but also surface opportunities for improvement, which can only happen if the performance on the subset of data is known.

Finally, disability-centered approaches must contend with which measures of ``success'' best represent disabled people's concerns. For example, \citet{kapur2024reference} demonstrates bias in reference-based metrics against BLV people, calling for evaluation methods based on user groups' specific needs. Towards this end, the research team manually reviewed and coded 7,436 model captions for accuracy and completeness, rather than relying on metrics that assess similarity and could lead to false positives (see Section \ref{ssec:data-metric-challenge}). That is, we aimed to emphasize BLV people's information needs by requiring models to generate both necessary and accurate product details rather than settling for general category identification (e.g., ``can of food'') or brand recognition (e.g., ``Campbell's''). Given the difficulty BLV people reported in assessing errors, let alone the risk of mis-identification, more nuanced and consistent frameworks for data annotation and error analysis are crucial for reliable VLMs, especially for high-stakes uses, such as identifying food products, medications, and household cleaners. Our annotation structure provides a pathway for annotating products, with similar structures being an important direction for future work on disability datasets.
%By considering this context of use as anything less than high-stakes, we have implicitly transferred the work of identifying and resolving machine errors to BLV users. 

%In line with \citeauthor{kacorri2017people}'s work on personalized object recognizers \cite{kacorri2017people}, our findings indicate that so-called low-quality images can still support strong model performance, as long as they contain sufficient distinguishing features. 

\subsection{Recommendations for Improving VLM Performance on Low-Quality Images}
While the studied closed-source models (i.e., GPT-4.1, Gemini) perform better on low-quality images, open-source models (i.e., Llama, Molmo) are likely more fruitful for developing reliable VLMs that meet BLV people's needs. Closed-source models are limited to prompt engineering---which is insufficient for handling distorted images---and fine-tuning to improve performance. While black-box APIs for closed-source VLMs allow limited fine-tuning on provided data, they offer far less flexibility, as details about the model architecture, training data, and the tuning process (e.g., which weights are frozen and the loss function used) are not disclosed. Moreover, closed-source models may leak private data \cite{loizos2025anthropic, emilyforlini2025backlash}, compromising data privacy that our survey respondents strongly desired. In contrast, open-source models make the model's architecture and training details available to researchers\footnote{Molmo goes further and makes training data available \cite{deitke2025molmo-cvpr}, while Llama only provides high-level descriptions of their dataset \cite{grattafiori2024llama}.}, while preserving privacy when run locally. To narrow the performance gap between open- and closed-source models, we propose three areas of research across the VLM pipeline: data curation, training objectives, and inference-time techniques.

\subsubsection{Improved Post-Training of VLMs Through Data Curation}
VLM performance is heavily shaped by post-training activities, including fine-tuning on specific tasks (e.g., PixMoCap for captioning \cite{deitke2025molmo-cvpr}) and diverse datasets \cite{li2025eagle}, or training to provide answers in specific formats (e.g., instruction tuning \cite{liu2023visual}). One way to improve models at this stage is to give examples when the model lacks knowledge about a task \cite{zhang2024why}. For recognizing products and their attributes, recent research suggests that VLMs require fine-tuning for good performance \cite{prabhakaran2025vitpro, trabelsi2025what}. However, our analysis shows that off-the-shelf VLMs perform well for U.S.-based products when product images are high-quality, suggesting that the issue is not due to the model's knowledge gaps. That said, such training could help adapt models for different user populations, such as BLV people in non-English-speaking countries, which we did not study. Products in those countries are infrequently found in the U.S. or on English-written webpages, which we hypothesize are the primary sources of training data for the VLMs studied.

% point 2: simulated data
Better datasets could be used to train VLMs to learn more robust representations of how products look when images are degraded. While performing well on high-quality images, all models had substantially lower performance on low-quality images, suggesting they could not find enough distinguishing characteristics in those images to support successful identification (as humans could).
% Even if models have knowledge about products (e.g., a Coca-Cola can is red), lower-quality images substantially reduce product identification performance. Notably, modern VLMs can accurately describe products in low-quality images if enough distinguishing characteristics are present (echoing prior work \cite{kacorri2017people}), suggesting a fruitful line of work could be helping them learn more robust representations of how products look when the image is degraded. 
To remedy this, future research could develop synthetic datasets in which high-quality images are systematically degraded with different image-quality issues (similar to  \cite{hendrycks2019benchmarking}), such as a can of soda with progressively greater blur or different framing issues, and fine-tune a VLM on them. Such work can draw inspiration from research in quality-agnostic learning (e.g., \cite{yu2023qualityagnostic, kim2021qualityagnostic}) that has demonstrated modest improvements in handling image distortions, yet still leaves significant room for improvement in modern VLMs. For instance, Molmo already applies an overlapping cropping procedure in its training \cite{deitke2025molmo-cvpr}, which we would expect to make it more resistant to misframed images, but our findings demonstrate that further development is needed to address its sensitivity to image framing. To that end, our findings can help focus these efforts when coupled with knowledge about model training. For example, in addition to misframed images, Molmo struggled most with rotated images, suggesting that providing pairs of correctly aligned, rotated images with high-quality annotations could help the model recognize object similarities despite different orientations. Likewise, Llama struggled the most with blurred images, suggesting that providing it with pairs of blurred and non-blurred images may help. Moreover, open-source training procedures allow us to focus on fine-tuning specific parts of the model for this task, such as the vision encoder, while freezing parts that work well, like the language encoder. Synthetic datasets, however, should still be tied to and evaluated alongside user-generated datasets to help preserve the nuanced qualities of authentic data. Our existing dataset serves as a good starting point for such initiatives, as it includes high-quality images that can be altered and low-quality images for naturalistic comparison. 

%\textbf{Develop meaningful loss functions:}
\subsubsection{Better Learning Objectives for Post-Training}
Alongside the data used for training, effective post-training may require reconsidering commonly used loss functions if they do not capture correctness well for the domain-specific task, such as product identification. Our study revealed that VLMs frequently produce believable product descriptions that are subtly incorrect, affecting their meaning (e.g., ``Coke Zero'' versus ``Diet Coke''). While VLM loss functions differ, many use cross-entropy loss between the distribution of the model's logits and the true labels of tokens from the training data. To more directly assess whether different attributes of product annotations are preserved during fine-tuning, future work may develop evaluation metrics based on semantic relationships within the annotations. Inspiration could be taken from evaluation metrics like SPICE \cite{anderson2016spice} that evaluates overlaps between scene graphs (e.g., can $\rightarrow$ on $\rightarrow$ countertop) or Cap F1 \cite{deitke2025molmo-cvpr}, which evaluates overlap between atomic concepts (e.g., ``A can of soda''; ``Soda is on the kitchen countertop''). Such loss functions could better steer models towards learning what constitutes good product annotations. 

\subsubsection{Addressing Captioning Errors During Inference}
% point 3: modifications to inference procedure (abstention, inpainting techniques for repairing images)
While improved model training can help, it is unlikely to fully resolve the issues our study reveals; instead, we hypothesize that additional inference-time techniques can enhance VLM output without burdening the BLV user to take additional photos. One way is to leverage image reconstruction techniques that repair images before captioning. For instance, with misframed images, researchers can explore inpainting techniques that produce multiple possible versions of a repaired image for captioning \cite{chung2022diffusion,agarwal2024vipaint}, eliminating the need to take additional photos. Another is to ensure key product details are included or excluded, for which we can look to related work on reducing toxicity or enforcing lexical constraints in LLM outputs, in which constraint-based optimization can have advantages over conventional fine-tuning \cite{lu2023inference,qin2022cold}. Furthermore, these techniques can often be applied to large VLMs without costly model training, or can be combined with training smaller VLMs (which require less hardware) to improve their output beyond that of larger models.
%These techniques can help produce better outputs from the model, rather than necessarily requiring additional photos when captions are of poor quality.

% abstention techniques
Even after applying reconstruction techniques, a VLM may still make errors; in such cases, it should abstain from providing a caption. Simple techniques involving prompt engineering to abstain are of limited efficacy, with no guarantees that the instruction to abstain will be followed (e.g., best abstention prompting yields only 0.78 accuracy on question-answer tasks with similarly low-quality images \cite{huh2024longform}). In contrast, recent work on LLM abstention explores techniques based on self-consistency, in which the model evaluates its own outputs and level of uncertainty before returning a response, demonstrating good performance in question-answering settings \cite{yadkori2024mitigating, kuhn2023semantic, manakul2023selfcheckgpt, cole2023selectively}. However, abstention for open-ended image captions is harder. In our study, we observed numerous cases in which image captions contained correct {\em parts} of our product annotations, even when the caption as a whole was incorrect. While recent work for VLMs has explored techniques to {\em repair} captioning errors prior to returning them during the generation process (e.g., controlling what objects are mentioned \cite{zhai2024hallecontrol}; strategically adjusting model weights \cite{yoon2025stop, sarkar2025mitigating, leng2024mitigating,yang2025nullu} or fine-tuning \cite{carragher2025segsub,zhang2024reflective}; sampling multiple patches \cite{chen2024halc}; guided decoding \cite{zhao2025mitigating}; backtracking when uncertain \cite{duan2025truthprint, wu2025generate}) or post-hoc verification \cite{zhou2024analyzing, yin2024woodpecker}, these techniques can induce further errors during correction, rather than providing a higher-precision output that only includes details that are likely correct. Instead,  systems for {\em partial abstention}, which abstain only on inconsistent caption parts, should be explored. These could help the user understand what the model knows and is confident about, allowing them to decide whether to retake a photo to gather more information about the image or to confirm the information with someone else. Together, these techniques help make VLMs more reliable by providing high-quality responses when possible and only sharing what it is confident in when not.

\subsection{Recommendations for Supporting Better User Understanding of Image Quality Issues}
While we emphasize multiple ways to improve VLM performance on low-quality images, BLV people may still need to re-take photos, which participants in our study wanted better guidance on. Thus, we must continue to design applications that provide richer feedback on the photo-taking experience, helping users understand their environment and potential image quality issues, and guiding them in resolving them. For example, as our participants suggested, a multi-faceted approach could provide feedback \emph{before} taking the photo, pointing out lighting conditions and environmental details that may affect the process; \emph{during} photo taking, offering continuous feedback to the user about the camera angle and object positioning to capture relevant parts of products (e.g., product logo, back of the box, nutrition label) \cite{lee2019revisiting, jayant2011supporting, vazquez2014assisted, ahmetovic2020recog}; and \emph{after} taking the photo, informing users about image quality issues to help them learn what might affect captioning and how to make adjustments. However, survey participants also raised concerns that people with multiple disabilities may find such interventions more difficult. For example, participants mentioned difficulty holding the camera steady enough and carefully controlling their breathing to prevent blur. Others mentioned their dexterity makes it difficult to orient the camera in particular ways. While improving the photo-taking experience is important, the complexities of photo-taking for disabled users underscore the need for technical improvements first and foremost, rather than placing the labor of taking good photos on the users.

\subsection{Limitations and Future Work}
% focus on product identification ignores other details in the caption
Our study has a few limitations that future work should address. First, we focus on evaluating product identification {\em accuracy} rather than the {\em caption quality} of VLMs generally. We focus on products because BLV respondents in our survey strongly wanted to know which products they had photographed. However, VLMs provide numerous details in image captions, including key product information (e.g., a can of Coca-Cola), plus visual details of the product and nearby objects (e.g., the can is red; the can is on the counter), which BLV people want in captions (shown by our survey and prior work \cite{morrison2023understanding, kapur2024reference}). Moreover, how information is presented can change its interpretation. For instance, humans often use {\em hedging language} to indicate uncertainty about information (e.g., ``likely is'' Diet Coca-Cola); as VLMs can also use such language, understanding how it affects BLV people's interpretation of uncertain information with respect to helpfulness and safety---such as if key dietary information is missing, leading to less trust in the output---may inform how a VLM should present captions. Existing work shows that expressions of uncertainty can meaningfully influence users’ reliance on model outputs \cite{yona2024can}. However, current VLMs struggle to communicate their internal uncertainty through natural language \cite{kim2024im, steyvers2025what}. This misalignment becomes particularly problematic for BLV people when models use overly confident language despite uncertainty, or, conversely, when they hedge even when the information is accurate. Future studies should examine caption quality in this more holistic manner.

% image quality is a spectrum, but we make it binary
A second limitation is reducing image quality issues to a binary variable. Our dataset included a count of crowdworkers who identified an image quality issue, but treating the count as continuous or ordinal over-interprets it (i.e., 5 is not necessarily more blurry than 2), which is why we converted it to a binary variable. In reality, image degradation occurs on a spectrum, likely affecting VLMs differently as it worsens. For instance, low blur may cause no issues with captioning, while higher blur is problematic. Future work can draw from computer vision research to quantify image degradations (e.g., blur kernel estimation \cite{fergus2006removing, sun2015learning, zhang2021exposure}; occlusion-robust object detection and segmentation \cite{zhan2022trilayer, qi2022occluded}; rotation-robust object and text detection \cite{saxena2009learning, yao2012detecting, ma2018arbitrary}) and, for instance, use these values in regression analysis similar to ours. 

% cultural impacts of study -- non us-centric view 
Finally, our experiment focused on VLMs and data with a U.S. and English-speaking bias. These VLMs would likely perform worse on product photos from a non-English-speaking country. Previous research has identified cross-cultural bias as a significant limitation of VLMs perceived by BLV users \cite{alharbi2024misfitting}. Future work should consider how well the VLMs we studied perform in cross-cultural contexts and may also explore other open-source models that explicitly train on other languages (e.g., Qwen \cite{yang2025qwen3a} or Deepseek \cite{deepseek-ai2025deepseekv3} for Chinese).

% conclusion
\section{Conclusion}
As blind and low-vision (BLV) people increasingly rely on Vision-Language Model (VLM)-based tools to generate image captions for product identification, we need a more nuanced understanding of how these systems handle the image-quality issues common in BLV people's photographs. Our survey of 86 BLV people reveals their perspectives on understanding image-quality issues and errors when using VLM-based tools for product captioning, and the difficulties BLV people face in recovering from those errors. We then constructed an annotated dataset of 1,859 images taken by BLV people (729 high-quality, 1,130 low-quality images that are blurred, misframed, or rotated) with detailed product annotations---including product type (e.g., soup), brand (e.g., Campbell's), and variety (e.g., tomato, low-sodium)---and evaluated four different VLMs on it. We found that all VLMs experience a decline in product identification accuracy when image quality issues are present, with performance worsening when multiple issues are present. Moreover, we showed that each VLM is more or less susceptible to the studied image quality issues, suggesting ways to prioritize improving its performance. Making VLM-based captioning tools reliable will require collaboration among HCI and ML researchers and tool designers. Together, we will need to revisit the datasets used to evaluate these models; improve model performance through fine-tuning or inference-time techniques, especially for privacy-preserving open-source models; and design systems to provide richer feedback on VLM errors.

% acknowledgements
\begin{acks}
	We thank the Accessibility Research Collective at the University of California, Irvine, and the CollabLab at Northwestern University for helpful discussions. Research funding was provided by the \grantsponsor{SES2326023}{National Science Foundation}{https://www.nsf.gov/awardsearch/show-award?AWD_ID=2326023} through awards \grantnum{SES2326023}{SES-2326023} and \grantnum{SES2326024}{SES-2326024}.
\end{acks}

%% References
\bibliographystyle{ACM-Reference-Format}
\bibliography{references}

%% Appendices
%TC:ignore
\appendix
% presented in order of reference
\section{Crowdworker Ratings for Captionability of Images and Image Quality Issues in Dataset}
\label{appendix:detailed-dataset-stats}
Table~\ref{tab:study2-dataset} details how many crowdworkers found images captionable (from \citet{gurari2018vizwiz}) and the presence of image quality issues (from \citet{chiu2020assessing}) for our subset of 1,859 images.

\begin{table*}[ht]
	\centering
	\caption{The final dataset for Study 2 included 1,859 images taken by BLV people, with 729 images being high-quality images and 1,130 being low-quality images. Each image has at least three captions from crowdworkers (i.e., no more than 3 people said the image was Unrecognizable). High-quality images have no image quality issues $>1$; low-quality images have at least one issue for which $\geq4$ crowdworkers reported it. The 17 high-quality images with rotation $\geq4$ were images that only had a rotation issue (noted by the crowdworkers) but were actually not rotated, as checked by two researchers (see Section~\ref{study-2:data-annotation}); since they had no other issues, we moved these into the high-quality subset. Each row indicates the number of crowdworkers who reported that an image was unrecognizable or had the specified image-quality issue. Percentages are column-wise.}
	\label{tab:study2-dataset}
	\resizebox{\textwidth}{!}{%
		\begin{tabular}{@{}lr*{9}{r@{\hspace{0.1em}}r}@{}}
			\toprule
			\textbf{Image Type} &
			\textbf{\begin{tabular}[c]{@{}l@{}}Num.\\ Crowdworkers\end{tabular}} &
			\multicolumn{2}{l}{\textbf{Unrecog.}} &
			\multicolumn{2}{l}{\textbf{Blur}} &
			\multicolumn{2}{l}{\textbf{Framing}} &
			\multicolumn{2}{l}{\textbf{Rotation}} &
			\multicolumn{2}{l}{\textbf{Obstruction}} &
			\multicolumn{2}{l}{\textbf{Too Dark}} &
			\multicolumn{2}{l}{\textbf{Too Bright}} &
			\multicolumn{2}{l}{\textbf{Other}} &
			\multicolumn{2}{l@{}}{\textbf{No Issue}} \\ 
			\midrule
						
			% ---------------- HIGH QUALITY ----------------
			\multirow{6}{*}{\begin{tabular}[c]{@{}l@{}}High-Quality\\ (729 Images)\end{tabular}} 
			  & 0 & 706 & (96.8\%) & 620 & (85.1\%) & 529 & (72.6\%) & 687 & (94.2\%) & 718  & (98.5\%) & 729 & (100.0\%) & 726 & (99.6\%) & 721  & (98.9\%) & 17   & (2.3\%)   \\
			  & 1 & 18  & (2.5\%)  & 109 & (15.0\%) & 200 & (27.4\%) & 25  & (3.4\%)  & 11   & (1.5\%)  & 0   &           & 3   & (0.4\%)  & 8    & (1.1\%)  & 0    &           \\
			  & 2 & 5   & (0.7\%)  & 0   &          & 0   &          & 0   &          & 0    &          & 0   &           & 0   &          & 0    &          & 0    &           \\
			  & 3 & 0   &          & 0   &          & 0   &          & 0   &          & 0    &          & 0   &           & 0   &          & 0    &          & 0    &           \\
			  & 4 & 0   &          & 0   &          & 0   &          & 14  & (1.9\%)  & 0    &          & 0   &           & 0   &          & 0    &          & 346  & (47.5\%)  \\
			  & 5 & 0   &          & 0   &          & 0   &          & 3   & (0.4\%)  & 0    &          & 0   &           & 0   &          & 0    &          & 366  & (50.2\%)  \\ 
			\midrule
						
			% ---------------- LOW QUALITY ----------------
			\multirow{6}{*}{\begin{tabular}[c]{@{}l@{}}Low-Quality\\ (1,130 Images)\end{tabular}} 
			  & 0 & 936 & (82.8\%) & 311 & (27.5\%) & 168 & (14.9\%) & 585 & (51.8\%) & 1018 & (90.1\%) & 973 & (86.1\%)  & 960 & (85.0\%) & 1081 & (95.7\%) & 1130 & (100.0\%) \\
			  & 1 & 194 & (17.2\%) & 173 & (15.3\%) & 137 & (12.2\%) & 112 & (9.9\%)  & 88   & (7.8\%)  & 133 & (11.8\%)  & 131 & (11.6\%) & 48   & (4.3\%)  & 0    &           \\
			  & 2 & 0   &          & 106 & (9.4\%)  & 132 & (11.7\%) & 90  & (8.0\%)  & 6    & (0.5\%)  & 18  & (1.6\%)   & 23  & (2.0\%)  & 1    & (0.1\%)  & 0    &           \\
			  & 3 & 0   &          & 76  & (6.7\%)  & 136 & (12.0\%) & 75  & (6.6\%)  & 4    & (0.4\%)  & 2   & (0.2\%)   & 8   & (0.7\%)  & 0    &          & 0    &           \\
			  & 4 & 0   &          & 278 & (24.6\%) & 395 & (35.0\%) & 209 & (18.5\%) & 10   & (0.9\%)  & 3   & (0.3\%)   & 5   & (0.4\%)  & 0    &          & 0    &           \\
			  & 5 & 0   &          & 186 & (16.5\%) & 162 & (14.3\%) & 59  & (5.2\%)  & 4    & (0.4\%)  & 1   & (0.1\%)   & 3   & (0.3\%)  & 0    &          & 0    &           \\ 
			\bottomrule
		\end{tabular}%
	}
	\Description{The table has eleven columns and is organized with horizontal lines separating the header row and two sections: high-quality and low-quality. In each section, there are 6 rows corresponding to the number of captions from 0-5, which is also written in the second column. The 3rd-11th columns include counts and percentages of responses for each image quality issue.}
\end{table*}
\section{Additional Examples of Product Captioning Performance on Low-Quality Images}
\label{appendix:additional-examples}

Tables \ref{fig:blur-failure-examples}, \ref{fig:faming-failure-examples}, \ref{fig:rotation-failure-examples}, \ref{fig:cooccurrance-failure-examples-mucinex}, and \ref{fig:text-panels-rounded-labels} provide additional examples of how the studied VLMs fail for various image quality issues.

% blurred images
%TC:ignore
\begin{table*}[ht]
	\centering
	\Large
	\caption{Examples of blurred product images where VLMs may only provide high-level information or incorrectly infer what the product is. Captions were shortened for presentation purposes only, indicated by [...].}
	\label{fig:blur-failure-examples}
	\resizebox{\linewidth}{!}{%
		\begin{tabular}{@{}p{0.20\linewidth}p{0.11\linewidth}*{4}{p{0.2925\linewidth}}@{}}
			\toprule
			\textbf{Image} & \textbf{Annotation} & \textbf{GPT} & \textbf{Gemini} & \textbf{Llama} & \textbf{Molmo} \tabularnewline \midrule
			% oatmeal
            \parbox[t]{\linewidth}{\vspace{-0.75em}\includegraphics[width=\linewidth]{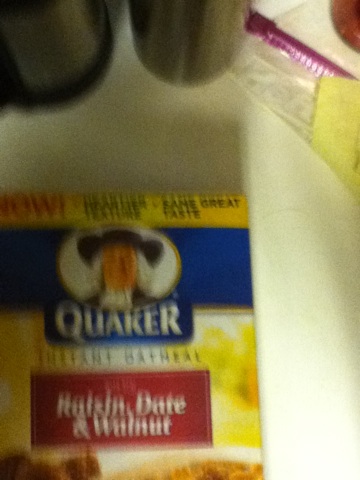}\Description{A blurry box of Quaker's instant oatmeal with raisin, date, and walnut.}}
			& \raggedright \textbf{\textcolor{col_product}{Product:\newline instant oatmeal\newline\newline}}\textbf{\textcolor{col_brand}{Brand:\newline Quaker\newline\newline}}\textbf{\textcolor{col_variety}{Variety:\newline raisin; date; walnut}} 
			& \raggedright \textbf{[Correct]} Box of \textbf{\textcolor{col_brand}{Quaker}} brand \textbf{\textcolor{col_product}{instant oatmeal}} in the \textbf{\textcolor{col_variety}{Raisin, Date \& Walnut}} variety, with a yellow and blue color scheme and the \textbf{\textcolor{col_brand}{Quaker}} logo featuring a man in a hat.
			& \raggedright \textbf{[Correct]} A yellow box of \textbf{\textcolor{col_brand}{Quaker}} \textbf{\textcolor{col_product}{Instant Oatmeal}} in the \textbf{\textcolor{col_variety}{Raisin, Date, and Walnut}} flavor. The \textbf{\textcolor{col_brand}{Quaker}} logo, a man in a hat, is shown above the word \textbf{\textcolor{col_brand}{``Quaker''}} on a blue stripe. Below this, a red rectangle states \textbf{\textcolor{col_variety}{``Raisin, Date \& Walnut''}} in white text. 
			& \raggedright \textbf{[Incorrect]} The product is a box of \textbf{\textcolor{col_brand}{Quaker}} Old Fashioned Oats, featuring a blue and yellow color scheme with a red rectangle on the front. The box has a rectangular shape and a medium size.\newline\newline
			\textbf{Missing:}\newline
			- \textbf{\textcolor{col_product}{product: instant oatmeal}}\newline
			- \textbf{\textcolor{col_variety}{variety: raisin; date; walnut}}
			& \raggedright \textbf{[Incorrect]} A rectangular box of \textbf{\textcolor{col_brand}{Quaker}} oats, specifically oatmeal packets. The brand name \textbf{\textcolor{col_brand}{``QUAKER''}} is visible on the packaging, with additional text indicating the flavor \textbf{\textcolor{col_variety}{``raisin, date, and walnut''}}. \newline\newline
			\textbf{Missing:}\newline
			- \textbf{\textcolor{col_product}{product: instant oatmeal}}
			\tabularnewline \midrule
			
			% yoplait yogurt
            \parbox[t]{\linewidth}{\vspace{-0.75em}\vspace{0pt}\includegraphics[width=\linewidth]{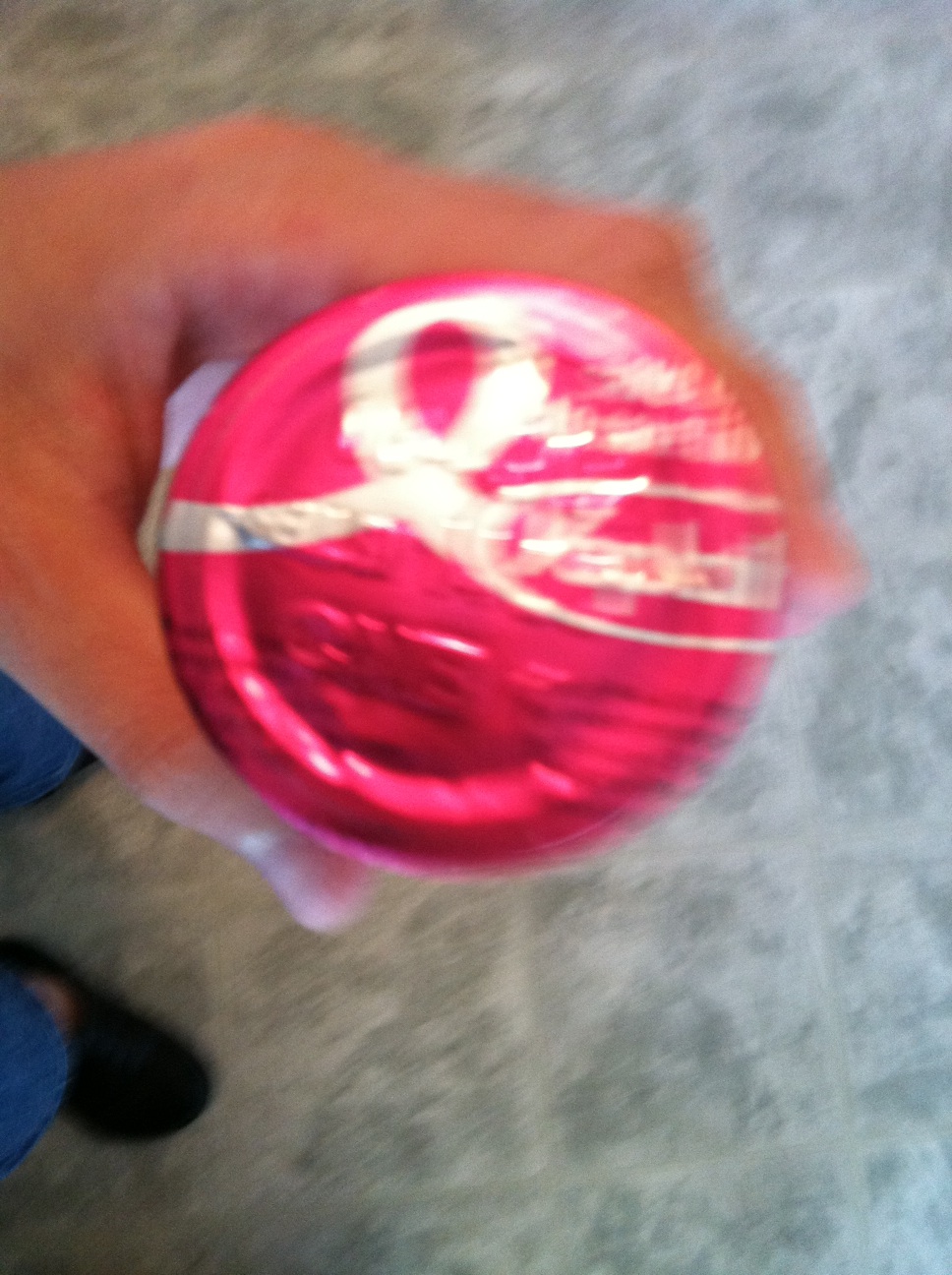}\Description{The blurry top of a Yoplait yogurt container. The container has red foil on it.}}
			& \raggedright \textbf{\textcolor{col_product}{Product:\newline yogurt\newline\newline}}\textbf{\textcolor{col_brand}{Brand:\newline Yoplait\newline\newline}}
			& \raggedright \textbf{[Incorrect]} Aluminum can with a pink top, featuring a white awareness ribbon design, commonly associated with breast cancer awareness campaigns. Specific brand and variety are not identifiable from visible features.\newline\newline
			\textbf{Missing:}\newline
			- \textbf{\textcolor{col_product}{product: yogurt}}\newline
			- \textbf{\textcolor{col_brand}{brand: Yoplait}}
			& \raggedright \textbf{[Incorrect]} A pink, spherical chocolate candy in a shiny metallic wrapper. A white ribbon design is on the front, along with the word ``GODIVA'' in white lettering.\newline\newline
			\textbf{Missing:}\newline
			- \textbf{\textcolor{col_product}{product: yogurt}}\newline
			- \textbf{\textcolor{col_brand}{brand: Yoplait}}
			& \raggedright \textbf{[Incorrect]} A cylindrical, red can with a white logo and text. The logo is a stylized letter ``D'' with the brand name ``Dr Pepper'' written in white text next to it. [...] \newline\newline
			\textbf{Missing:}\newline
			- \textbf{\textcolor{col_product}{product: yogurt}}\newline
			- \textbf{\textcolor{col_brand}{brand: Yoplait}}
			& \raggedright \textbf{[Incorrect]} Coffee bag from Eight O'Clock Coffee brand. Red packaging with white branding visible.\newline\newline
			\textbf{Missing:}\newline
			- \textbf{\textcolor{col_product}{product: yogurt}}\newline
			- \textbf{\textcolor{col_brand}{brand: Yoplait}}
			\tabularnewline \bottomrule
		\end{tabular}%
	}
    \Description{Organized in six columns, separated by horizontal lines, the table presents two image examples, their annotation supplied by the researchers, and the outputs from four VLM models. The first column has a preview of the image. The second has annotations of products, including product, brand, and variety. The third through sixth include caption outputs from each VLM, with an indicator of whether it is correct and color coding for which annotations matched or were missed. The two images in the table are as follows: (1) A blurry box of Quaker's instant oatmeal with raisin, date, and walnut; and (2) The blurry top of a Yoplait yogurt container. The container has red foil on it.}
\end{table*}
%TC:endignore

% misframed images
%TC:ignore
\begin{table*}[ht]
	\centering
	\Large
	\caption{Examples of images illustrating how framing affects product identification and resulting captions. In the Corn Pops and McCormick Great Guacamole examples, all VLMs fail to fill in the missing information needed for correct identification. The Honey Nut Cheerios example provides two alternate framings, with varying amounts of the text visible. Despite the cereal's mascot being visible on both, Llama and Molmo fail to correctly identify the product when more of the product text is hidden. Captions were shortened for presentation purposes only, indicated by [...].}
	\label{fig:faming-failure-examples}
	\resizebox{\linewidth}{!}{%
		\begin{tabular}{@{}p{0.20\linewidth}p{0.11\linewidth}*{4}{p{0.2925\linewidth}}@{}}
			\toprule
			\textbf{Image} & \textbf{Annotation} & \textbf{GPT} & \textbf{Gemini} & \textbf{Llama} & \textbf{Molmo} \tabularnewline \midrule
            % corn pops
			\parbox[t]{\linewidth}{\vspace{-0.75em}\includegraphics[width=\linewidth]{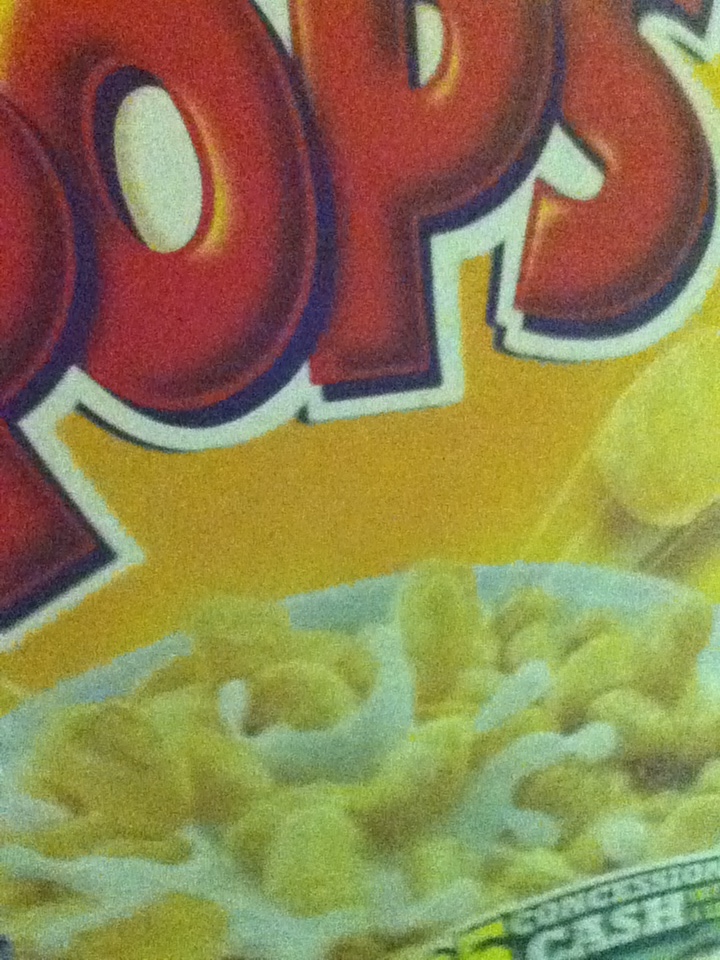}\Description{A zoomed-in picture of Kellogg's Corn Pops cereal. ``OPS'' are visible, but the first ``P'' is hidden. There is a bowl of cereal on the bottom of the box that is visible.}} 
			& \raggedright \textbf{\textcolor{col_product}{Product:\newline cereal\newline\newline}}\textbf{\textcolor{col_brand}{Brand:\newline Kellogg's\newline\newline}}\textbf{\textcolor{col_variety}{Variety:\newline Corn Pops}}
			& \raggedright \textbf{[Incorrect]} Cardboard box of breakfast \textbf{\textcolor{col_product}{cereal}}, likely corn flakes, with yellow flakes pictured in a bowl of milk and bold red and blue outlined letters on a yellow background.\newline\newline
			\textbf{Missing:}\newline
			- \textbf{\textcolor{col_brand}{brand: Kellogg's}}\newline
			- \textbf{\textcolor{col_variety}{variety: Corn Pops}}
			& \raggedright \textbf{[Incorrect]} A rectangular box of Cap'n Crunch's Crunch Berries \textbf{\textcolor{col_product}{cereal}}. The box is yellow and red with an image of crunchy \textbf{\textcolor{col_product}{cereal}} pieces.\newline\newline
			\textbf{Missing:}\newline
			- \textbf{\textcolor{col_brand}{brand: Kellogg's}}\newline
			- \textbf{\textcolor{col_variety}{variety: Corn Pops}}
			& \raggedright \textbf{[Incorrect]} The product is a box of Kraft macaroni and cheese. The box is yellow and features the brand name in red letters, with a black and white logo in the top-left corner. The background of the box is yellow, and it has a white border around the edges.\newline\newline
			\textbf{Missing:}\newline
			- \textbf{\textcolor{col_product}{product: cereal}}\newline
			- \textbf{\textcolor{col_brand}{brand: Kellogg's}}\newline
			- \textbf{\textcolor{col_variety}{variety: Corn Pops}}
			& \raggedright \textbf{[Incorrect]} Coffee-Mate French Vanilla coffee creamer bottle, white with yellow label, on yellow background. 32 fl oz (946 ml) size. Red lettering on left side. Partial yellow and black text on right, ``CASH'' visible. \newline\newline
			\textbf{Missing:}\newline
			- \textbf{\textcolor{col_product}{product: cereal}}\newline
			- \textbf{\textcolor{col_brand}{brand: Kellogg's}}\newline
			- \textbf{\textcolor{col_variety}{variety: Corn Pops}}
			\tabularnewline \midrule

            % mccormick great guac
			\parbox[t]{\linewidth}{\vspace{-0.75em}\includegraphics[width=\linewidth]{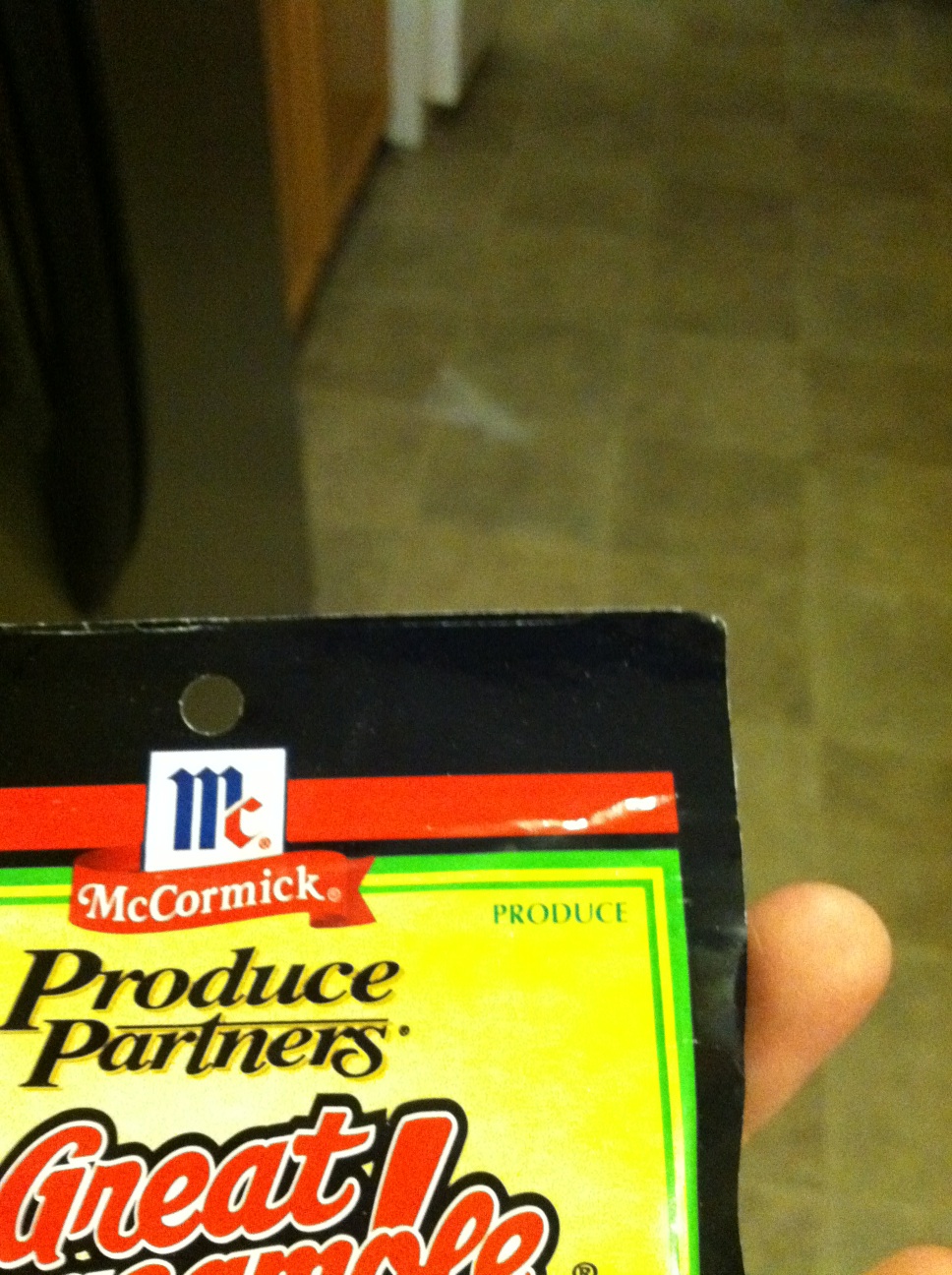}\Description{A packet of McCormick Produce Partners Great Guacamole seasoning mix. The McCormick logo, ``Produce Partners'', and ``great'' are all visible; only the bottom half of the word ``guacamole'' is visible.}} 
			& \raggedright \textbf{\textcolor{col_product}{Product: seasoning mix\newline\newline}}\textbf{\textcolor{col_brand}{Brand:\newline McCormick Produce Partners\newline\newline}}\textbf{\textcolor{col_variety}{Variety: Great Guacamole}}
			& \raggedright \textbf{[Incorrect]} Packet of \textbf{\textcolor{col_brand}{McCormick Produce Partners}} Great Sample \textbf{\textcolor{col_product}{seasoning mix}}, featuring a yellow label with black and red text, outlined in green and black.\newline\newline
			\textbf{Missing:}\newline
			- \textbf{\textcolor{col_variety}{variety: Great Guacamole}}
			& \raggedright \textbf{[Incorrect]} A rectangular pouch of \textbf{\textcolor{col_brand}{McCormick Produce Partners}} Great American Potato \textbf{\textcolor{col_product}{seasoning mix}}. The top is black, and a red banner with the white McCormick logo is below it. The main part of the pouch is yellow with green borders.\newline\newline
			\textbf{Missing:}\newline
			- \textbf{\textcolor{col_variety}{variety: Great Guacamole}}
			& \raggedright \textbf{[Incorrect]} The product is a bag of \textbf{\textcolor{col_brand}{McCormick's}} ``Great Sample'' \textbf{\textcolor{col_brand}{Produce Partners}} \textbf{\textcolor{col_product}{seasoning mix}}, featuring the brand name \textbf{\textcolor{col_brand}{``McCormick''}} and the product name \textbf{\textcolor{col_brand}{``Produce Partners''}} in black text on a yellow background. [...]\newline\newline
			\textbf{Missing:}\newline
			- \textbf{\textcolor{col_variety}{variety: Great Guacamole}}
			& \raggedright \textbf{[Incorrect]} \textbf{\textcolor{col_brand}{McCormick Produce Partners}} Great (partially visible) package. Black top, yellow background with green outline. Red and black text, McCormick logo with blue M and red C. Red banner with white text ``Produce Partners''. \newline\newline
			\textbf{Missing:}\newline
			- \textbf{\textcolor{col_product}{product: seasoning mix}}\newline
			- \textbf{\textcolor{col_variety}{variety: Great Guacamole}}
			\tabularnewline \midrule

            % cherrios 1
			\parbox[t]{\linewidth}{\vspace{-0.75em}\includegraphics[width=\linewidth]{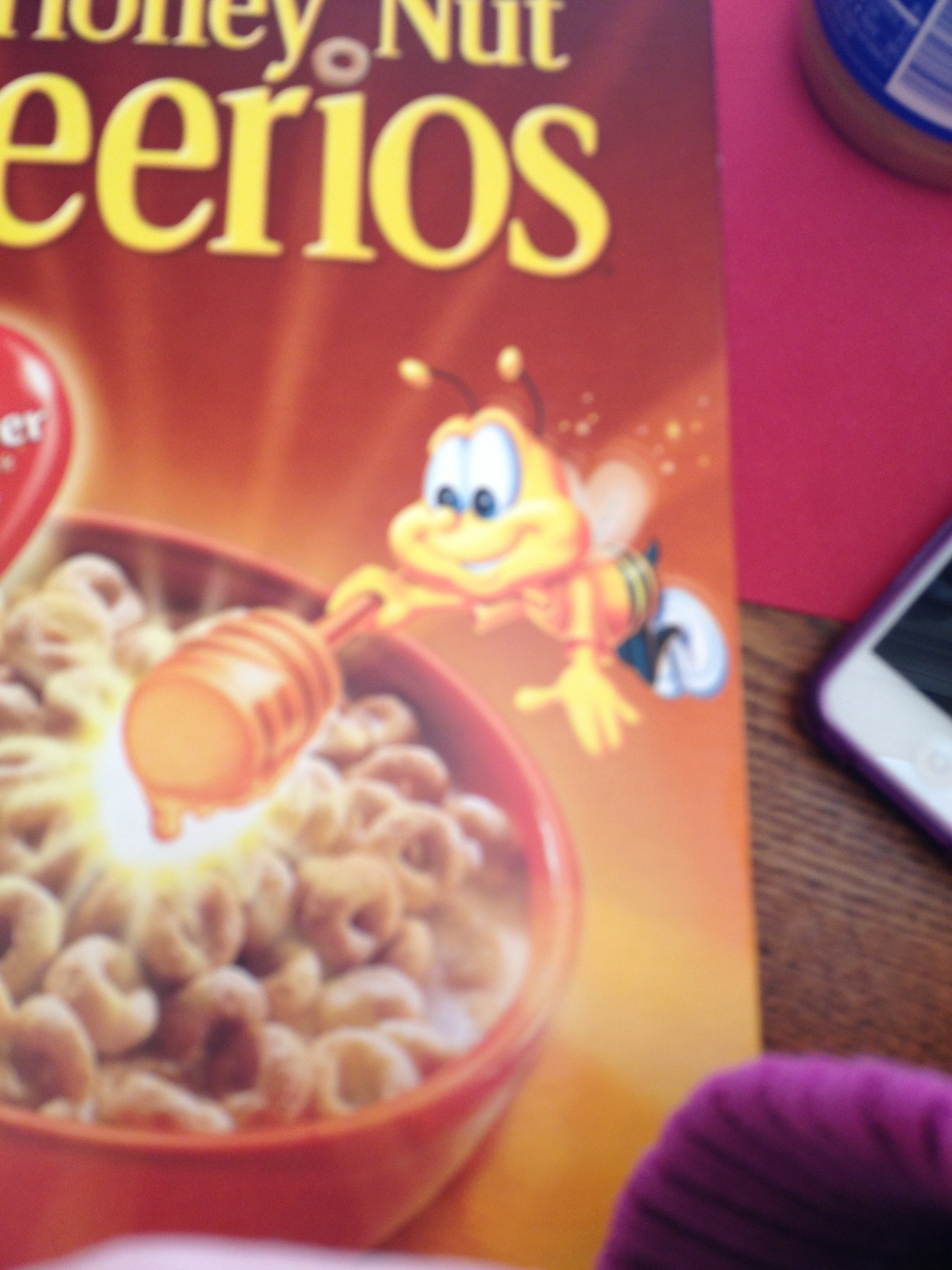}\Description{A box of Honey Nut Cheerios with the top left hidden. The bottom of ``Honey Nut'' is visible, as is ``eerios''. The brand's bee mascot is clearly visible in full.}} 
			& \raggedright \textbf{\textcolor{col_product}{Product:\newline cereal\newline\newline}}\textbf{\textcolor{col_brand}{Brand:\newline Cheerios\newline\newline}}\textbf{\textcolor{col_variety}{Variety:\newline Honey Nut}}
			& \raggedright \textbf{[Correct]} \textbf{\textcolor{col_product}{Cereal}} box, \textbf{\textcolor{col_variety}{Honey Nut}} \textbf{\textcolor{col_brand}{Cheerios}}, General Mills brand, \textbf{\textcolor{col_variety}{honey nut}} flavor, features a cartoon bee mascot and a bowl of cereal with honey being drizzled, predominantly orange and yellow color scheme.
			& \raggedright \textbf{[Correct]} A box of General Mills \textbf{\textcolor{col_variety}{Honey Nut}} \textbf{\textcolor{col_brand}{Cheerios}} \textbf{\textcolor{col_product}{cereal}}. The box is orange, with a cartoon bee character holding a honey dipper over a red bowl of cereal with milk.
			& \raggedright \textbf{[Correct]} The product is a red and orange \textbf{\textcolor{col_product}{cereal}} box featuring the \textbf{\textcolor{col_variety}{Honey Nut}} \textbf{\textcolor{col_brand}{Cheerios}} brand and variety. The box is adorned with a yellow bee character and contains a bowl of \textbf{\textcolor{col_product}{cereal}} with a honey dipper on top.
			& \raggedright \textbf{[Correct]} \textbf{\textcolor{col_product}{Cereal}} box:
			Type: Box
			Product: \textbf{\textcolor{col_product}{Cereal}}
			Brand: Kellogg's
			Variety: \textbf{\textcolor{col_variety}{Honey Nut Cheerios}}
			Shape: Rectangular
			Color scheme: Orange and yellow
			Prominent graphic: Honeybee character
			Main graphic: Bowl of \textbf{\textcolor{col_brand}{Cheerios}} cereal with milk, honey drizzled on top \tabularnewline \midrule

            % cherrios 2
			\parbox[t]{\linewidth}{\vspace{-0.75em}\includegraphics[width=\linewidth]{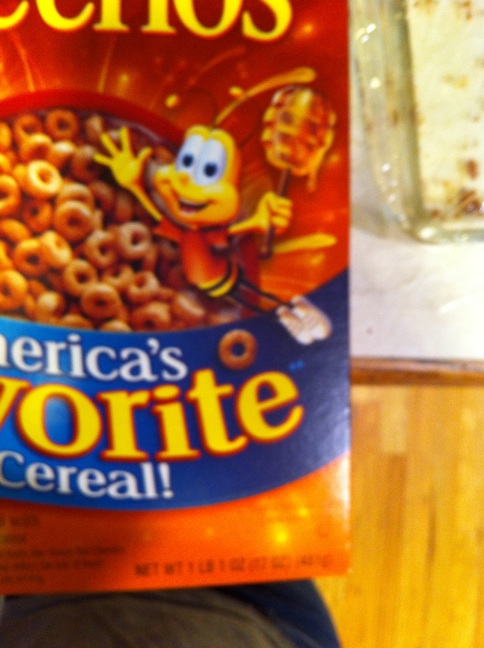}\Description{A box of Honey Nut Cheerios with the top half. Only the bottom of ``eerios'' is visible. The brand's bee mascot is clearly visible in full.}} 
			& \raggedright \textbf{\textcolor{col_product}{Product:\newline cereal\newline\newline}}\textbf{\textcolor{col_brand}{Brand:\newline Cheerios\newline\newline}}\textbf{\textcolor{col_variety}{Variety:\newline Honey Nut}}
			& \raggedright \textbf{[Correct]} Cardboard box of \textbf{\textcolor{col_variety}{Honey Nut}} \textbf{\textcolor{col_brand}{Cheerios}} \textbf{\textcolor{col_product}{cereal}} by General Mills, featuring the honey bee mascot holding a honey dipper, with round oat \textbf{\textcolor{col_product}{cereal}} pieces and text highlighting it as ``America's Favorite \textbf{\textcolor{col_product}{Cereal}}.'' The box has an orange and blue color scheme. Net weight is 1 lb 5 oz (595 g). 
			& \raggedright \textbf{[Correct]} \textbf{\textcolor{col_product}{Cereal}} box of \textbf{\textcolor{col_variety}{Honey Nut}} \textbf{\textcolor{col_brand}{Cheerios}}. The box is red with blue and yellow accents. It features a cartoon bee holding a honey dipper, and a bowl of cereal. The text ``America's Favorite \textbf{\textcolor{col_product}{Cereal}}!'' is written on the bottom. The net weight is 1 pound, 1 ounce (481 grams).
			& \raggedright \textbf{[Incorrect]} Product Description\newline
			- Object Type: Plastic container\newline
			- Product Type: Prepared meal\newline
			- Brand: Heinz\newline
			- Variety: Baked beans[...]
			\newline\newline
			\textbf{Missing:}\newline
			- \textbf{\textcolor{col_product}{product: cereal}}\newline
			- \textbf{\textcolor{col_brand}{brand: Cheerios}}\newline
			- \textbf{\textcolor{col_variety}{variety: Honey Nut}}
			& \raggedright \textbf{[Incorrect]} \textbf{\textcolor{col_product}{Cereal}} box: Kellogg's Froot Loops. Red background. Yellow text on blue banner: ``America's Favorite Cereal''. Animated bee character holding a Froot Loop. Visible net weight: 11oz. Standing on wooden floor.\newline\newline
			\textbf{Missing:}\newline
			- \textbf{\textcolor{col_brand}{brand: Cheerios}}\newline
			- \textbf{\textcolor{col_variety}{variety: Honey Nut}}
			\tabularnewline \bottomrule
		\end{tabular}%
	}
    \Description{Organized in six columns, separated by horizontal lines, the table presents four image examples, their annotation supplied by the researchers, and the outputs from four VLM models. The first column has a preview of the image. The second has annotations of products, including product, brand, and variety. The third through sixth include caption outputs from each VLM, with an indicator of whether it is correct and color coding for which annotations matched or were missed.  The four images in the table are as follows: (1) A zoomed-in picture of Kellogg's Corn Pops cereal. ``OPS'' are visible, but the first ``P'' is hidden. There is a bowl of cereal on the bottom of the box that is visible; (2) A packet of McCormick Produce Partners Great Guacamole seasoning mix. The McCormick logo, ``Produce Partners'', and ``great'' are all visible; only the bottom half of the word ``guacamole'' is visible; (3) A box of Honey Nut Cheerios with the top left hidden. The bottom of ``Honey Nut'' is visible, as is ``eerios''. The brand's bee mascot is clearly visible in full; (4) A box of Honey Nut Cheerios with the top half. Only the bottom of ``eerios'' is visible. The brand's bee mascot is clearly visible in full.}
\end{table*}
%TC:endignore

% rotated images
%TC:ignore
\begin{table*}[ht]
	\centering
	\Large
	\caption{Examples of images with rotation issues where different VLMs may only provide high-level information or incorrectly infer what the product is. Captions were shortened for presentation purposes only, indicated by [...].}
	\label{fig:rotation-failure-examples}
	\resizebox{\linewidth}{!}{%
		\begin{tabular}{@{}p{0.20\linewidth}p{0.11\linewidth}*{4}{p{0.2925\linewidth}}@{}}
			\toprule
			\textbf{Image} & \textbf{Annotation} & \textbf{GPT} & \textbf{Gemini} & \textbf{Llama} & \textbf{Molmo} \tabularnewline  \midrule 
			
			% deli napoli
			\parbox[t]{\linewidth}{\vspace{-0.75em}\includegraphics[width=\linewidth]{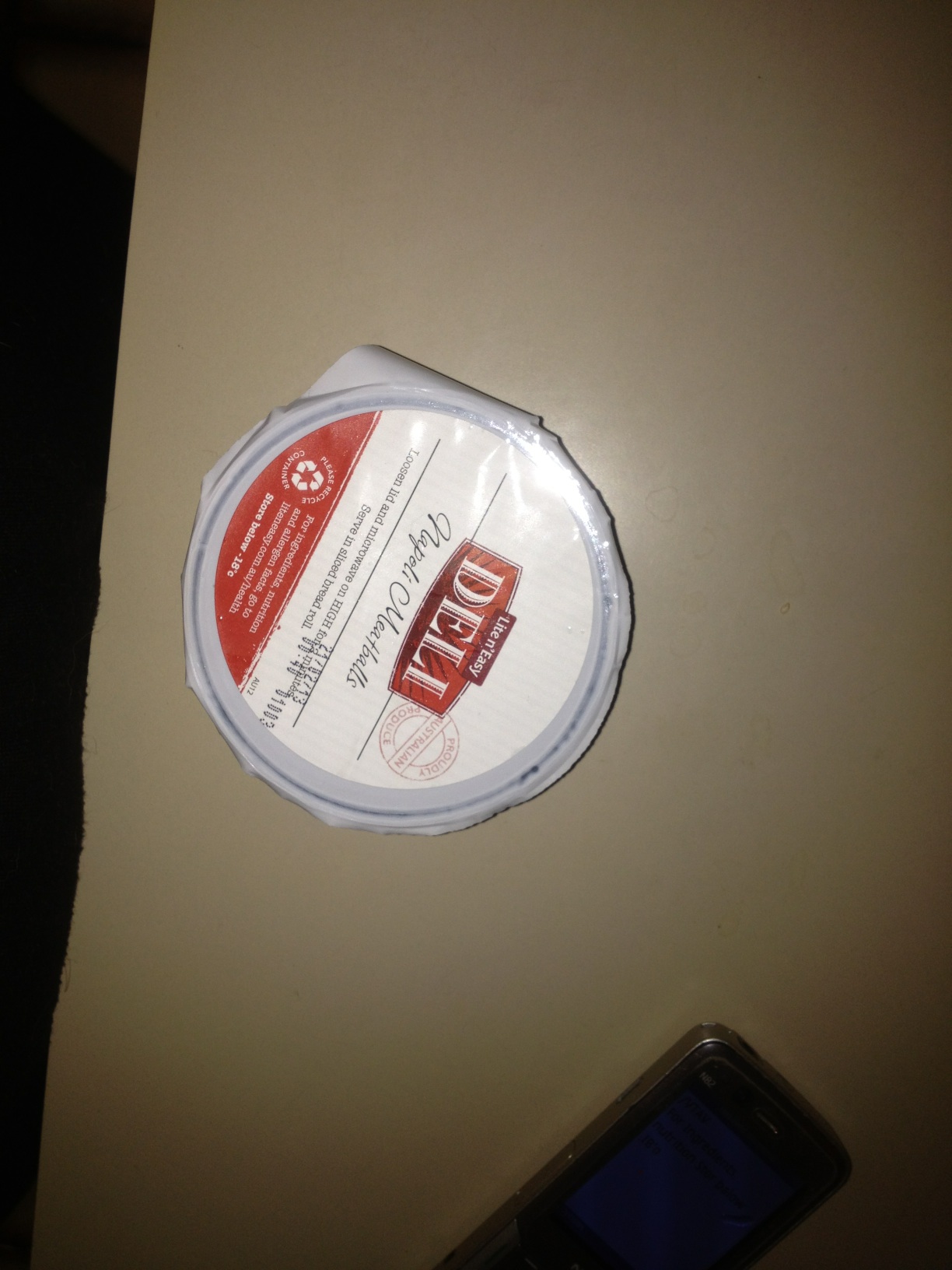}\Description{A container of deli Napoli meatballs from Lite 'n Easy.}} 
			& \raggedright \textbf{\textcolor{col_product}{Product: Deli Napoli Meatballs\newline \newline\newline}}\textbf{\textcolor{col_brand}{Brand:\newline Lite n' Easy\newline\newline}}
			& \raggedright \textbf{[Incorrect]} Round plastic container of DEKA brand Angel Cookies. The lid is white with a red and white label that features the brand logo and product name. \newline\newline
			\textbf{Missing:}\newline
			- \textbf{\textcolor{col_product}{product: Deli Napoli Meatballs}} \newline
			- \textbf{\textcolor{col_brand}{brand: Lite n' Easy}}
			& \raggedright \textbf{[Incorrect]} A round, plastic, white container of \textbf{\textcolor{col_product}{Deli Napoli Meatballs}}, 100g. The lid is white with a red stripe across the top and the words ``Deli Easy'' in white font. Below that, the words ``Napoli Meatballs'' are written in black cursive font. The bottom of the lid has a red recycling symbol and the words ``Store below 4°C.''\newline\newline
			\textbf{Missing:}\newline
			- \textbf{\textcolor{col_brand}{brand: Lite n' Easy}}
			& \raggedright \textbf{[Incorrect]} This is a plastic container of meatballs from the brand IDEAL, with a white label and red accents. The label features black text, including the product name ``Meatballs'' and a handwritten note. The container has a round shape and is placed on a beige surface, accompanied by a black object in the bottom right corner. \newline\newline
			\textbf{Missing:}\newline
			- \textbf{\textcolor{col_product}{product: Deli Napoli Meatballs}} \newline
			- \textbf{\textcolor{col_brand}{brand: Lite n' Easy}}
			& \raggedright \textbf{[Incorrect]} Kraft Philadelphia cream cheese container, 8 oz.\newline\newline
			\textbf{Missing:}\newline
			- \textbf{\textcolor{col_product}{product: Deli Napoli Meatballs}} \newline
			- \textbf{\textcolor{col_brand}{brand: Lite n' Easy}}
			\tabularnewline  \midrule
			
			% smooth n juicy
			\parbox[t]{\linewidth}{\vspace{-0.75em}\includegraphics[width=\linewidth]{images/study-2/rotation/smoothnjuicy.jpg}\Description{A carton of Dawn Smooth 'n Juicy juice.}} 
			& \raggedright \textbf{\textcolor{col_product}{Product:\newline juice\newline\newline}}\textbf{\textcolor{col_brand}{Brand:\newline Dawn\newline\newline}}\textbf{\textcolor{col_variety}{Variety: Smooth 'n Juicy\newline }}
			& \raggedright \textbf{[Incorrect]} Carton of Dalin \textbf{\textcolor{col_variety}{Smooth n' Juicy}} orange \textbf{\textcolor{col_product}{juice}}, featuring an image of orange slices and juice splash on a blue and orange background.
			\newline\newline
			\textbf{Missing:}\newline
			- \textbf{\textcolor{col_brand}{brand: Dawn}}
			& \raggedright \textbf{[Incorrect]} A blue and orange \textbf{\textcolor{col_product}{juice}} box is oriented vertically.[...] The juice box is Dairymen brand orange juice. The top of the juice box is blue and has a logo in white that says Dairymen. [...] Beneath the logo, the words ``\textbf{\textcolor{col_variety}{Smooth 'n Juicy}}'' are written in yellow and orange. [...]\newline\newline
			\textbf{Missing:}\newline
			- \textbf{\textcolor{col_brand}{brand: Dawn}}
			& \raggedright \textbf{[Incorrect]} The product is a rectangular-shaped carton of orange \textbf{\textcolor{col_product}{juice}} with the top-left corner of the carton facing the viewer. [...] The brand name ``Dair'' is written in white text on the top-right corner of the carton, while the product name ``\textbf{\textcolor{col_variety}{Smooth 'n Juicy}}'' is displayed in yellow and white 3D-style text on the center-left side. [...]\newline\newline
			\textbf{Missing:}\newline
			- \textbf{\textcolor{col_brand}{brand: Dawn}}
			& \raggedright \textbf{[Incorrect]} A blue rectangular container of orange \textbf{\textcolor{col_product}{juice}} smoothie. The container displays an image of oranges and has ``Smooth'' in white text and ``Juice'' in yellow text with a drop shadow. The brand ``Dewon'' is written in cursive script, with the ``W'' featuring a leaf design. The container is placed at an angle on a light brown-orange countertop.
			\newline\newline
			\textbf{Missing:}\newline
			- \textbf{\textcolor{col_brand}{brand: Dawn}}\newline
			- \textbf{\textcolor{col_variety}{variety: Smooth 'n Juicy}}
			\tabularnewline \bottomrule
		\end{tabular}%
	}
	\Description{Organized in six columns, separated by horizontal lines, the table presents two image examples, their annotation supplied by the researchers, and the outputs from four VLM models. The first column has a preview of the image. The second has annotations of products, including product, brand, and variety. The third through sixth include caption outputs from each VLM, with an indicator of whether it is correct and color coding for which annotations matched or were missed. The two images in the table are as follows: (1) A container of deli Napoli meatballs from Lite 'n Easy; and (2) A carton of Dawn Smooth 'n Juicy juice.}
\end{table*}
%TC:endignore

% co-occurring
%TC:ignore    
\begin{table*}[ht]
	\centering
	\Large
	\caption{Examples of images with multiple issues (blur, framing, and rotation) with variations in product captions. Two slightly different frames of the same product, a package of Mucinex Expectorant Maximum Strength medication, are shown along with generated captions. Captions were shortened for presentation purposes only, indicated by [...].}
	\label{fig:cooccurrance-failure-examples-mucinex}
	\resizebox{\linewidth}{!}{%
		\begin{tabular}{@{}p{0.20\linewidth}p{0.11\linewidth}*{4}{p{0.2925\linewidth}}@{}}
			\toprule
			\textbf{Image} & \textbf{Annotation} & \textbf{GPT} & \textbf{Gemini} & \textbf{Llama} & \textbf{Molmo} \tabularnewline \midrule 
			% mucinex 1
			\parbox[t]{\linewidth}{\vspace{-0.75em}\includegraphics[width=\linewidth]{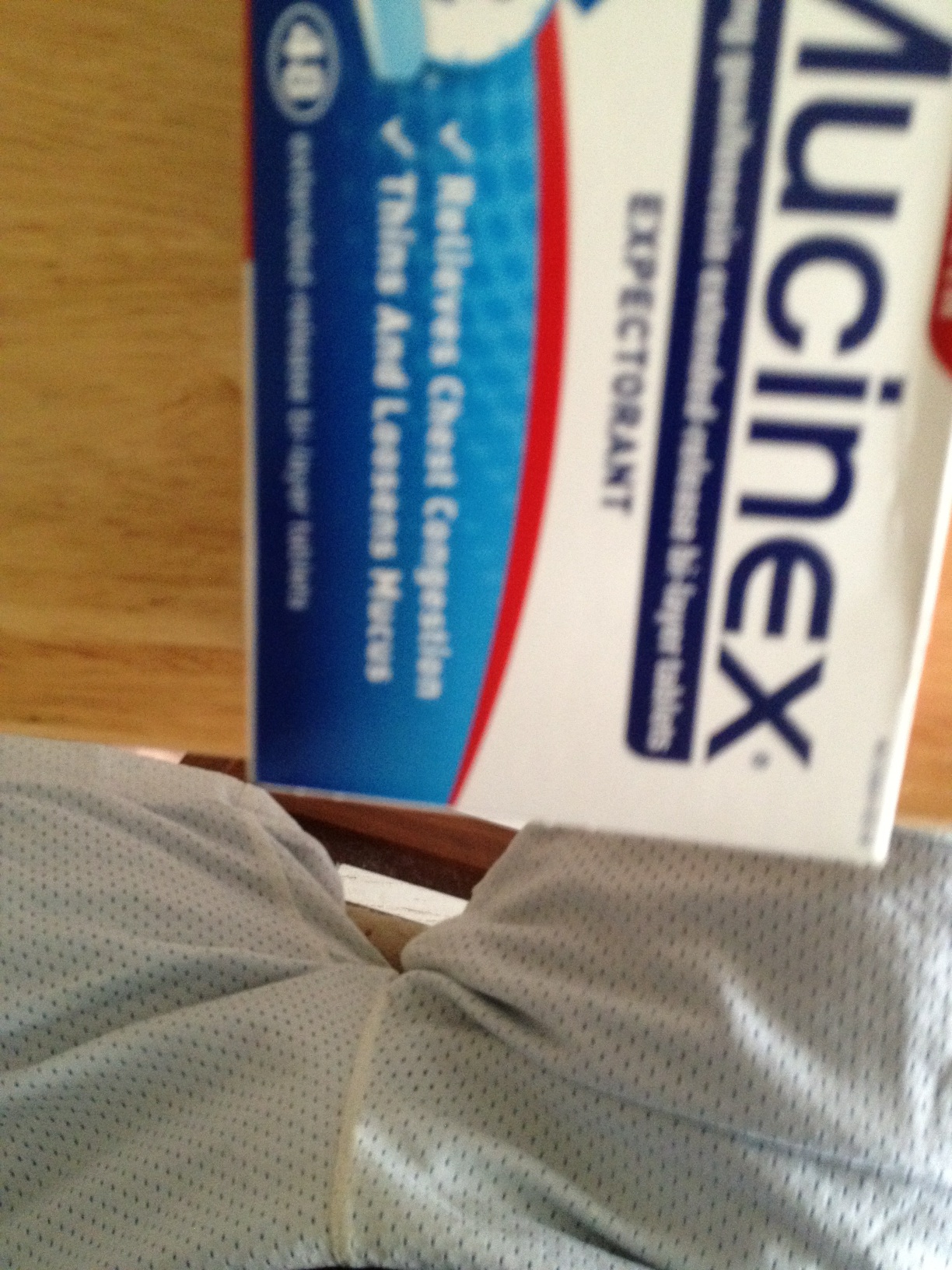}\Description{A box of Mucinex Expectorant, maximum strength, with the M slightly out of frame.}} 
			& \raggedright \textbf{\textcolor{col_product}{Product: expectorant}}\newline\newline \textbf{\textcolor{col_brand}{Brand:\newline Mucinex\newline\newline}}\textbf{\textcolor{col_variety}{Variety: maximum strength}}
			& \raggedright \textbf{[Incorrect]} Box of \textbf{\textcolor{col_brand}{Mucinex}} \textbf{\textcolor{col_product}{expectorant}} tablets, blue and white packaging, designed to relieve chest congestion, thin and loosen mucus, and intended for adults and children over 12 years.
			\newline\newline
			\textbf{Missing:}\newline
			- \textbf{\textcolor{col_variety}{variety: maximum strength}}
			& \raggedright \textbf{[Incorrect]} A rectangular box of \textbf{\textcolor{col_brand}{Mucinex}} \textbf{\textcolor{col_product}{expectorant}}, extended-release bi-layer tablets. The box is white with the ``Mucinex'' brand name printed in blue letters. It contains 40 tablets and lists ``Relieves chest congestion'' and ``Thins and loosens mucus'' in blue text on a blue background on the side of the box.
			\newline\newline
			\textbf{Missing:}\newline
			- \textbf{\textcolor{col_variety}{variety: maximum strength}}
			& \raggedright \textbf{[Incorrect]} The product is a white box with a red stripe and blue accents, featuring the brand name ``Aucinex'' in dark blue text. [...] along with additional text in smaller white letters on a blue background. The background of the image is a light-colored wood surface, likely a table or countertop.
			\newline\newline
			\textbf{Missing:}\newline
			- \textbf{\textcolor{col_product}{product: expectorant}}\newline
			- \textbf{\textcolor{col_brand}{brand: mucinex}}\newline
			- \textbf{\textcolor{col_variety}{variety: maximum strength}}
			& \raggedright \textbf{[Incorrect]} Aucinex \textbf{\textcolor{col_product}{expectorant}} box. It's a medication product, likely used for respiratory conditions. The box has a white background with a blue strip and red curvy line. The brand name ``Aucinex'' is visible in blue text.
			\newline\newline
			\textbf{Missing:}\newline
			- \textbf{\textcolor{col_brand}{brand: mucinex}}\newline
			- \textbf{\textcolor{col_variety}{variety: maximum strength}}
			\tabularnewline \midrule
			
			% mucinex 2
			\parbox[t]{\linewidth}{\vspace{-0.75em}\includegraphics[width=\linewidth]{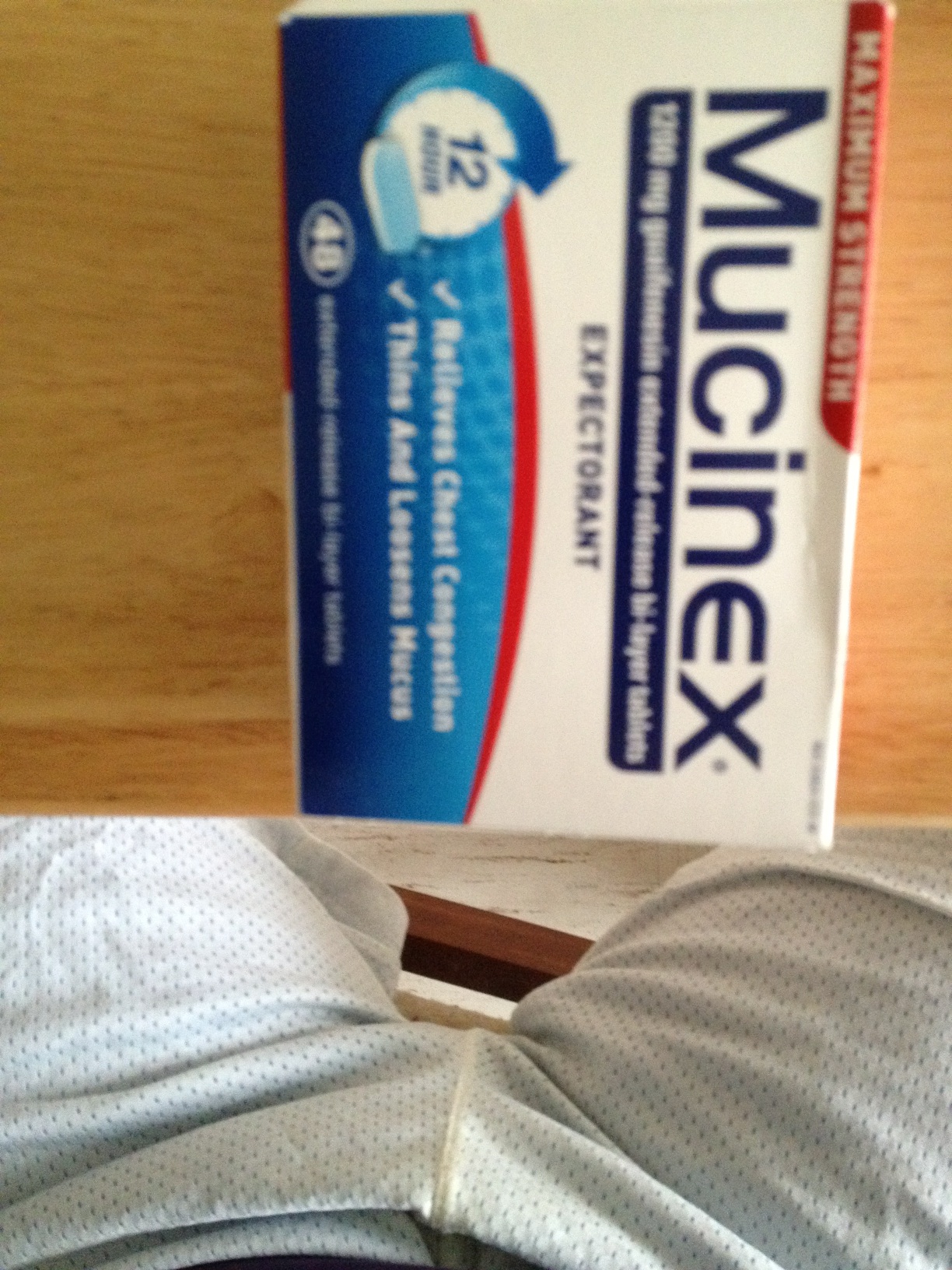}\Description{A box of Mucinex Expectorant, maximum strength, with the M more visible.}} 
			& \raggedright \textbf{\textcolor{col_product}{Product: expectorant}}\newline\newline \textbf{\textcolor{col_brand}{Brand:\newline Mucinex\newline\newline}}\textbf{\textcolor{col_variety}{Variety: maximum strength}}
			& \raggedright \textbf{[Correct]} Box of \textbf{\textcolor{col_brand}{Mucinex}} \textbf{\textcolor{col_variety}{Maximum Strength}} \textbf{\textcolor{col_product}{Expectorant}}, 1200 mg guaifenesin extended-release bi-layer tablets, labeled for 12-hour relief, with 14 tablets indicated on the blue section, primarily white packaging with blue and red accents.
			& \raggedright \textbf{[Correct]} A white and blue box of \textbf{\textcolor{col_variety}{Maximum Strength}} \textbf{\textcolor{col_brand}{Mucinex}} \textbf{\textcolor{col_product}{Expectorant}} contains 1200 mg guaifenesin extended-release bi-layer tablets. [...] A white-outlined circle with a blue arrow pointing rightward and the words ``12 HOUR'' [...] It reads: ``Relieves Chest Congestion'' and ``Thins And Loosens Mucus.'' Below this in white, it states: ``10 extended-release bi-layer tablets.''
			& \raggedright \textbf{[Incorrect]} The product is a box of \textbf{\textcolor{col_brand}{Mucinex}} \textbf{\textcolor{col_product}{Expectorant}} tablets. The box is white with blue and red accents, featuring the brand name ``Mucinex'' in large blue letters on the front. [...] \newline\newline
			\textbf{Missing:}\newline
			- \textbf{\textcolor{col_variety}{variety: maximum strength}}
			& \raggedright \textbf{[Correct]} Boxed \textbf{\textcolor{col_product}{expectorant}} medication. Brand: \textbf{\textcolor{col_brand}{Mucinex}}. Variety: \textbf{\textcolor{col_variety}{Maximum Strength}} 12-Hour Extended Release Chest Congestion. packaging features red and blue color scheme on a white background.\tabularnewline
						
			\bottomrule
		\end{tabular}%
	}
	\Description{Organized in six columns, separated by horizontal lines, the table presents two image examples, their annotation supplied by the researchers, and the outputs from four VLM models. The first column has a preview of the image. The second has annotations of products, including product, brand, and variety. The third through sixth include caption outputs from each VLM, with an indicator of whether it is correct and color coding for which annotations matched or were missed. The two images in the table are as follows: (1) A box of Mucinex Expectorant, maximum strength, with the ``M'' slightly out of frame; and (2) A box of Mucinex Expectorant, maximum strength, with the ``M'' more visible.}
\end{table*}
%TC:endignore

% text and rounded panels
%TC:ignore    
\begin{table*}[ht]
	\centering
	\Large
	\caption{Examples of images with text panels (rows 1--2) and rounded labels (rows 3--4). VLMs can read text panels, but often incorrectly or miss key information. Rounded objects often obscure the label, requiring more inference about the product, which humans do well, but VLMs still struggle with. Captions were shortened for presentation purposes only, indicated by [...].}
	\label{fig:text-panels-rounded-labels}
	\resizebox{\linewidth}{!}{%
		\begin{tabular}{@{}p{0.20\linewidth}p{0.11\linewidth}*{4}{p{0.2925\linewidth}}@{}}
			\toprule
			\textbf{Image} & \textbf{Annotation} & \textbf{GPT} & \textbf{Gemini} & \textbf{Llama} & \textbf{Molmo} \tabularnewline \midrule 
			
			% o organics milk
			\parbox[t]{\linewidth}{\vspace{-0.75em}\includegraphics[width=\linewidth]{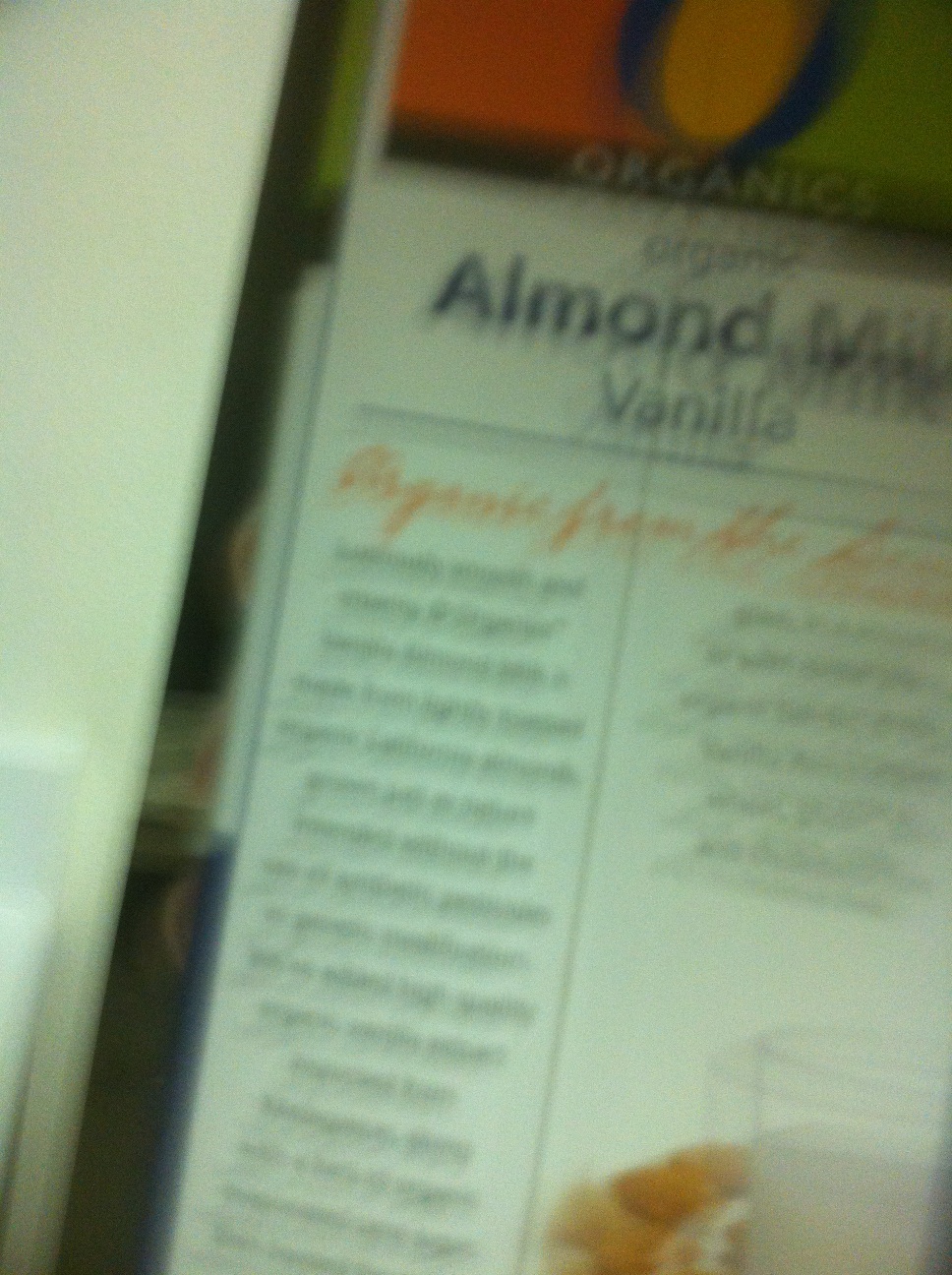}\Description{A carton of O Organics vanilla-flavored almond milk, with the side of the carton shown. The top of the carton is cropped such that ``O'' from O Organics is only half visible.}} 
			& \raggedright \textbf{\textcolor{col_product}{Product:\newline almond milk\newline\newline}}\textbf{\textcolor{col_brand}{Brand:\newline O Organics\newline\newline}}\textbf{\textcolor{col_variety}{Variety:\newline vanilla}}
			& \raggedright \textbf{[Correct]} Carton of \textbf{\textcolor{col_brand}{O Organics}} \textbf{\textcolor{col_product}{Almond Milk}}, \textbf{\textcolor{col_variety}{Vanilla variety}}. The package includes the \textbf{\textcolor{col_brand}{O Organics}} brand logo and text indicating it is an organic \textbf{\textcolor{col_product}{almond milk}} product with \textbf{\textcolor{col_variety}{vanilla flavor}}. The carton features a combination of white and other colors, with descriptive product information on the side.
			& \raggedright \textbf{[Incorrect]} A rectangular white carton contains organic \textbf{\textcolor{col_variety}{vanilla}} \textbf{\textcolor{col_product}{almond milk}}. The carton has a light brown banner with the words \textbf{\textcolor{col_product}{``Almond Milk''}} in white sans-serif font. Below the banner, the word \textbf{\textcolor{col_variety}{``Vanilla''}} is printed in white text. At the top of the carton, an orange and blue logo is visible, with the word ``ORGANICS'' printed underneath.  \newline\newline
			\textbf{Missing:}\newline
			- \textbf{\textcolor{col_brand}{brand: O Organics}}
			& \raggedright \textbf{[Incorrect]} The image is out of focus, but it appears to depict a product label for \textbf{\textcolor{col_variety}{vanilla}} \textbf{\textcolor{col_product}{almond milk}}. [...] It features a rectangular shape with a prominent orange banner at the top, displaying the brand name ``Almond Milk'' in large black letters. Below the banner, the flavor \textbf{\textcolor{col_variety}{``Vanilla''}} is written in smaller black text. [...] \newline\newline
			\textbf{Missing:}\newline
			- \textbf{\textcolor{col_brand}{brand: O Organics}}
			& \raggedright \textbf{[Incorrect]} \textbf{\textcolor{col_product}{Almond milk}} \textbf{\textcolor{col_variety}{vanilla}} carton, possibly by Horizon Organic. White carton with black text and color accents.  \newline\newline
			\textbf{Missing:}\newline
			- \textbf{\textcolor{col_brand}{brand: O Organics}}
			\tabularnewline \midrule
			            
			% lean ground beef
			\parbox[t]{\linewidth}{\vspace{-0.75em}\includegraphics[width=\linewidth]{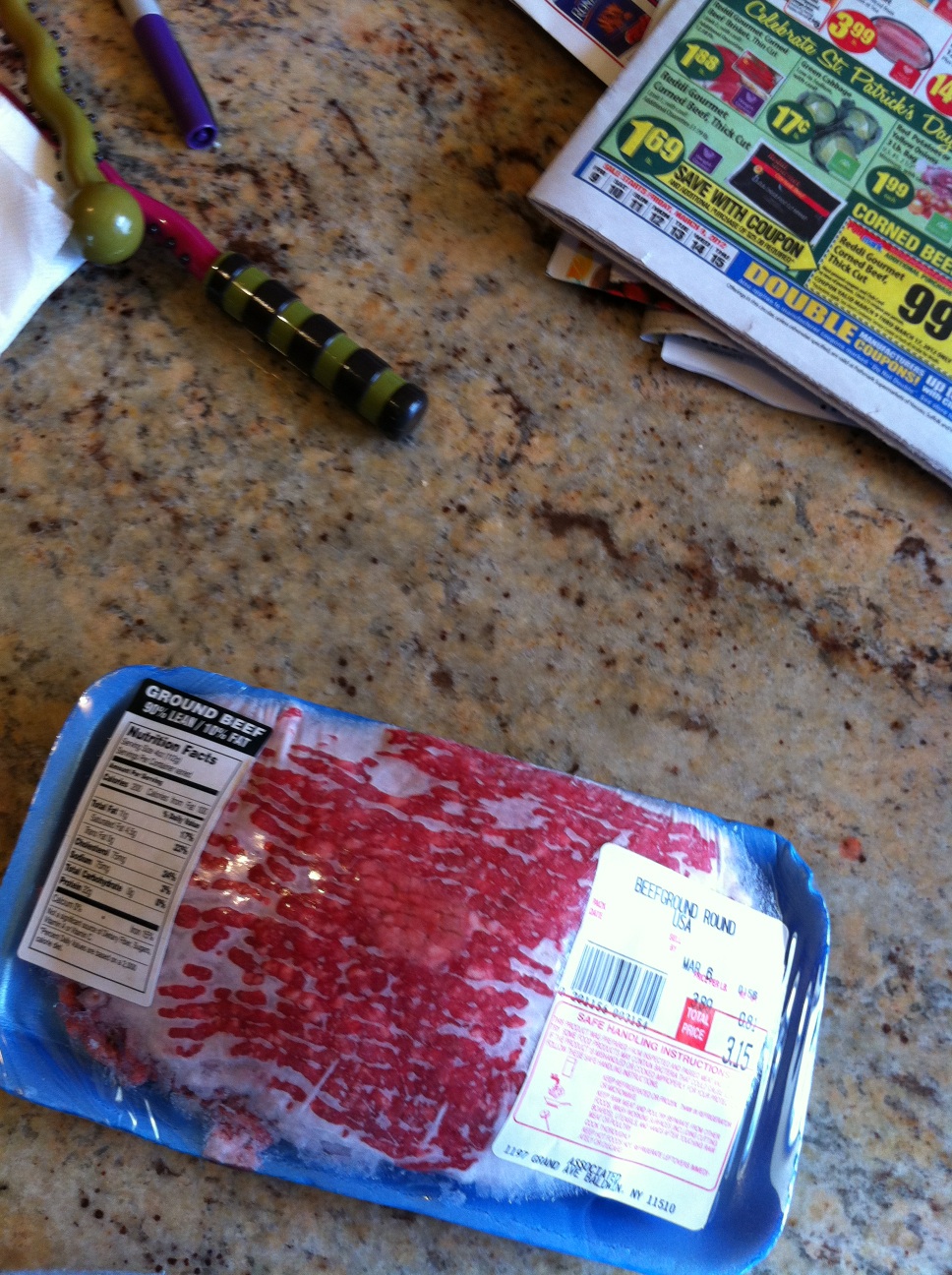}\Description{A plastic-wrapped package of ground beef, 90\% fat and 10\% lean.}} 
			& \raggedright \textbf{\textcolor{col_product}{Product:\newline ground beef\newline\newline}}\textbf{\textcolor{col_variety}{Variety:\newline 90\% lean; 10\% fat}}
			& \raggedright \textbf{[Incorrect]} Plastic-wrapped foam tray of \textbf{\textcolor{col_product}{ground beef}}, labeled as ``Ground Beef 93\% Lean 7\% Fat'' and ``Beef Ground Round USA.'' [...] \newline\newline
			\textbf{Missing:}\newline
			- \textbf{\textcolor{col_variety}{variety: 90\% lean; 10\% fat}}
			& \raggedright \textbf{[Incorrect]} \textbf{\textcolor{col_product}{Ground beef}}, 96\% lean/4\% fat, in a plastic and Styrofoam tray with a Nutrition Facts label and safe handling instructions.
			\newline\newline
			\textbf{Missing:}\newline
			- \textbf{\textcolor{col_variety}{variety: 90\% lean; 10\% fat}}
			& \raggedright \textbf{[Incorrect]} The product is a package of \textbf{\textcolor{col_product}{ground beef.}} The package is blue and rectangular with a clear plastic window showing the red ground beef inside. [...] \newline\newline
			\textbf{Missing:}\newline
			- \textbf{\textcolor{col_variety}{variety: 90\% lean; 10\% fat}}
			& \raggedright \textbf{[Incorrect]} \textbf{\textcolor{col_product}{Ground beef}} in blue plastic container with clear plastic top. Nutrition facts label on left side shows 93\% lean/7\% fat. [...] \newline\newline
			\textbf{Missing:}\newline
			- \textbf{\textcolor{col_variety}{variety: 90\% lean; 10\% fat}}
			\tabularnewline \midrule
			
			% canned sloppy joe sauce
			\parbox[t]{\linewidth}{\vspace{-0.75em}\includegraphics[width=\linewidth]{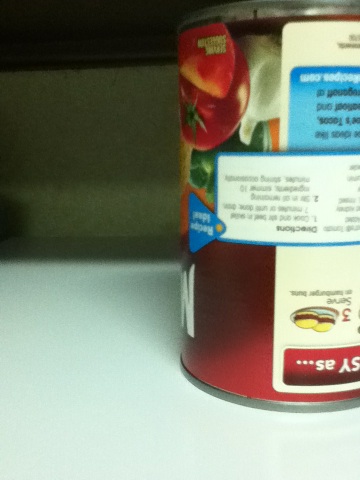}\Description{An upside-down can of sloppy joe sauce from Mancich. Only part of the ``M'' is visible. A picture of prepared sloppy joes is visible.}} 
			& \raggedright \textbf{\textcolor{col_product}{Product:\newline sloppy joe sauce \newline\newline}}\textbf{\textcolor{col_brand}{Brand:\newline manwich\newline\newline}}
			& \raggedright \textbf{[Incorrect]} Metal can of what appears to be a tomato-based product, featuring an image of a tomato and other vegetables on the label. [...] The specific brand and variety are not fully visible.
			\newline\newline
			\textbf{Missing:}\newline
			- \textbf{\textcolor{col_product}{product: sloppy joe sauce}}\newline
			- \textbf{\textcolor{col_brand}{brand: Manwich}}
			& \raggedright \textbf{[Incorrect]} A red metal can of Campbell's brand condensed tomato soup. The label features a large, realistic image of a red tomato.
			\newline\newline
			\textbf{Missing:}\newline
			- \textbf{\textcolor{col_product}{product: sloppy joe sauce}}\newline
			- \textbf{\textcolor{col_brand}{brand: Manwich}}
			& \raggedright \textbf{[Incorrect]} The product is a can of tomato sauce. The can is red with a blue and white label. The label has a picture of a tomato on it. The brand name is not visible. [...] \newline\newline
			\textbf{Missing:}\newline
			- \textbf{\textcolor{col_product}{product: sloppy joe sauce}}\newline
			- \textbf{\textcolor{col_brand}{brand: Manwich}}
			& \raggedright \textbf{[Incorrect]} Small metal can of Hunt's tomato sauce. White label with red color block. Recipes on label for dishes using product. Image of red tomatoes on label.
			\newline\newline
			\textbf{Missing:}\newline
			- \textbf{\textcolor{col_product}{product: sloppy joe sauce}}\newline
			- \textbf{\textcolor{col_brand}{brand: Manwich}}
			\tabularnewline \midrule
						
			% canned sliced peaches
			\parbox[t]{\linewidth}{\vspace{-0.75em}\includegraphics[width=\linewidth]{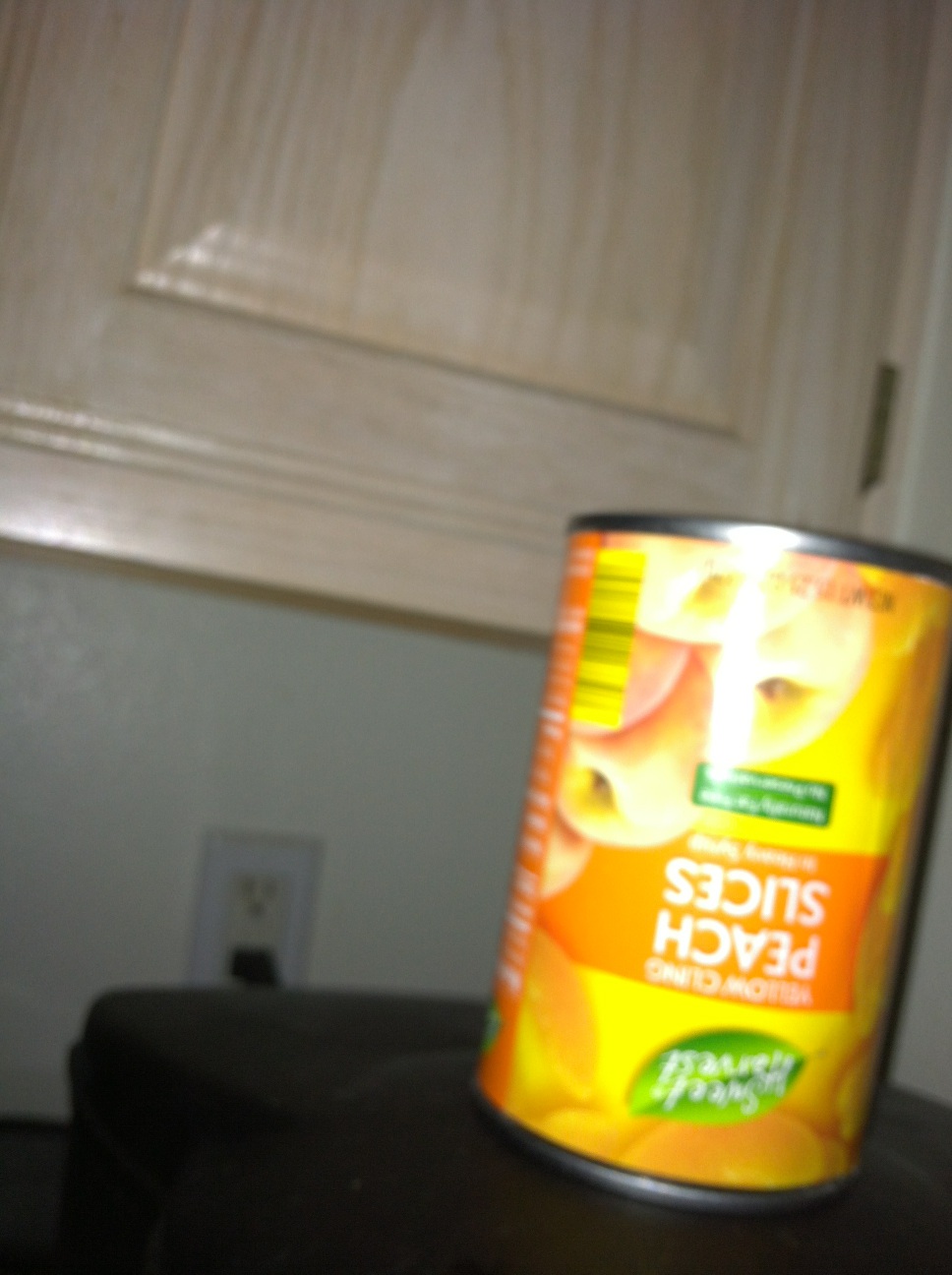}\Description{An upside-down can of sliced peaches in heavy syrup from Sweet Harvest. The brand and label are clearly visible.}} 
			& \raggedright \textbf{\textcolor{col_product}{Product:\newline peach; slices\newline\newline}}\textbf{\textcolor{col_brand}{Brand:\newline Sweet Harvest\newline\newline}}\textbf{\textcolor{col_variety}{Variety:\newline heavy syrup}}
			& \raggedright \textbf{[Correct]} Metal can of \textbf{\textcolor{col_brand}{Sweet Harvest}} brand yellow cling \textbf{\textcolor{col_product}{peach slices}} in \textbf{\textcolor{col_variety}{heavy syrup}}. The can has an orange and yellow label with images of peach slices and green accents.
			& \raggedright \textbf{[Correct]} A can of \textbf{\textcolor{col_brand}{Sweet Harvest}} Yellow Cling \textbf{\textcolor{col_product}{Peach Slices}} in \textbf{\textcolor{col_variety}{heavy syrup}}. The can is primarily yellow and orange, with a picture of peach slices at the top. The product name is displayed in white text against an orange background.
			& \raggedright \textbf{[Incorrect]} Product Description\newline
			- Object Type: Can\newline
			- Product Type: Prepared meal\newline
			- Brand: Great Value\newline
			- Variety: \textbf{\textcolor{col_product}{Peach Slices}} in Syrup\newline
			- Visual Features: Yellow and orange can with a green leaf logo on the front.\newline\newline
			\textbf{Missing:}\newline
			- \textbf{\textcolor{col_brand}{brand: Sweet Harvest}}\newline
			- \textbf{\textcolor{col_variety}{variety: heavy syrup}}
			& \raggedright \textbf{[Incorrect]} A yellow can of Del Monte Sweet Peas. \newline\newline
			\textbf{Missing:}\newline
			- \textbf{\textcolor{col_product}{product: peach; slices}}\newline
			- \textbf{\textcolor{col_brand}{brand: Sweet Harvest}}\newline
			- \textbf{\textcolor{col_variety}{variety: heavy syrup}}
			\tabularnewline \bottomrule
		\end{tabular}%
	}
    \Description{Organized in six columns, separated by horizontal lines, the table presents four image examples, their annotation supplied by the researchers, and the outputs from four VLM models. The first column has a preview of the image. The second has annotations of products, including product, brand, and variety. The third through sixth include caption outputs from each VLM, with an indicator of whether it is correct and color coding for which annotations matched or were missed. The four images in the table are as follows: (1) A carton of O Organics vanilla-flavored almond milk, with the side of the carton shown. The top of the carton is cropped such that ``O'' from O Organics is only half visible; (2) A plastic-wrapped package of ground beef, 90\% fat and 10\% lean; (3) An upside-down can of sloppy joe sauce from Mancich. Only part of the ``M'' is visible. A picture of prepared sloppy joes is visible; (4) An upside-down can of sliced peaches in heavy syrup from Sweet Harvest. The brand and label are clearly visible.}
\end{table*}
%TC:endignore
\section{Image Captioning Prompt for All VLMs}
\label{appendix:captioning-prompt}
You are a helpful assistant who identifies products in images for blind and low-vision individuals. Identify the product in the image while following these guidelines:
\begin{enumerate}
    \item Identify crucial features about the product, including:
    \begin{enumerate}
        \item Object type (can, bag, plastic container, etc.)
        \item Product type (prepared or frozen meal, seasoning mix, soda, coffee)
        \item Brand (Heinz, Coca-Cola, Starbucks, etc.)
        \item Variety (specific flavors, sizes, count of items, etc.)
        \item Visual features (color, shape, size, etc.)
    \end{enumerate}
    \item Use clear, direct, and objective language. Do not use vague adjectives like `large' or `small', or vague adverbs like `prominently' or `clearly'.
    \item DO NOT mention camera artifacts (e.g., blur) or if an object is partially visible.
    \item DO NOT use introductory phrases (e.g., `The image shows', `The object is', `The primary object is').
\end{enumerate}
%TC:endignore

\end{document}